\documentclass[a4paper]{article}

 \usepackage[sort&compress,numbers,colon,merge]{natbib}		
 
 \usepackage[margin=2cm]{geometry}	

 \usepackage{hyperref}
 \hypersetup{
  colorlinks   = true, 	
  urlcolor     = blue, 	
  linkcolor    = blue, 	
  citecolor   = red 		
}

\usepackage{amsmath}		
\usepackage{amssymb}	
\usepackage{slashed}
\usepackage{bbold}		
\usepackage{relsize }		

\usepackage{fixltx2e}		

\usepackage{graphicx}		
\usepackage[font=small,labelfont=bf]{caption}
\usepackage{subcaption}	

\title{Electroweak Dark Matter}
\author{Ramtin Amintaheri
\footnote{Ramtin.Amintaheri@gmail.com}}	
\date{\normalsize \emph{School of Physics, Physics Road, The University of Sydney, NSW 2006 Camperdown, Australia.}}

\begin{document}

\maketitle

\begin{abstract}
In the absence of any hints of new physics in LHC, TeV dark matter candidates interacting through electroweak force (EWDM) are still highly motivated. We extend the Standard Model by adding an arbitrary SU(2) DM multiplet in non-chiral representation. In addition to the well-known real representation which has no coupling to the nuclei at tree level, the complex representation can lead to a new DM candidate providing that one includes a higher dimensional mass-splitting operator, which survives the current direct detection bounds. Since the masses of gauge mediators are light compared to the dark particles, Sommerfeld effect is dominant and affects the value of annihilation cross-section in both the early universe and current time. We computed the relic abundance through freeze-out mechanism in order to determine DM mass. Gamma ray fluxes in our galaxy and dwarf satellites provide a promising tool to probe EWDM theory. We confronted the four fermionic representations of the model with the latest astrophysical observations. It can be concluded that the model passes the current experimental constraints successfully, and it is accessible to future observations. 
\end{abstract}


\section{Introduction}

A range of observational evidence indicate existence of anomalies at galactic and cosmic scales. One of the most compelling explanations is the presence of huge amount of non-luminous \emph{dark matter} (DM) \cite{DM1, DM2} constituting about 83\% of the total mass of the universe \cite{Planck}. However, the Standard Model of particles (SM) fails to provide any justification about the nature of DM candidate and its interactions with other known particles \cite{BSM}. 

The only remaining possibility to describe interactions of DM using one of the known subatomic forces, is through couplings mediated by the weak force. In this case, there would be no need to introduce gauge groups that are new to physics, and our theory requires changes with minimal impact on the SM.This idea is well motivated by the so-called \emph{weak miracle} \cite{WIMP1, WIMP2}. Based on the measured value of DM abundance, if dark matter is thermally produced via \emph{freeze-out mechanism}, then its  thermal cross-section should be appropriate for pair annihilation via the weak force for a particle with a mass around the weak scale.

We extend the SM by adding one fermionic multiplet that is invariant under electroweak gauge symmetry $SU(2)_L \times U(1)$, and refer to it as electroweak dark matter (EWDM). In the fermionic version of EWDM theory all free parameters are fixed unambiguously through observations which means that the phenomenological predictions of this model can be tested accurately by dark matter experiments.

Extensions of the SM involving an additional weak multiplet have been studied in a range of contexts such as supersymmetry \cite{Susy_DM}, little Higgs model \cite{Little_Higgs}, inert Higgs models \cite{Inert_2et, Inert_3et}, neutrino mass generation \cite{SeesawIII}, Kaluza-Klein theory \cite{KK1},  etc. These extensions mostly contain lower dimensional representations of SU(2) doublet and triplet. However, larger multiplets have been proposed in recent works such as inert Higgs doublet-septuplet \cite{Inert_2et_7et}, exotic Higgs quintuplet \cite{Higgs_5et}, quintets in neutrino mass mechanism \cite{neutrino_5et_1, neutrino_5et_2}, and finally fermionic quintuplets and scalar septuplet in minimal dark matter model. The latter remains stable due to an accidental symmetry in the SM gauge groups and Lorentz representations leaving no renormalisable decay mode for DM \cite{MDM, MDM_2009}.

Past literature has mostly focus on the real representations of SU(2) electroweak gauge group probably because the phenomenology is more straightforward for a Majorana DM candidate. In contrast to the pseudo-real formalism where Dirac DM couples to the neutral Z boson at tree-level. This results in a an elastic cross-section with target nuclei which is a few order of magnitude above the current Direct Detection (DD) constraints and the model is therefore experimentally ruled out \cite{PandaX_2017, LUX_2017}. We recover the pseudo-real representations, with special emphasis on the least-known case of quadruplet module, by introducing an effective operator that breaks the U(1)\textsubscript{D} symmetry of the dark sector after electroweak symmetry breaking (EWSB). The extra higher-dimensional operator decomposes the neutral Dirac DM state into two Majorana fermions which are not allowed to have a vector coupling to Z gauge field so that the coherent scattering with nucleon is avoided.

The interesting point is that even if such model of electroweak dark matter is ruled out, it would be equally important, since this would imply existence of a new gauge group, and hence a fifth force of nature.

Indirect Detection (ID) aims at searching for by-products of DM decay or annihilation in a diverse range of astrophysical targets. Among these products, gamma rays benefit from some unique features that make them a suitable choice to study DM. Light rays are not deflected by magnetic fields unlike charged particles, and therefore can assist in identifying location of the emission source. In addition, photon radiation have negligible energy attenuation, so the original spectrum information remains intact. These properties make it possible to look for signatures of DM beyond the proximal objects, in the distant sources like satellite galaxies and clusters.

In this paper, we provide accurate phenomenological analysis of indirect signals of EWDM theory. Latest observational constraints have been used to test viability of the models in a variety of astronomical targets including the Milky Way (MW) black hole, continuum emission from inner galaxy, dwarf satellite and spectral lines.

This article is organised as follows:

In the next section, we provide a generalised framework to assess whether dark matter particles can couple to the SM weak boson, and introduce the electroweak theory of dark matter. The chiral and non-chiral including real and pseudo-real representations of such DM theory are reviewed. It will be explained that, as a general rule, any pseudo-real model can avoid bounds from direct detection experiments, by including a dimension five operator that splits the neutral Dirac state into two Majorana components.

In high-mass scales, the perturbative quantum field formalism beaks down due to long-range forces mediated by weak vectors. In section 3, the Sommerfeld enhancement of annihilation amplitudes due to non-perturbative effects will be studied. We compute the complete set of formulae for available fermionic EWDM potentials and annihilation cross-sections in the following subsections.

The detailed cosmological and astrophysical constraints imposed on electroweak multiplets are discussed in section 4. In first subsection, we compute the cosmological abundance by solving the effective Boltzmann equation. The current measured yield fixes the last unknown parameter that is DM mass. In the remainder of this paper, we compare the predicted continuum and monochromatic components of the spectrum with the most recent data from different observatories, and discuss the experimental constraints on electroweak theory of dark matter.


\section{EWDM Model}

In general, EWDM could be a fermion or scalar multiplet. In this paper, we consider extension of the SM by adding an arbitrary fermionic n-tuplet charged under $SU(2) \times U(1)_y$.


\subsection{Chiral Dark Matter}
\label{sec:Chiral}

In general, a gauge theory is called \emph{chiral}, if the fields with different chirality are not charged the same under the symmetry group. Since the left-handed and the right handed fermions transform differently under SU(2) gauge group, the electroweak interactions of the Standard Model of particles are described by a chiral theory. 

Electroweak precision measurements highly constrain an additional exotic chiral particle. \emph{Peskin-Takeuchi parameters}, namely, $S$, $T$, and $U$ describe the effect of new physics on electroweak radiative corrections. The S parameter accounts for the difference in number of the left-handed and right-handed fermions with weak isospin $t^{(3)}_{L/R}$ \cite{STU90, STU92}, and for a mass-degenerate n-tuplet yields:
\begin{equation}
	S = \frac{1}{3\pi}	\sum_{i=1}^n 	\left( t^{(3)}_{L,i} - t^{(3)}_{R,i} \right)^2
\end{equation}

 It puts tight constrains on the number of extra generation of chiral fermions. Based on \emph{S parameter}, the number of fermionic families corresponds to $N_F = 2.75 \pm 0.14$, and accordingly a new generation of chiral fermions is disfavoured at 9-$\sigma$ confidence level \cite{PDG}.

In addition to this, direct detection experiments almost rule out the chiral dark matter candidate. That is due to DM vector coupling to the neutral weak gauge field at tree-level so that it can scatter coherently off the nuclei by Z-boson exchange. The resultant cross-section is so large that such DM particles would already have been detected. To understand the reason, we need to take a closer look at DM-nucleon interaction. Assuming scattering amplitude $\mathcal{M}$ has spherical symmetry at low energies, the elastic scattering cross section is given by \cite{Witten_DM}:
\begin{equation}
	\frac{1}{\pi} m_{\chi N}^2 |\mathcal{M}|^2
\end{equation}

with $m_{\chi N}= m_\chi m_N / (m_\chi + m_N)$ being DM-nucleus reduced mass. In the non-relativistic limit, weak scattering amplitude reduces to \cite{Witten_DM}:
\begin{equation}
	\mathcal{M}= 4\sqrt{2} G_F J_\chi^0 J_N^0
\end{equation}

where $G_F$ is Fermi coupling constant, and $J_\chi^0$ and $J_N^0$ are temporal components of the neutral weak current for DM and nucleus respectively. For a chiral fermion DM with $y_{L/R}$ left/right hypercharge  $J_\chi^0  = (y_L +y_R)/4$, and for a nucleus having $Z$ protons and $\mathcal{N}$ neutrons $J^0_N = [\mathcal{N}- (1-4 s^2_w)Z]/4$ 
\footnote{We use the notation $s_w \equiv \sin\theta_w$, $c_w \equiv \cos\theta_w$, and $t_w \equiv \tan\theta_w$ where $\theta_w$ is the Weinberg mixing angle.}. 
So the cross section per nucleon for a nucleus of atomic number $A$ becomes:
\begin{equation}
	\label{eq:DM_N}
	\sigma_{\chi N} =	 \frac{1}{8\pi}	\frac{ G_F^2 m_{\chi N}^2 }{A^2}	\left[ \mathcal{N}- (1-4 s^2_w)Z \right]^2	(y_L +y_R)
\end{equation}

The term proportional to $Z$ accounts for proton-DM cross-section, and is suppressed as $1-4 s^2_w$ is negligible. The other term proportional to $\mathcal{N}$ expresses DM scattering off the neutron, and is the dominant term in the cross-section. For a Xenon atom ${}_{54}^{127} Xe$ which is widely used in DD experiments, and large enough dark matter mass of above 10 GeV, the cross-section is about $\sigma \approx 10^{-40}$ cm\textsuperscript{2}. Based on the current experimental results, it is clear that
chiral dark matter is ruled out \cite{XENON_2018}.

Direct detection experiments still allow room for smaller cross-section of a lighter chiral DM, below GeV scale. Nevertheless, because of tree-level coupling, Z-boson could easily decay to the light dark matter. Since dark matter is a neutral weakly interacting particles, it cannot be detected in a collider detector. But imbalance of total momentum (energy) known as missing transverse momentum (energy) can imply the presence of such invisible particles. Current measurements of Z-boson invisible decay width, has ruled out the light chiral dark matter scenario \cite{CERN}.


\subsection{Non-chiral Dark Matter}

\emph{Non-chiral} or \emph{vector} fermion has the property that its right and left chirality components transform in the same way under gauge symmetry \cite{Vector1, Vector2}. As a result a gauge-invariant mass term $\bar{\chi}\chi$ is allowed. Their masses are unbounded as the mass is not obtained through EWSB mechanism \cite{Vector3}. Their coupling to the electroweak bosons $Z$ and $W^\pm$ are purely vector $\bar{\chi}\gamma^\mu\chi$ hence having the left and right currents on an equal footing.

Vector fermions are not subject to bounds resulting from the electroweak precision data. A degenerate non-chiral multiplet gives no contribution to the value of $S$, $T$ and $U$ oblique parameters \cite{STU_multiplet}. 

The multiplets are labelled by dimension of the representation and the hypercharge $(n,y)$. 
Considering the normalisation convention $q= y +t^{(3)}$ for Gell-Mann–Nishijima relation, the hyper-charge is the offset of the electric charge from the range of the weak isospin values. 
So the electric charge of the $i^\mathrm{th}$ element in the multiplet reads:
\begin{equation}
	q_i= \frac{1}{2} (n+1)-i+y
\end{equation}

Since dark matter is electrically neutral, then its weak isospin must have the same value as the hyper charge $-t^{(3)}_0 =y$. 
Therefore a generic gauge-invariant n-tuplet takes the form: 
\begin{equation}
	\label{eq:nplet}
	X=\left( \chi^{t+y}, \ldots \chi^q,\dots \chi^0, \ldots \chi^{-t+y} \right)^T
\end{equation}

where $\chi^q$ denotes the element with electric charge $q$. The $\frac{1}{2}(n+1)+y$ neutral component $\chi^0$ is the actual dark matter candidate.%
\footnote{It can be seen that the element n+1-i, in the multiplet with hypercharge -y, has the opposite charge of -(n+1)/2+i-y. So, the opposite hypercharge multiplet can be considered as the reversed order multiplet with elements having opposite electric charges. As an example compare the dark matter quadruplet 
$\left( \chi^{++}, \chi^+, \chi^0, \chi^- \right)^T$
with hypercharge 3/2 with the one 
$\left( \chi^+, \chi^0, \chi^-, \chi^{--} \right)^T$ 
with $y=-3/2$. This becomes a useful property when the conjugate multiplet is defined in \ref{eq:Xc} as the later is proportional to the conjugate of the former.}

This means that for representations of odd dimension, $y$ takes integer values; while even-dimensional multiplets have half-integer hypercharge.
In addition, the total number of n-tuplets containing a DM candidate equals the dimension of the representation $n=2t+1$ where $t$ is the highest isospin weight.%
\footnote{For example a DM quadruplet with highest weight $t=3/2$ could refer to four possible multiplets 
$\left( \chi^{+++}, \chi^{++}, \chi^+, \chi^0 \right)^T$, 
$\left( \chi^{++}, \chi^+, \chi^0, \chi^- \right)^T$, 
$\left( \chi^+, \chi^0, \chi^-, \chi^{--} \right)^T$, and 
$\left( \chi^0, \chi^-, \chi^{--}, \chi^{---} \right)^T$ 
with hypercharges $y=$ 3/2, 1/2, -1/2 and -3/2 respectively.}

The Standard Model is believed to be a self-consistent description of the physics up to the \emph{Plank scale} $M_{pl}$. As a result the gauge couplings need to remain perturbative up to that cut-off scale. Adding extra SU(2) multiplets will accelerate running of these marginal couplings to the non-perturbative regime which might lead into appearance of \emph{Landau pole} (LP) before the Plank scale. Landau pole is thought to be associated with some new physics mechanisms that violate accidental symmetries of the SM. So we demand LP to be below $M_{pl}$ which will lead into an upper bound on dimensionality of the fermionic EWDM multiplet to $n \le 5$. This result holds true for renormalisation group equations solved up to two-loop level \cite{MDM, RGE}.

Due to different theoretical properties and phenomenological consequences, non-chiral dark matter is usually classified into real and complex representations. We intend to keep focus of this discussion on the pseudo-real models. Extra information about the real modules are included in the Appendix \ref{app:R}.


\subsection{Pseudo-real Representation}

For DM in pseudo-real representation of $SU(2) \times U(1)_y$, the hypercharge is non-vanishing $y \neq 0$. In this case, all components of the multiplet including the neutral DM are Dirac fermions.

The SM is extended by adding the dark sector Lagrangian:
\begin{align}
\label{eq:L_C}
	\mathcal{L}_D 	&= \overline{X} \left( i \slashed{D} - m \right) X	
				-\frac{\lambda_c}{\Lambda} ( H^\dagger \mathcal{T}^a H) (\overline{X}T^a X)	\\
				& -\frac{\lambda_0}{\Lambda} ( H^\dagger \mathcal{T}^a H^\mathfrak{C}) (\overline{X^\mathfrak{C}}T^a X )
					- \frac{\lambda_0^*}{\Lambda} (H^{\mathfrak{C}\dagger} \mathcal{T}^a H) (\overline{X} T^a X^\mathfrak{C})		
							\nonumber
\end{align}

where $\Lambda$ is the mass scale, $\lambda_0$ and $\lambda_c$ are coefficients, and $T^a$ and $\mathcal{T}^a=\sigma^a /2$ are SU(2) generators in the n-dimensional and fundamental representations respectively.
The covariant derivative reads 
$ D_\mu =	\partial_\mu 	+ iy g_y B_\mu 	+ i g_w	W_\mu ^a T^a$.

The conjugate representation is defined as $X^\mathfrak{C} \equiv C X^c$ where $X^c$ denotes the multiplet with charge conjugated fields. $C$ is an anti-symmetric off-diagonal matrix with alternating $\pm1$ entries, normalised so that it equals $-i\sigma^2$ in 2 dimensions i.e. 
$H^\mathfrak{C} = -i\sigma^2 H^*$. 
More precisely:
\begin{equation}
	C_{i,j} \equiv \delta_{i, n+1-j}	\ (-1)^{\frac{n+1}{2} -i -y}
\end{equation}

So for the generic multiplet \ref{eq:nplet} the conjugate multiplet reads:
\begin{equation}
	\label{eq:Xc}
	X^\mathfrak{C}= \left( (-1)^{-t+y} (\psi^{-t+y})^c, \ldots 	(\psi^0)^c, \ldots 	(-1)^q (\psi^q)^c,\dots 	 (-1)^{t+y} (\psi^{t+y})^c \right)^T
\end{equation}

Note that the scalar product of the adjoint vectors in the the second line, will preserve electric charge only for hyper-charge $y=\frac{1}{2}$, (c.f. equation \ref{eq:neutral}).%
\footnote{ In addition to this, note that the mass-splitting term proportional to $\lambda_c$ will vanish for the real DM. So, it can be concluded that the pseudo-real Lagrangian  \ref{eq:L_C} will identically reduce to the real Lagrangian \ref{eq:L_R}.}
This also means that such an operator only exists  for even-dimensional multiplets. 
It can also be seen that the theory initially preserves a $U(1)_D$ symmetry.
\footnote{ Such $U(1)_D$ or $Z_2$ invariances can potentially originate from some unknown fundamental local gauge symmetry masqueraded as discrete global symmetries to an observer probing at low energies \cite{Discrete_Sym}.
These discrete symmetries provide interesting phenomenological applications, for example in supersymmetric extensions of the SM with $Z_n$ generalised matter parities and $R$-parity \cite{Zn_SUSY1, Zn_SUSY2, R_SUSY} or grand unified theories with an extra U(1) symmetry which guarantees the stability \cite{Z_GUT, U1_GUT}.}


\subsubsection{Mass splitting}

\begin{figure} [t] 				
	\begin{subfigure}{.5\linewidth}
		\centering
		\includegraphics[width=\linewidth]{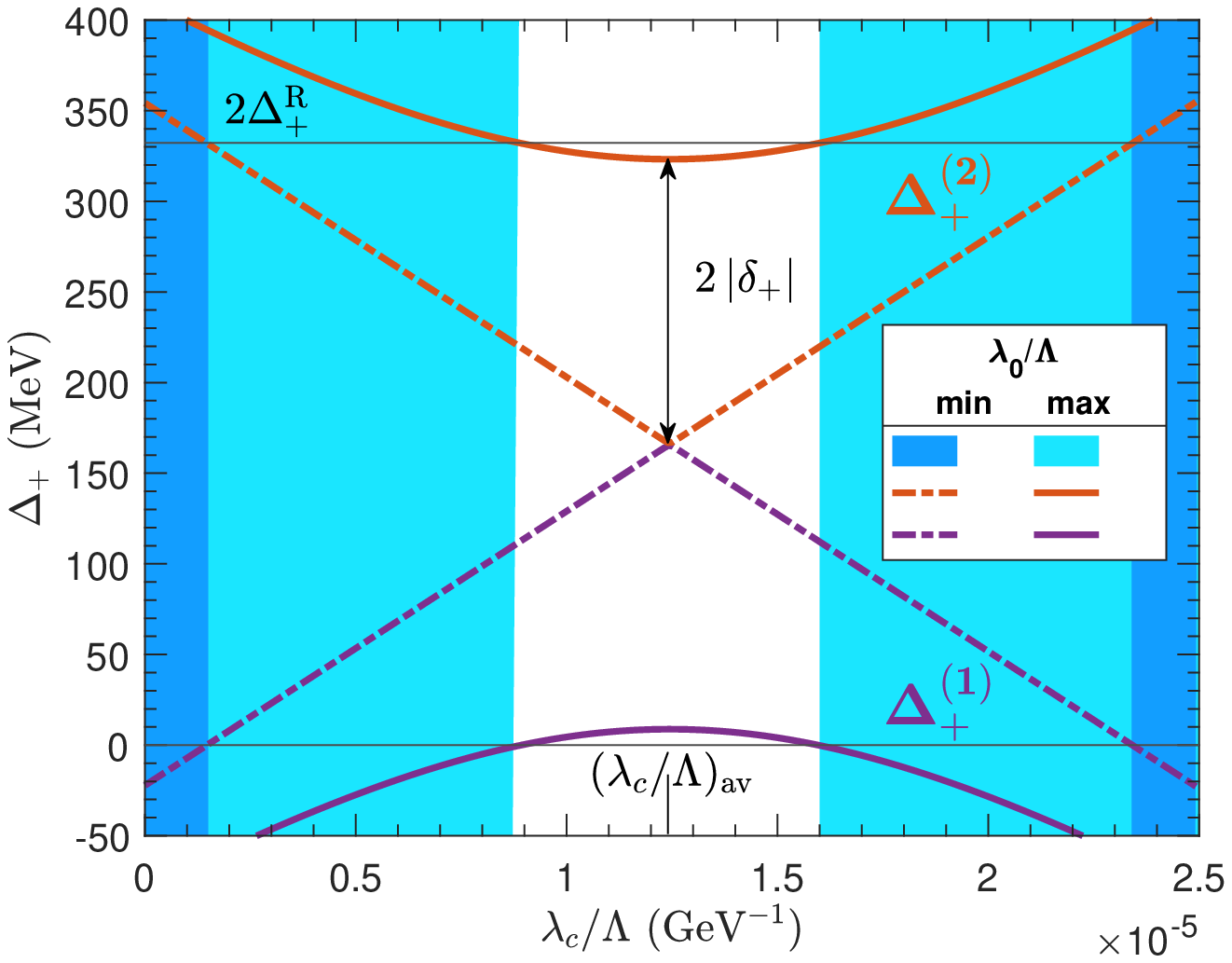}
		\subcaption{}
		\label{fig:Dm_lc}
	\end{subfigure}%
	\begin{subfigure}{.5\linewidth}
		\centering
		\includegraphics[width=\linewidth]{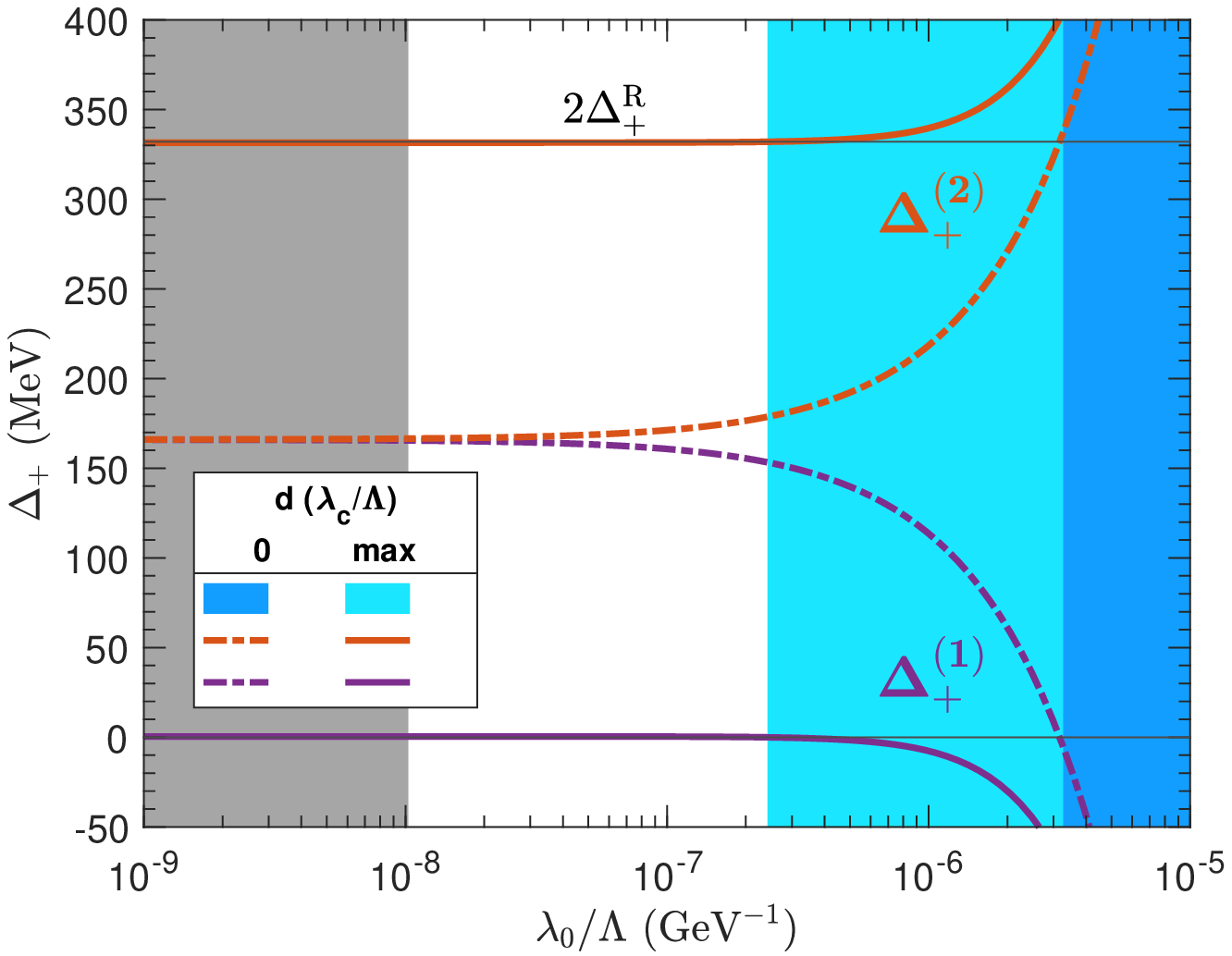}
		\subcaption{}
		\label{fig:Dm_l0}
	\end{subfigure}
	\caption{Left \ref{fig:Dm_lc}: Mass splitting between the dark matter candidate $\chi^0$ and the singly charged states $\chi^+_1$ (in blue) and $\chi^+_2$ (in red) as a function of the non-renormalisable coupling $\lambda_c/\Lambda$ for the complex quadruplet. The figure shows mass-splitting for two extremum values of the other non-renormalisable coefficient $\lambda_0/\Lambda$. The curves corresponding to the maximum value of 
$(\lambda_0/\Lambda)_\mathrm{max} = 3\times10^{-6} \,\mathrm{GeV}^{-1}$
\ref{eq:l0_Max} are in solid lines, while those at the minimum 
$(\lambda_0 /\Lambda)_\mathrm{min} = 10^{-8} \, \mathrm{GeV^{-1}}$ 
\ref{eq:l0_min} are in plotted dashed-dotted style. The light (dark) blue shaded regions are excluded at maximum (minimum) value of $\lambda_0/\Lambda$ as the theory cannot be considered as a dark mater model 
$\Delta_+^{(1)} <0$.\\
Right \ref{fig:Dm_l0}: The right panel illustrates the mass splitting of these fields $\Delta_+^{(1)}$ (blue) and $\Delta_+^{(2)}$ (red) with respect to the effective coupling $\lambda_0/\Lambda$ for the same quadruplet. The dashed-dotted curves show the result in case the other coupling has maximum deviation from the mean value 
$\mathrm{d}(\lambda_c/\Lambda)_\mathrm{max} = 1.1\times10^{-5} \,\mathrm{GeV}^{-1}$ 
\ref{eq:lc}, and the curves in solid line correspond to the average value of 
$(\lambda_c / \Lambda)_\mathrm{av}	=	1.2\times10^{-5} \mathrm{GeV}^{-1}$ 
\ref{eq:lc}. 
In the light (dark) blue coloured region, the $\chi^+_1$ field becomes the lightest particle $\Delta_+^{(1)} <0$, at maximum (zero) deviation of $\mathrm{d}(\lambda_c/\Lambda)$.The grey shaded area is ruled out by DD experiments due to inelastic scattering of DM off target nucleus via exchange of $Z$ boson at tree-level \ref{eq:l0_min}. \\
Note that in both figures, the blue coloured regions still remain a valid beyond the Standard Model (BSM) proposal. }
	\label{fig:Dm}
\end{figure}

The effective operator proportional to $\lambda_c$ is responsible for mass splitting between the neutral and charged components at tree-level. After electroweak symmetry breaking,  $H\to \left(0, (h+\nu)/\sqrt{2}) \right)^T$, the mass splitting can be cast as:
\begin{equation}
	\Delta_q^{(\mathrm{t})}	\equiv	m_q^{(\mathrm{t})} -m_0^{(\mathrm{t})}		=	- \frac{\lambda_c}{4\Lambda} \nu^2 q
\end{equation}

where $m_q^{(\mathrm{t})}$ indicates the tree-level mass of the component $\psi^q$, and $m_0^{(\mathrm{t})}=m + \frac{\lambda_c}{4\Lambda} \nu^2 y$.

In addition, similar to the real case, self energy corrections via coupling to the SM gauge bosons, at one loop order, induce radiative mass splitting between charged and neutral particles which takes the form ~\cite{MDM_2009}:
\begin{equation}
	\Delta_q^{(\ell)}	\equiv	m_q^{(\ell)} -m_0^{(\mathrm{t})}		=	q\, ( q +\frac{2}{c_w} y )\, \Delta
\end{equation}

here $m_q^{(\ell)}$ is the loop-induced mass of $\psi^q$ field, and radiative mass splitting $\Delta$ is defined in \ref{eq:Dm_R}.

Interactions of DM with gauge fields are encoded in the kinetic term of the Lagrangian \ref{eq:L_C}. The covariant derivative after EWSB takes the form 
$D_\mu = \partial_\mu 	+i eQA_\mu 	+i \frac{g_w}{c_w} \left( T^{(3)} -s_w^2 Q \right) Z_\mu 	+ i \frac{g_w}{\sqrt{2}}  \left( W^-_\mu T^- +W^+_\mu T^+ \right)$.

The DM-nucleon cross-section can be computed using equation \ref{eq:DM_N} by replacing $y=y_L=y_R$. Following the same logic as discussed in section \ref{sec:Chiral}, complex DM would be excluded as a result of large cross-section via coherent neutral weak boson exchange with the target nuclei.

However, complex dark matter scenario can be resurrected if DM-Z boson interaction at tree level can be avoided, by adding a new mechanism. This can be done through introducing an effective operator which enables coupling of DM with Higgs boson, as will be explained. 

After EWSB, the non-renormalisable operator proportional to $\lambda_0$ reduces to:
\begin{equation}
\label{eq:neutral}
	-\frac{\lambda_0}{\Lambda} ( H^\dagger \mathcal{T}^a H^\mathfrak{C}) (\overline{X^\mathfrak{C}}T^a X )		+ \mathrm{cc}	
	\quad = \quad
	-\frac{\lambda_0}{\Lambda}	\frac{(h+\nu)^2}{4}		\sum_{q=-t+\frac{1}{2}}^{t-\frac{1}{2}}	(-1)^q		\sqrt{ n^2 -4q^2} \ 
		(\overline{\psi}{}^{-q})^c  \,\psi^q 		+ \mathrm{cc}
\end{equation}

The U(1)\textsubscript{D} symmetry is spontaneously broken down to Z\textsubscript{2} under which dark matter particles are odd.

Vacuum expectation value (VEV) of Higgs field makes an additional contribution to the mass matrix of the neutral components, and their mass term will change to:
\begin{equation}
	\begin{pmatrix}		(\overline{\psi}{}^0)^c		&\overline{\psi}{}^0 \ 	\end{pmatrix}
	\begin{pmatrix}		\delta_0	&\frac{m_0^{(\mathrm{t})}}{2} 		\\	\frac{m_0^{(\mathrm{t})}}{2} 	&\delta_0^*	\end{pmatrix}
	\begin{pmatrix}		\psi^0	\\ 	(\psi^0)^c		\end{pmatrix}	=	
	\frac{1}{2}	\begin{pmatrix}		\overline{\widetilde{\chi}^0} 	& 	\overline{\chi}^0	\end{pmatrix}
				\begin{pmatrix}		\widetilde{m}_0  	&0	 \\	0	&m_0	\end{pmatrix}
				\begin{pmatrix}		\widetilde{\chi}^0	\\	\chi^0	\end{pmatrix}.
\end{equation}

So, the the scalar product of the adjoint vectors \ref{eq:neutral} splits the masses of the neutral components through the 
$\delta_0 = n \, \nu^2 \lambda_0  / 8\Lambda$ 
term. As will be discussed $|\delta_0| \ll  m$, and DM mass eigenstates up to zeroth order in $\mathcal{O}(\Im \, \delta_0/m)$ can be written as:
\begin{equation}
	\widetilde{\chi}^0 =	\frac{1}{\sqrt{2}}		\left(	  \psi^0   + (\psi^0)^c	 \right)\,,			\qquad
	\chi^0 = 				\frac{1}{\sqrt{2}i}		\left( \psi^0   - (\psi^0)^c 		\right)\,.
\end{equation}

with masses $\widetilde{m}_0 = m_0^{(\mathrm{t})} + 2\Re\,\delta_0$, and $m_0 = m_0^{(\mathrm{t})} - 2\Re\,\delta_0$ respectively. The mass eigenstates $\widetilde{\chi}^0$ and $\chi^0$ are Majorana fermions into which pseudo-Dirac state $\psi^0$ is split. Without loss of generality, we take the imaginary field $\chi^0$ to be the lightest DM candidate.

Equation \ref{eq:neutral} shows that $\lambda_0$ term also introduces an off-diagonal contribution proportional to \\
$\delta_q \equiv (-1)^q \, \nu^2 \sqrt{ n^2 -4q^2} (\lambda_0/8\Lambda)$ 
causing mixing of the charged states $\psi^q$ and $(\psi^{-q})^c$.
\footnote{It should be emphasised that for a Dirac particle $\psi^q$ and $(\psi^{-q})^c$ are not the same.}
Then using the relations $\overline{\psi}{}^{-q} \psi^{-q} = (\overline{\psi}{}^{-q})^c (\psi^{-q})^c$ and $(\overline{\psi}{}^q)^c \psi^{-q} = (\overline{\psi}{}^{-q})^c \psi^q$, the mass term for the charged particles with $q>0$ can be cast as:
\begin{equation}
	\begin{pmatrix}		\overline{\psi}{}^q		&(\overline{\psi}{}^{-q})^c \ 	\end{pmatrix}
	\begin{pmatrix}		\mathfrak{m}_q + 	d_q		&2\delta^*_q	\\
					2\delta_q		&\mathfrak{m}_q - d_q	 	\end{pmatrix}
	\begin{pmatrix}		\psi^q	\\ 	(\psi^{-q})^c		\end{pmatrix}	=	
		\begin{pmatrix}		\overline{\chi}^q_2 	& 	\overline{\chi}^q_1	\end{pmatrix}
		\begin{pmatrix}		m^{(2)}_q  	&0	 \\	0	&m^{(1)}_q	\end{pmatrix}
		\begin{pmatrix}		\chi^q_2	\\	\chi^q_1	\end{pmatrix}.
\end{equation}

After diagonalisation of the mass matrix, the following new eigenstates are obtained:
\begin{equation}
	\chi^q_1 = -e^{i\widehat{\lambda}_0/2} s_q  \ \psi^q		+ e^{-i\widehat{\lambda}_0/2}	c_q  \ (\psi^{-q})^c \,,			\qquad
	\chi^q_2 = e^{i\widehat{\lambda}_0/2}	c_q	\ \psi^q	+e^{-i\widehat{\lambda}_0/2} s_q	\ (\psi^{-q})^c \,.
\end{equation}

where the hat in the exponent should be understood as argument $\widehat{\lambda}_0  \equiv  \arg \lambda_0$.

With corresponding masses:
\begin{equation}
	m_q^{(1),(2)}= \mathfrak{m}_q	\mp \sqrt{ 4 |\delta_q|^2	+d^2_q}	
				\ \approx \ 	 \mathfrak{m}_q  \mp   |d_q|
\end{equation}

where $\mathfrak{m}_q \equiv m^{(\mathrm{t})}_0 +  \Delta_q^\mathbb{R}$ \ref{eq:Dm_R}, $d_q \equiv \left[ 2y c_w^{-1} \Delta -(\lambda_c /4 \Lambda) \nu^2 \right] q$, 
and defining $m_q^{(1)}<m_q^{(2)}$. 

The mixing angle $\phi_q$ is defined such that $\sin \phi_q = |\delta_q| / \sqrt{  |\delta_q|^2	+d^2_q /4 }$. 
For ease of notation we denote $c_q \equiv \cos(\phi_q /2)$ and $s_q \equiv \sin(\phi_q /2)$.

The highest weight particle $\chi^{n/2}$ is unique and has a mass of:
\begin{equation}
	m_\frac{n}{2}	=	m_0^{(\mathrm{t})} 	+\Delta_\frac{n}{2}^{(\mathrm{t})}	+\Delta_\frac{n}{2}^{(\ell)}
\end{equation}

The dark matter candidate $\chi^0$ needs to be the lightest member of the multiplet $m_q < m_0$. 
This condition always holds true for all particles except the lighter charged field $\chi^q_1$. Demanding $m^{(1)}_q$ to be less than DM mass i.e. 
$4 |\delta_q|^2	+d^2_q		>	 (\Delta_q^\mathbb{R})^2$,	
further imposes constraints on the value of the non-renormalisable coupling constants $\lambda_c$ and $\lambda_0$. The necessary conditions are that $\lambda_c$ is expected to stay within the range of:
\begin{equation}
	\label{eq:lc}
	4 \left(c_w^{-1} -1 \right)	 \frac{\Delta}{\nu^2}	\approx	1.5 \times 10^{-6} \, \mathrm{GeV}^{-1}
		\quad	<	 \frac{\lambda_c}{\Lambda}	<	\quad
	4 \left(c_w^{-1} +1 \right) \frac{\Delta}{\nu^2}	\approx	2.3 \times 10^{-5} \, \mathrm{GeV}^{-1}	\,,
\end{equation}

with an average of 
$(\lambda_c / \Lambda)_\mathrm{av}	=	4 c_w^{-1} \Delta / \nu^2		\approx 1.2\times10^{-5} \mathrm{GeV}^{-1}$ 
for vanishing $\lambda_0 \approx 0$.

This requirement further sets the limit for the scale of new physics that gives rise to the charged-neutral splitting to $\Lambda/\lambda_c \sim 10^5$ GeV.

The other condition for the dark matter candidate to be the lightest component of the electro-weak multiplet, is existence of an upper bound on the value of $\lambda_0$:
\begin{equation}
	\label{eq:l0_Max}
	\left| \frac{\lambda_0}{\Lambda}	\right|	<	3\times10^{-6} \,\mathrm{GeV}^{-1}
\end{equation}

for $\lambda_c = ( \lambda_c )_\mathrm{av}$.

Figure \ref{fig:Dm_lc} illustrates the mass splitting of the charged states $\chi^+_1$ and $\chi^+_2$ as a function of the non-renormalisable coupling $\lambda_c$, for pseudo-real quadret. It can be seen that the mass difference between the states 
$\Delta_+^{(2)} - \Delta_+^{(1)}$ 
increases as the coupling deviates from the mean value. It has a minimum at the average $\lambda_c^\mathrm{av}$, and the particles with the same charge $q$ become degenerate 
$ m_q^{(1)} = m_q^{(2)} = \Delta_q^\mathbb{R}$ 
if $\lambda_0 \approx 0$.
The maximum mass of the heavier field is independent of the couplings 
${m_q^{(2)}}_\mathrm{max} = m_0^{(\mathrm{t})} + 2\Delta_q^\mathbb{R}$.
As a result the mass splitting between the charged particles has a value of about $\mathcal{O}(100)$ MeV at the most. Note that the range of acceptable values of 
$\mathrm{d}(\lambda_c/\Lambda) =	\sqrt{ (\Delta_+^\mathbb{R})^2		-4 |\delta_+|^2}$ 
decreases as the other coupling $\lambda_0$ gets stronger.

The left panel \ref{fig:Dm_l0} of the figure shows changes in the mass splitting of theses states with respect to the coupling $\lambda_0$. It can be observed that difference between the masses of particles increases with coupling strength going up. The maximum allowed value of 
$\lambda_0^\mathrm{max}	=	\sqrt{ (\Delta_+^\mathbb{R})^2		-d_+^2}$ 
declines as the other coupling $\lambda_c$ stays away from the average.


\subsubsection{Inelastic Scattering}

Interactions of the dark sector is derived from expansion of the gauge-fermion kinetic term in \ref{eq:L_C}:
\begin{align}
	\mathcal{L}_D & \supset
	i \frac{g_w}{2c_w} \ \overline{\widetilde{\chi}^0} \gamma^\mu \chi^0 	Z_\mu 		\\	\nonumber
	& + \frac{g_w}{4} \left[ 	
	\left(   n \ e^{-\frac{i}{2} \widehat{\lambda}_0} c_+   - \sqrt{ n^2 -4} \ e^{\frac{i}{2} \widehat{\lambda}_0} s_+ \right)	\overline{\widetilde{\chi}^0} \gamma^\mu \chi_2^+
	-i 	\left(   n \ e^{-\frac{i}{2} \widehat{\lambda}_0} c_+   + \sqrt{ n^2 -4} \ e^{\frac{i}{2} \widehat{\lambda}_0} s_+ \right)	\overline{\chi^0} \gamma^\mu \chi_2^+  
					\right.	\\	\nonumber
	& \left. - \left(   n \ e^{-\frac{i}{2} \widehat{\lambda}_0} s_+   + \sqrt{ n^2 -4} \ e^{\frac{i}{2} \widehat{\lambda}_0} c_+ \right)	\overline{\widetilde{\chi}^0} \gamma^\mu \chi_1^+
	+i 	\left(   n \ e^{-\frac{i}{2} \widehat{\lambda}_0} s_+   - \sqrt{ n^2 -4} \ e^{\frac{i}{2} \widehat{\lambda}_0} c_+ \right)	\overline{\chi^0} \gamma^\mu \chi_1^+  	
	\right]	W^-_\mu 		+ \mathrm{cc}
\end{align}

The full Lagrangian of the dark sector is lengthy and is presented in the Appendix \ref{app:Feynman}, together with the Feynman rules for couplings to EW gauges and scalar Higgs.

It can be seen that dark matter particle $\chi^0$ has no interaction with Z-boson, so elastic scattering off the nuclei is forbidden. However, there exists a coupling between $\chi^0$ and $\widetilde{\chi}^0$ via Z-boson exchange which leaves a possibility for DM-nucleon inelastic scattering $\chi^0 N\to \widetilde{\chi}^0 N$.

Inelastic scattering of dark matter off nucleus is an \emph{endothermic} process in which DM kinetic energy loss 
$ \Delta K_\chi = ( m_0 v_\chi^2 - \widetilde{m}_0 {v'_\chi}^2 ) /2$ 
is converted into the mass difference of the outgoing odd particle $\Delta_{\widetilde{0}}$ and recoil energy of the target nucleus $E_R$:
\begin{equation}
	\label{eq:iDM}
	\Delta K_\chi	=	\Delta_{\widetilde{0}}	+ E_R
\end{equation}

One needs to Impose momentum conservation 
$ 2m_\mathcal{N} E_R 	=	(m_0 v_\chi)^2	+(\widetilde{m}_0 v'_\chi)^2	 	-2 m_0 \widetilde{m}_0 v_\chi v'_\chi \cos \theta$, 
in \ref{eq:iDM}, to eliminate $E_R$ which is a difficult to measure quantity. Solving for the unknown parameter $v'_\chi$ we obtain:
\begin{equation}
	\widetilde{m}_0 ( \widetilde{m}_0 + m_\mathcal{N} ) \,{v'_\chi}^2		-2 \,m_0 \widetilde{m}_0 v_\chi \cos \theta \,v'_\chi
				+2 m_\mathcal{N} \Delta_{\widetilde{0}}		+ m_0 (m_0 -m_\mathcal{N} )  v_\chi^2		=	0
\end{equation}

For an endothermic reaction, there exist a \emph{threshold energy} $K_\chi^{(\mathrm{th})}$ below which no solution is available for the equation above \cite{dm_inelastic, iDM}, and thus scattering is not possible:
\begin{equation}
	K_\chi^{(\mathrm{th})}		=		\left( 1 +	\frac{ m_\chi }{ \Delta_{\widetilde{0}} +m_\mathcal{N} }	\right)	\Delta_{\widetilde{0}}
\end{equation}

So keeping the kinetic energy of the incoming particle less than the threshold 
$m_0 v_\chi^2 /2	<	K_\chi^{(\mathrm{th})}$,
dark matter $\chi_0$ cannot up-scatter to the excited state $\widetilde{\chi}_0$. In the high mass regime, we finally arrive at the condition:
\begin{equation}
	\Delta_{\widetilde{0}}	>	\frac{1}{2} m_{\mathcal{N} \chi}	v_\chi^2
\end{equation}

It means if the mass splitting $\Delta_{\widetilde{0}}$ is sufficiently large, then the inelastic nucleonic scattering will be kinematically forbidden. It means that by setting 
$ \Delta_{\widetilde{0}}		 > 	\mathcal{O}(100)$ keV 
the tree-level coupling to nuclei is completely suppressed.

In addition, this restricts the value of the neutral pseudo-Dirac splitting coefficient to 
\begin{equation}
	\label{eq:l0_min}
	\Re \, \lambda_0 /\Lambda > 10^{-8} \, \mathrm{GeV^{-1}} \,. 
\end{equation}

In other words, the scale of the new physics responsible for breaking $U(1)_D$ symmetry is set to $\Lambda/\lambda_0 \sim 10^8$ GeV.


\section{Sommerfeld Effect}

Electroweak dark matter annihilates into pairs of SU(2) gauge vectors. When DM velocity is smaller than weak structure constant $\alpha_w$, and its mass very heavy compared to the annihilating bosons, the perturbative calculations used in usual quantum field theory breaks down. It proves that the annihilation amplitude for each loop order has a factor which goes like $\alpha_w m_0/m_W$, so contribution from higher order loop diagrams becomes more significant. This means that weak boson exchange acts as a long-range force, and wave-function of EWDM particles get distorted from plane wave by this long-lasting Yukawa potential \cite{QCD_Sommerfeld}. A phenomenon known as \emph{Sommerfeld Effect} in the literature \cite{Sommerfeld}.

In order to consider the mentioned non-perturbative effects in high mass and low velocity regime, one needs to construct a non-relativistic effective field theory for two-particle states of EWDM. Light particles like gauge bosons generated in annihilation process are relativistic, therefore to obtain the non-relativistic action, we ought to integrate out theses high energy fields, factorising the short range effects from the long-range interactions \cite{Unitarity, Non_pertve} .

Exchange of electroweak gauge bosons between different dark sector particles leads to formation of DM pairs. Therefore, we need to formulate an effective action that describes dynamics of theses two-body states.

Equation of motion of the system is derived from the two-particle effective action. Starting from Dyson-Schwinger equation, expanding the two-body s-wave Green function in absorptive term, one obtains the following standard Schrodinger equation at the leading order \cite{Non_pertve}:
\footnote{The same equation could be obtained in non-relativistic quantum mechanics by writing the one-dimensional Schroedinger equation for the radial part of the reduced wave-function of two interacting particles. However, still one needs the non-relativistic effective field prescription explained in the main text to work out the potential $V$ and annihilation $\Gamma$ matrices \cite{MDM_2015}.}
\begin{equation}
\label{eq:SE}
	-\frac{1}{m_0} \,	\partial^2_r \mathlarger\varphi(r)_{_{IJ}}		+ V(r)_{_{IK}} \, \mathlarger\varphi(r)_{_{KJ}}		= K_{_{IK}} \, \mathlarger\varphi(r)_{_{KJ}}
\end{equation}

where $r$ is the distance between two pairs of DM, and the matrix $\mathlarger\varphi$ is the two-body non-relativistic s-wave-function in the central potential matrix $V$. The equation is solved for the continuous energy eigenvalue of $K_{IJ}=m_0 v^2 \delta_{_{IJ}}$ which is the kinetic energy of $\chi^0\chi^0$ pair, with $v$ being DM velocity in the centre of mass. 
\footnote{In addition to the continuum solution, there are discrete negative eigenvalues for Schrodinger equation \ref{eq:SE} which correspond to binding energies of bound states of dark matter \cite{BS}.}

The collective index $I$ for a pair of particles $\bar{\chi}_i \chi_{i'}$ indicates the single particle indices $ii'$, and indices run over all the possible states formed by interactions. Given the dark matter in the representation $R$ of SU(2), the bound state lies in $ \overline{R} \otimes R$, or $R\otimes R$ if $R$ is real.

There are two boundary conditions for the equation of motion \ref{eq:SE}. The first one normalises the wave-functions at the origin:	
\begin{equation}
	\mathlarger\varphi(0)_{_{IJ}} =		\delta_{_{IJ}}
\end{equation}

and the second condition states that if DM pair has enough kinetic energy for mass splitting $ K_{_{IJ}} > V(\infty)_{_{IJ}}$ then the solution behaves as an outgoing wave at infinity, otherwise the two-body wave-function decays exponentially:
\begin{equation}
	\lim_{r \to \infty} \partial_r \mathlarger\varphi(r)_{_{IJ}}	=	\lim_{r \to \infty} i \sqrt{ m ( K_{_{IK}} - V(r)_{_{IK}} )}	\ \mathlarger\varphi(r)_{_{KJ}}
\end{equation}

Factorising out the oscillating phase that describes the pairs of dark matter at infinity $r\gg 1/M_W$, we are left with a constant matrix $D$. If there is no long distance interactions i.e. $V=0$, then $D=\mathbb{1}$. So, elements of this matrix are defined as \emph{Sommerfeld enhancement factors}.
\begin{equation}
	\lim_{r \to \infty} \mathlarger\varphi(r) 	= \lim_{r \to \infty} D . \exp \left\{ i \sqrt{ m ( K - V(r) )} \right\}  		
\end{equation}

In the galactic halo,  the mass splitting between the DM candidate and other heavier odd particles is bigger than their kinetic energy, therefore the Schrodinger-like equation \ref{eq:SE} implies that only the actual neutral DM pair will not vanish at large distances. As a result, the second boundary condition allows for only one non-zero column at $00$ entry in the constant matrix.
\begin{equation}
	\lim_{r \to \infty} \mathlarger\varphi(r) 	 		
		= e^{imvr} 	\begin{pmatrix}		0 & \ldots & \vdots	\\ 	\vdots & \ddots & \ d_{\widetilde{0}\widetilde{0}} 	\\	0 & \ldots & \ d_{00}	\end{pmatrix}
\end{equation}

Factorising out the outgoing wave $e^{imvr}$, Sommerfeld enhancement factors, in this case, are reduced to a vector 
$d_0 \equiv (\ldots, \ d_{\bar{+}+}, \ d_{\widetilde{0}\widetilde{0}}, \ d_{00})^T$.
\footnote{$d$ contains real elements up to an overall global phase \cite{MDM_Ibarra}.}
So, $\chi^0 \chi^0$ is the only relevant initial state for indirect detection purposes.

Other incoming pairs participate in coannihilation processes and are taken into account when computing the relic abundance. That is due to the fact that high temperature in the early universe allows formation of bound states out of all components of the dark sector. In a similar way, one can define $d_I$ vector corresponding to $\chi^i \chi^{i'}$ state.

Non-perturbative effects are dominant in the non-relativistic velocities and also for large DM mass $m_0 \gg m_W$.
At small velocities $v \ll m_W/m_0$, Sommerfeld factor becomes independent of velocity, and in case of an attractive effective potential a series of resonant enhancement happens when bond states of dark matter are formed \cite{BS}.

The potential matrix $V$ in \ref{eq:SE} includes the mass splitting between different components of the multiplet, in addition to the interactions mediated by exchange of the SM gauge boson among these pairs. It can be read from real part of the two-body effective propagator as \cite{Ramsauer_Townsend}:
\begin{equation}
	V_{IJ} (r) = 	\Delta_{IJ}		-\alpha_w		\ N_I N_J 	\ \sum_{A,B} K_{AB} \ (T^A_{ij'}) (T^B_{ji'})\ 	\frac{e^{-m_A r}}{r}	\ .
\end{equation}

where the mass difference is defined as $\Delta_{IJ} \equiv \left( m_i + m_{i'} -2m_0 \right)	\delta_{IJ}$. 
For a two-body state $| \bar{\chi}_i \chi_{i'} \rangle$, the normalisation factor $N_I=\sqrt{2}$ if it contains identical particles i.e. $i=i'$, and otherwise $N_I=1$ if $i \ne i'$. 
The sum runs over all the SM gauge bosons $\gamma$, $Z$, $W^-$ and $W^+$ associated with gauge generators 
$\hat{T}^\gamma = s_w \hat{Q}$, 
$\hat{T}^Z = ( c^2_w \hat{Q} - \hat{Y} ) /c_w$, and 
$\hat{T}^{W\pm} = \hat{T}^{\pm} /\sqrt{2}$ 
in the DM multiplet representation of SU(2). 
$K_{AB}$ encodes EW force between non-relativistic particles, and in the basis $\left( \gamma, Z, W^-, W^+\right)^T$ can be written in matrix form as:
\begin{equation*}
	K = \begin{pmatrix}	1&0&0&0	\\	0&1&0&0	\\	0&0&0&1	\\	0&0&1&0	\end{pmatrix}
\end{equation*}

When SU(2) symmetry gets broken, then the conserved quantum numbers will be total charge $Q$ and total spin $S$. Therefore, we classify each representation into sub-systems according to the values of theses conserved numbers. Within each sector, different initial states can mix by interactions mediated by gauge vectors.

Obliteration of non-relativistic DM pair to create other SM particles can be described by tree-level annihilation matrix $\Gamma$ which is the imaginary potential in the effective field action. The diagonal entries of this matrix intuitively represent the annihilation of the two-body DM state, and off-diagonal elements imply the interference between these different initial states. Dark matter pair with zero total spin $S=0$ can only annihilate into two SM gauge vectors, and one has \cite{MDM_Sommerfeld}:
\begin{equation}
	{(\Gamma^{AB}_{IJ})}^{S=0}	=	\frac{\pi \alpha_w^2}{4 m_0^2}		\frac{N_I N_J}{2(1+\delta_{AB})}	
								\left\{ T^A, T^B \right\}_I 		\left\{ T^A, T^B \right\}_J		\ .
\end{equation}

Note that $\Gamma$ is a rank-one matrix and can be readily written as the product $\Gamma^{AB} = (G_{AB} \ G_{AB}^\dagger) (\pi/4 m_0^2)$ where vector $G_{AB}$ corresponds to emission of the gauge boson pair $AB$. The total annihilation matrix is simply obtained by adding the contribution from all final boson states $\Gamma^{S=0} = \sum_{AB} \Gamma^{AB}$. 

It immediately follows that spinless $S=0$ dark matter pair is only allowed to have electric charge $Q=0, \pm1, \pm2$.

Depending on electric charge of the initial state, DM can annihilate through different gauge boson channels. The neutral pair $Q=0$ can decay into $\gamma\gamma$, $\gamma Z$, $ZZ$ and $W^-W^+$. The charged state $Q=1$ annihilates into vector pairs $\gamma W^+$ and $ZW^+$, and doubly charged $Q=2$ into $W^+W^+$ final state.

In the real representation, there are only two independent annihilation modes for the $Q=0$ neutral states i.e. $W^-W^+$ and $\gamma\gamma$. The two other channels can be obtained from the latter through $\Gamma^{ZZ} = \Gamma^{\gamma\gamma}/t^4_w$ and $\Gamma^{\gamma Z} = 2 \ \Gamma^{\gamma\gamma} / t^2_w$. Also the charged state $Q=1$ has one channel $\gamma W^-$ which is related to the other one via $\Gamma^{ZW^-} = \Gamma^{\gamma W^-} / t_w$.
While the pseudo-real EWDM provides three independent modes for the $Q=0$ neutral state that are $W^-W^+$, $ZZ$ and $\gamma\gamma$, and the remaining channel is given by $\Gamma^{\gamma Z}_{IJ} = 2 \sqrt { (\Gamma^{\gamma\gamma}_{IJ}) \ (\Gamma^{ZZ}_{IJ})}$.

Based on Landau-Yang's theorem, the massive spin-1 two-body state cannot decay into a pair of gauge vectors, but they are allowed to annihilate into SM fermions or scalars \cite{MDM_BS}. The corresponding annihilation matrix reads:
\begin{equation}
	{(\Gamma_{IJ})}^{S=1}	=	\frac{\pi \alpha_w^2}{4 m_0^2}		\frac{N_I N_J}{(1+\delta_{AB})}	
								\sum_{AB}  C_{AB}	\left\{ T^A, T^B \right\}_{IJ} 			\ .
\end{equation}

The matrix $C_{AB}$ describes the gauge interactions of SM fermions and scalar, and in the basis $\left( \gamma, Z, W^-, W^+\right)^T$ takes the form:
\begin{equation*}
	C_{AB}	=	\frac{25}{4} g_w^2
	\begin{pmatrix}	s^2_w	&	s_w c_w	&0&0	\\	s_w c_w	&	c^2_w	&0&0	\\	0&0&0&1	\\	0&0&1&0	\end{pmatrix}
			+	\frac{41}{4} g_y^2
	\begin{pmatrix}	c^2_w	&	-s_w c_w	&0&0	\\	-s_w c_w	&	s^2_w	&0&0	\\	0&0&0&0	\\	0&0&0&0	\end{pmatrix}	
\end{equation*}
		
As an immediate result, the electric charge of the spin-one $S=1$ dark matter state, is limited to $Q=0, \pm1$.

Furthermore, for a Majorana dark matter particle $\chi^0$, Pauli exclusion principle forbids formation of state $|\chi^0 \chi^0\rangle$ with $S=1$.

Annihilation cross section at low velocities is dominated by \emph{s-wave} scattering, with \emph{p-wave} contribution suppressed by a relative order of $v^2$ \cite{MDM_2009}. Using optical theorem, the s-wave cross-section for annihilation of DM pair $\bar{\chi}_i \chi_{i'}$ with vanishing angular momentum $\ell=0$ into the SM particles, can be obtained from the imaginary part of tree-level amplitude for the corresponding annihilation process \cite{Non_pertve}:
\begin{equation}
	\sigma_I	 = N_I^2  \ \left(D^\dagger .\, \Gamma .\, D\right)_{II}	\,,
\end{equation}

and for annihilation through a particular channel mediated by $AB$ bosons pairs, this simplifies to:
\begin{equation}
\label{eq:Xn_AB}
	\sigma^{AB}_I  =  \frac{\pi}{4 m_0^2}	N_I^2	\left| G^{AB} .\, d_I \right|^2	\,.
\end{equation}

The following methodology is employed to find the physical observables in the non-perturbative regime. In the first step, we need to derive the expression for the EW potential of two DM particle system using non-relativistic QFT machinery. Then Sommerfeld enhancement factors are obtained through solving the associated Schrodinger-like equation of motion. These factors allows one to compute the non-perturbative annihilation amplitudes in each sector which in turn will be used to find the total annihilation cross-section.

At this point, we apply our formalism to compute physical quantities for different fermionic EWDM models. As discussed before, there are four fermionic representations available, namely, pseudo-real doublet $(\mathbb{C}2)$, pseudo-real quadruplet $(\mathbb{C}4)$, real triplet $(\mathbb{R}3)$, and real quintuplet $(\mathbb{R}5)$. The pseudo-real models generalised through effective field framework especially the generalised complex quadruplet are new DM scenarios. For the real cases, our results are presented in the appendices \ref{sec:R3} and \ref{sec:R5} which update those previously studied in the literature.


\subsection{Generalised Pseudo-real Doublet}

\begin{figure} [t] 				
	\begin{subfigure}{.5\linewidth}
		\centering
		\includegraphics[width=\linewidth]{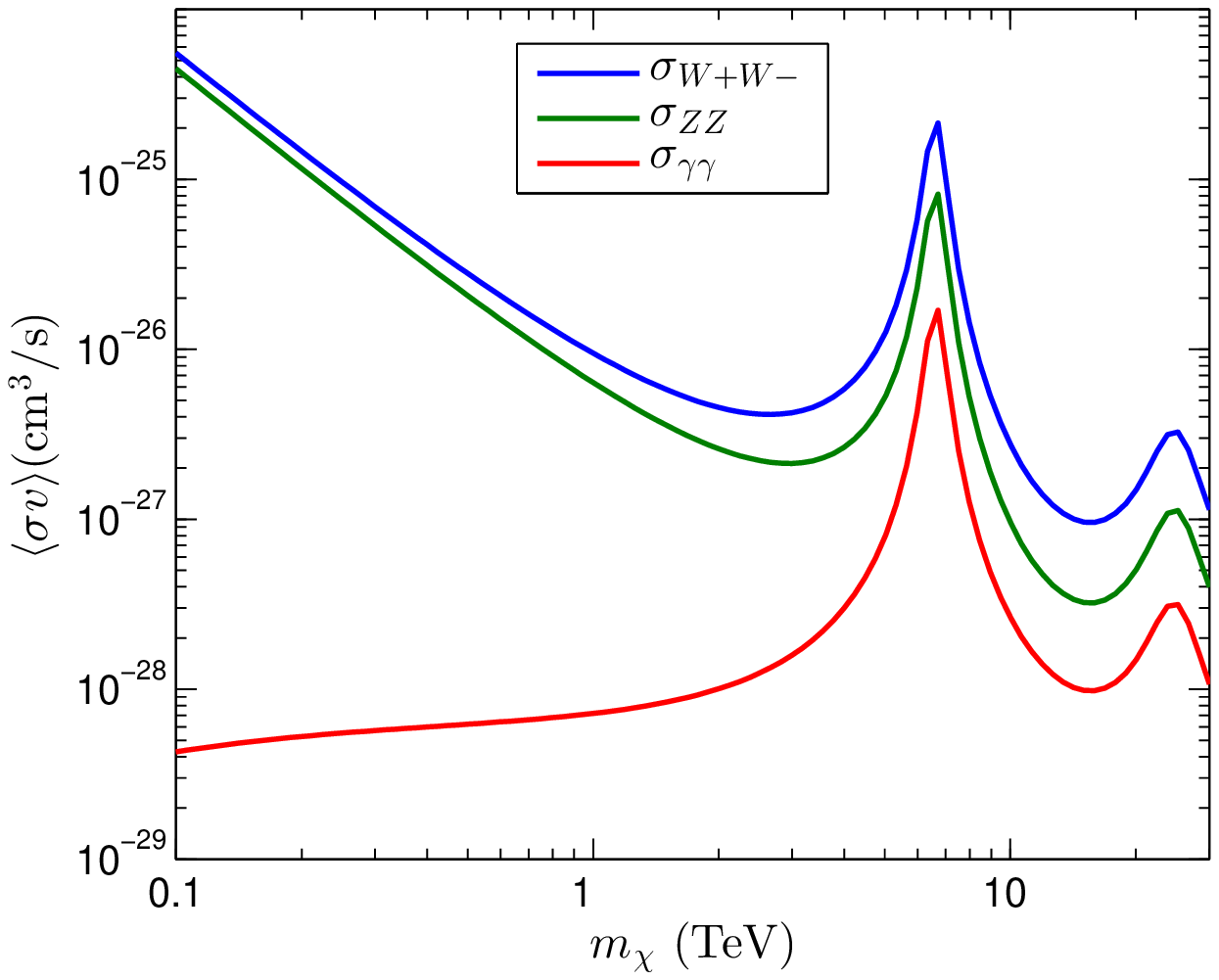}
		\subcaption{}
		\label{fig:C2eft:Xnv_m}
	\end{subfigure}%
	\begin{subfigure}{.5\linewidth}
		\centering
		\includegraphics[width=\linewidth]{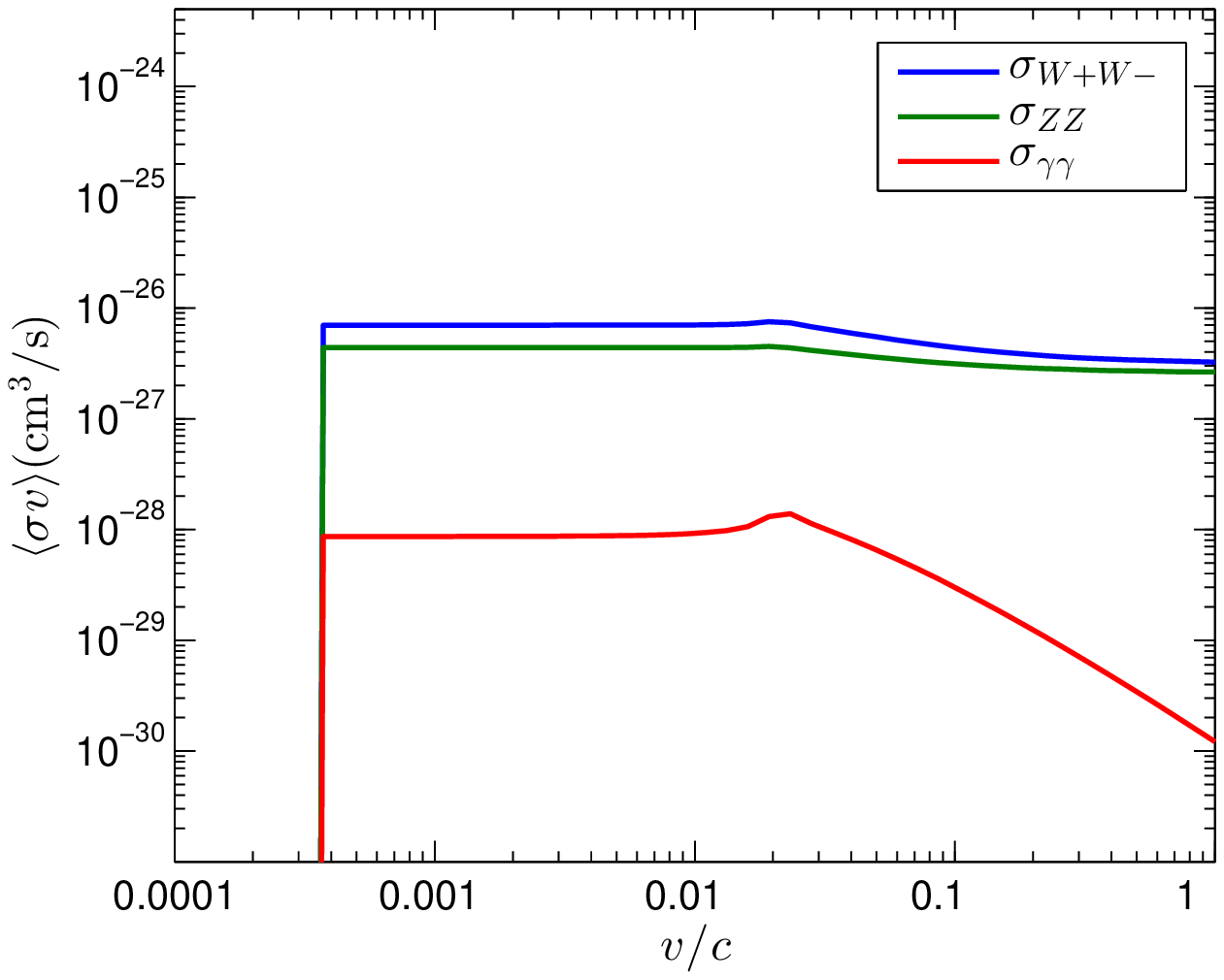}
		\subcaption{}
		\label{fig:C2eft:Xnv_v}
	\end{subfigure}
	\caption{Annihilation cross-section into independent channels $W^+W^-$ (blue), $ZZ$ (green) and $\gamma\gamma$ (red) in pseudo-real doublet representation. Left panel (\ref{fig:C2eft:Xnv_m}) illustrates the of $\chi^0\chi^0$ initial state spectrum with respect to the mass at MW halo velocity $v=10^{-3}c$, while the right panel (\ref{fig:C2eft:Xnv_v}) depicts the changes for $\widetilde{\chi}^0 \widetilde{\chi}^0$ pair as a function of DM velocity at thermal mass of $m_0 = 1.2$ TeV.}
	\label{fig:C2eft:Xnv}
\end{figure}

The DM doublet can be cast as:
\begin{equation}
	X = \begin{pmatrix}		\chi^+	\\ 	\frac{1}{\sqrt{2}}   (\widetilde{\chi}^0 +i \chi^0)		\end{pmatrix}
\end{equation}

The mass splitting $\Delta_i \equiv m_i -m_0$ between DM candidate $\chi^0$ and odd particle $\chi_i$ is computed to be $\Delta_+ =$ 310 MeV and $\Delta_{\widetilde{0}}^{\mathbb{C}2}$ = 60 keV.

This model resembles an almost pure Higgsino in the corners of the minimal supersymmetric theory (MSSM) where the lightest neutralino has a very small gaugino fraction \cite{Susy_DM}. In this theory, the problematic tree-level coupling to Z boson can be eliminated by mixing the pure Higgsino with the Majorana singlet bino \cite{MSSM_phny}. However, realisation of Higgsino as the Lightest supersymmetric particle (LSP) requires precise fine tuning to fit with observations as the Higgs potential depends on the mass parameter of Higgsino.

In this paper we proceed free from supersymmetric restrictions, and use the most general effective field operators to describe interactions of the complex doublet, and to avoid restrictions imposed by direct detection experiments.

Pairs of complex doublet dark matter decompose into four sectors each labeled by conserved quantum numbers total charge $Q$ and total spin $S$. The basis for these sectors reads:
\begin{subequations}
\begin{align}
	\label{eq:C2_Q0}
	& \Psi^{S=0}_{Q=0} = \left(  \overline{\chi}^+ \chi^+,
				  \ \frac{1}{\sqrt{2}} \, \widetilde{\chi}^0 \widetilde{\chi}^0,  \ \frac{1}{\sqrt{2}} \, \chi^0 \chi^0  \right)^T 	\,, 	\\
	& \Psi^{S=1}_{Q=0} = \overline{\chi}^+ \chi^+	\,,	\\
	& \Psi_{Q=1} = \left(  \widetilde{\chi}^0 \chi^+,  \ \chi^0 \chi^+  \right)^T 	\,.
\end{align}
\end{subequations}
		
The different normalisation factor in the neutral pairs of \ref{eq:C2_Q0}, comes from the fact that they are Majorana fermions and identical to their antiparticles.

The s-wave potential matrices describing exchange of gauge bosons between DM particles in spin-less neutral sector, are given by:
\begin{equation}
	V^{S=0}_{Q=0} = 	
	\begin{pmatrix}		
		\Delta_{+,+} -\mathcal{A}-(2c_w^2-1)^2 \mathcal{Z}	&\ -\frac{1}{2\sqrt{2}} \,\mathcal{W}		&\ -\frac{1}{2\sqrt{2}} \,\mathcal{W}	\\
		-\frac{1}{2\sqrt{2}} \,\mathcal{W}			& \Delta_{\widetilde{0}, \widetilde{0}}^{\mathbb{C}2}		&-\mathcal{Z}		\\
		-\frac{1}{2\sqrt{2}} \,\mathcal{W}				&-\mathcal{Z}						&0		
	\end{pmatrix}	
\end{equation}

where $\mathcal{A} = \alpha/r$, 
$\mathcal{Z} = \alpha_w \,e^{-M_z r}/ 4c_w^2 \,r$, 
and $\mathcal{W} = \alpha_w \,e^{-M_w r}/ r$;
and mass difference between DM pairs read 
$\Delta_{+,+} = 2 (\Delta_+ + 2 \Re \delta_0)$, and 
$\Delta_{\widetilde{0},\widetilde{0}}^{\mathbb{C}2} = 2 \Delta_{\widetilde{0}}^{\mathbb{C}2} = 8 \Re \delta_0$.

DM annihilation process is encoded in the following matrices:
\begin{equation}
	\Gamma^{S=0}_{Q=0} =	\frac{\pi\alpha_w^2}{32 m_0^2}	
		\begin{pmatrix}		 
			\Gamma_{++,++}				& \frac{1}{\sqrt{2}} \Gamma_{00,++}		& \frac{1}{\sqrt{2}} \Gamma_{00,++}		\\
			\frac{1}{\sqrt{2}} \Gamma_{00,++}		& \frac{1}{2} \Gamma_{00,00}		& \frac{1}{2} \Gamma_{00,00}		\\
			\frac{1}{\sqrt{2}} \Gamma_{00,++}		& \frac{1}{2} \Gamma_{00,00}		& \frac{1}{2} \Gamma_{00,00}		
		\end{pmatrix}	\,,
	\qquad
	\begin{array}{l}
		{G_{\gamma\gamma}}^{S=0}_{Q=0} =	\sqrt{2} \alpha \left( 1, 0, 0 \right)^T	,	\\		
		{G_{ZZ}}^{S=0}_{Q=0} = ( \alpha_w / 2\sqrt{2} c_w^2 )	\left( ( 2c_w^2 -1)^2,	\frac{1}{\sqrt{2}},	\frac{1}{\sqrt{2}}  \right)^T,	\\	
		{G_{WW}}^{S=0}_{Q=0}	= ( \alpha_w / 2 )	\ 	( 1, \frac{1}{\sqrt{2}},	\frac{1}{\sqrt{2}}  )^T.		
	\end{array}
\end{equation}

where $\Gamma_{++,++} = \Gamma_{00,00} \equiv	 t_w^4	  +2 t_w^2  +3$; 
and $\Gamma_{00,++} \equiv	t_w^4	  -2 t_w^2  +3$.

To find the enhancement factors for charged states, one needs the potential:

\begin{equation}
	V_{Q=1} =	
	\begin{pmatrix}
		\Delta_+	+6 \Re \delta_0			&\frac{i}{8}(2c_w^2-1)^2 \mathcal{Z}	\\
		\frac{i}{8}(2c_w^2-1)^2 \mathcal{Z}		&\Delta_+	 +2\Re \delta_0	
	\end{pmatrix}	\,.
\end{equation}

Annihilation amplitudes for spin-less charged pairs is computed from:
\begin{equation}
	\Gamma^{S=0}_{Q=1} =	\frac{\pi\alpha_w^2}{16 m_0^2}	
	\begin{pmatrix}		 
		1		&-i		\\		i 		&1		\\
	\end{pmatrix}	\,,	
	\qquad
	\begin{array}{l}
		{G_{\gamma W^-}}^{S=0}_{Q=1} =		( s_w \alpha_w /2)	\left( 1, -i \right)^T	,	\\		
		{G_{ZW^-}}^{S=0}_{Q=1} = 			( s_w^2 \alpha_w / 2 c_w )	\left( 1, -i \right)^T.		
	\end{array}	
\end{equation}
						
For spin-one part, one obtains the potential and annihilation below:
\begin{equation}
	V^{S=1}_{Q=0} =	\Delta_{+,+}	-\mathcal{A}	-(2c_w^2-1)^2 \mathcal{Z}\,,
	\qquad
	\Gamma^{S=1}_{Q=0} =	\frac{\pi\alpha_w^2}{32 m_0^2}	\left(41 t_w^4 +25 \right)	\,,
	\qquad
	\Gamma^{S=1}_{Q=1} =	\frac{25\pi\alpha_w^2}{32 m_0^2}	
	\begin{pmatrix}		 
		1		&-i		\\		i 		&1		\\
	\end{pmatrix}\,.
\end{equation}

Figure \ref{fig:C2eft:Xnv_m} illustrates the annihilation cross-section of DM pair $\chi^0\chi^0$ into independent final states $W^+W^-$, $ZZ$ and $\gamma\gamma$ as a function of mass, in pseudo-real doublet model.

Dark matter velocity is set to $10^{-3} c$ which is the typical value in the galactic halo environment.

It is noticeable that Sommerfeld mechanism can enhance the value of cross-section up to a few orders of magnitude. However, in this representation, non-perturbative effects are insignificant around the thermal mass of 1.2 TeV. For dark matter mass of about 6.5 TeV, amplitudes of different wave functions $d_{IJ}$ mixing states $\chi^+\bar{\chi}^+$, $\tilde{\chi}^0\tilde{\chi}^0$ and $\chi^0\chi^0$ add up. The cross-sections in different modes such as $\sigma_{WW} = (\pi \alpha^2_w / 16 m^2_0) \left| d_{\bar{+}+} + d_{\widetilde{0}\widetilde{0}} /\sqrt{2} + d_{00} /\sqrt{2} \right|^2$ reach to a maximum, so that one can observe a peak in the spectrum due to constructive resonance.

Perturbative QFT formulates a vanishing cross-section for annihilation into $\gamma\gamma$ mode at tree level. Nevertheless long-distance forces caused by Sommerfeld effect mixes the neutral states $\chi^0\chi^0$ with charged state $\chi^+\chi^+$ which interacts with electromagnetic field, resulting in a considerable cross-section of $\sigma_{\gamma\gamma} = (\pi \alpha^2 / 2 m^2_0) \left| d_{\bar{+}+} \right|^2$.

Figure \ref{fig:C2eft:Xnv_v} depicts annihilation cross-section of heavy neutral EWDM state $\widetilde{\chi}^0 \widetilde{\chi}^0$ as a function of velocity. Enhancement in the non-relativistic regime is the result of Sommerfeld factors dominating annihilation over the perturbative effects. It can be seen that for relativistic velocities $m_w/m_0 \ll v \le 1$, the enhancement factors and thus cross-section increase as $1/v$ until reaching the maximum value at small velocities $v \ll m_w/m_0 \approx 0.1-0.01$.

The states containing heavier odd particles $\bar{\chi}_i \chi_{i'}$ cannot exist below a cut-off velocity $v_{\mathrm{cut-off}} = \sqrt{\Delta_{ii'} /m_0}$ as their kinetic energy falls smaller than their mass splitting.


\subsection{Generalised Pseudo-real Quadruplet}

\begin{figure} [t] 				
	\begin{subfigure}{.5\linewidth}
		\centering
		\includegraphics[width=\linewidth]{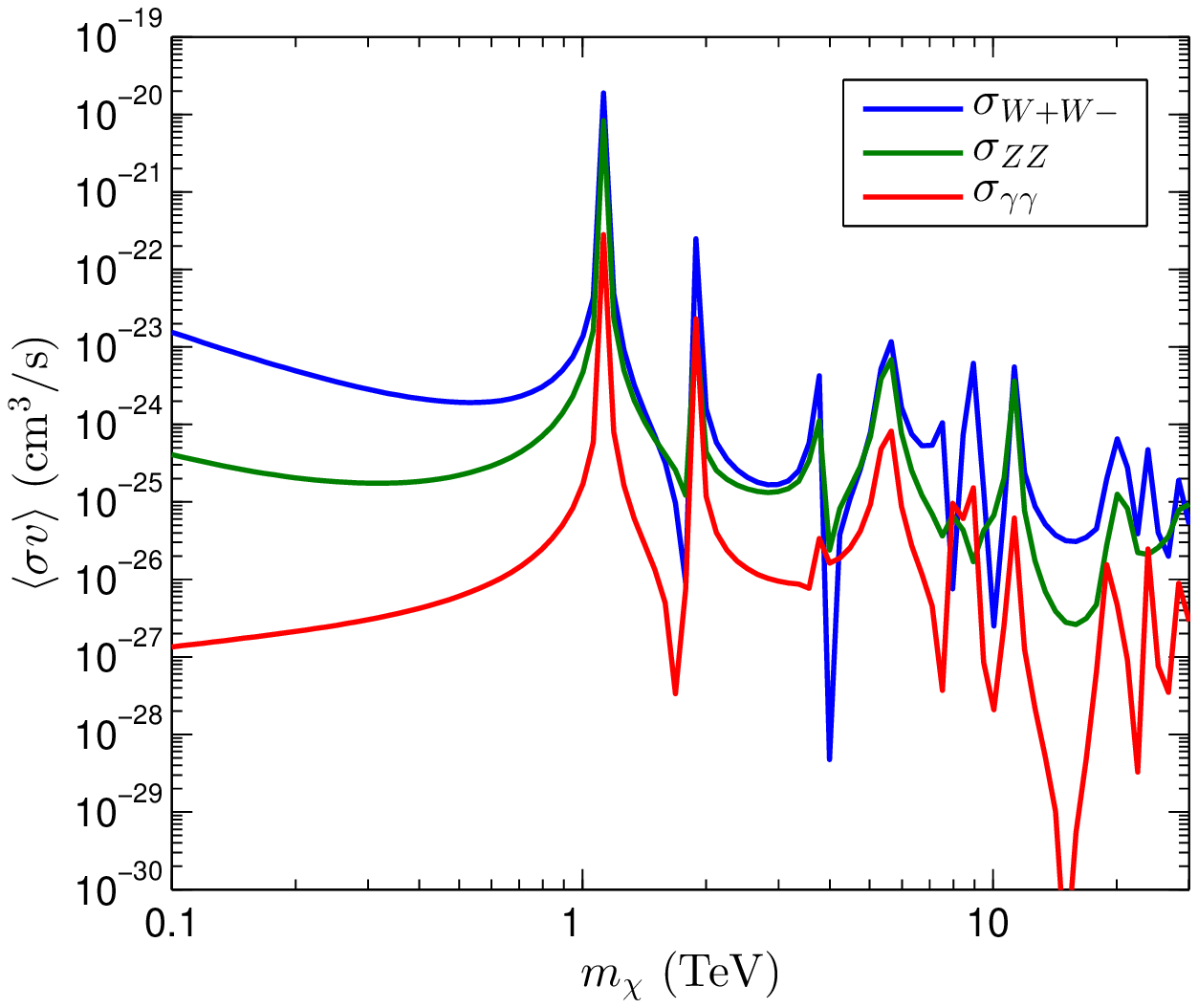}
		\subcaption{}
		\label{fig:C4eft:Xnv_m}
	\end{subfigure}%
	\begin{subfigure}{.5\linewidth}
		\centering
		\includegraphics[width=\linewidth]{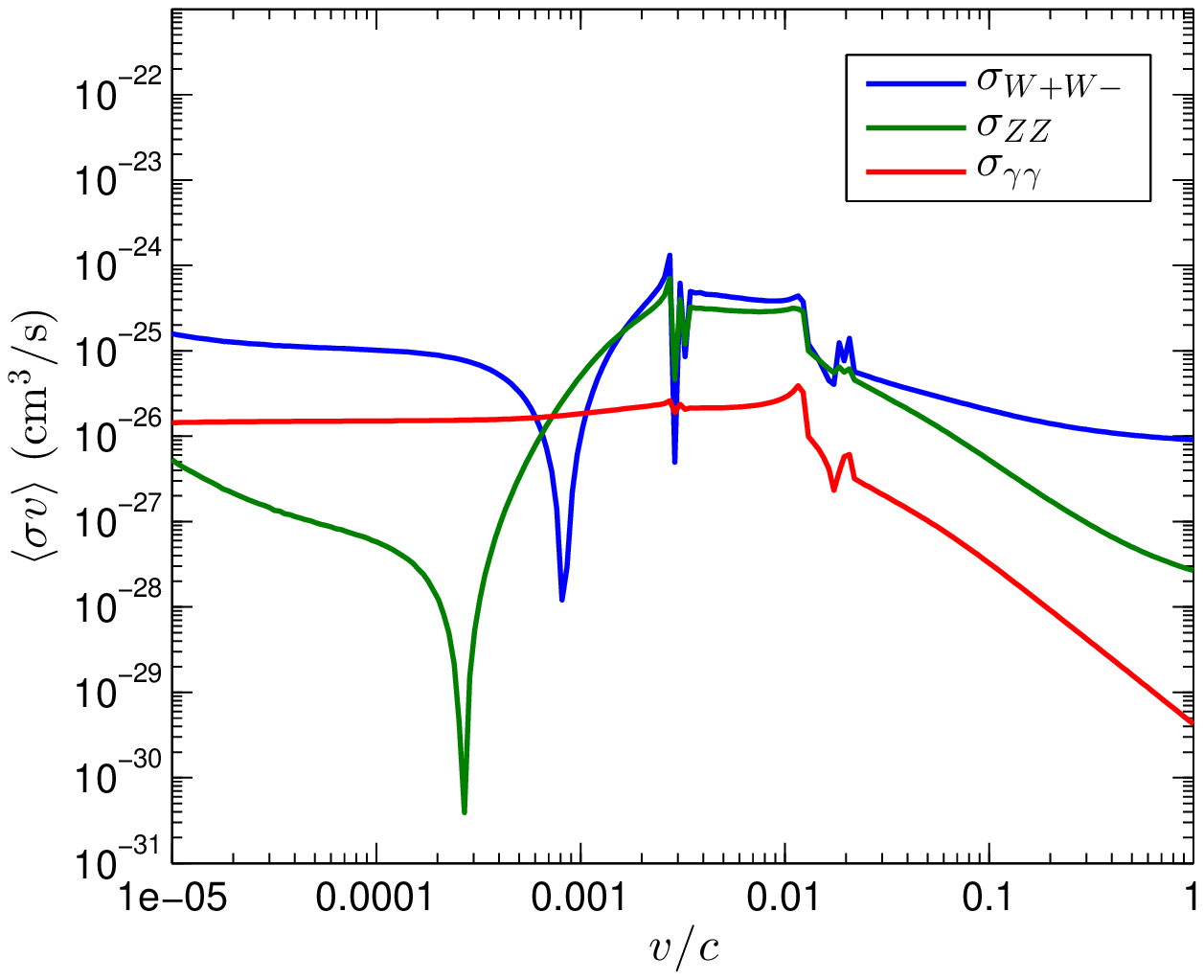}
		\subcaption{}
		\label{fig:C4eft:Xnv_v}
	\end{subfigure}
	\caption{Pseudo-real quadret annihilation cross section into $W^+W^-$ (blue), $ZZ$ (green) and $\gamma\gamma$ (red) gauge vectors. In left panel (\ref{fig:C4eft:Xnv_m}) as a function of DM mass at velocity $v=10^{-3}c$; and in right panel (\ref{fig:C4eft:Xnv_v}) as a function velocity at freeze-out mass $m_0 = 4.05$ TeV.}
	\label{fig:C4eft:Xnv}
\end{figure}

The quadruplet of dark matter has the form:
\begin{equation}
	X = \begin{pmatrix}		
		\chi^{++}	\\ 	e^{-i\widehat{\lambda}_0 /2} \left( c_+ \chi^+_2 -s_+ \chi^+_1 \right)	\\	
		\frac{1}{\sqrt{2}}   (\widetilde{\chi}^0 +i \chi^0)		\\	e^{-i\widehat{\lambda}_0 /2} \left( c_- \chi^-_2 -s_- \chi^-_1 \right)	
	\end{pmatrix}
\end{equation}

The mass splittings between DM and other heavier components of the multiplet read $\Delta_{++}$ = 950 MeV, $\Delta_-$ = 25 MeV and $\Delta_{\widetilde{0}}^{\mathbb{C}4}$ = 120 keV. The mixing angle is about $\phi_+ = 0.02^\circ$.

In this case, two-body DM state are split into five sectors according to their total charge and total spin. These sector are expanded in the following basis:
\begin{subequations}
\begin{align}
	& \Psi^{S=0}_{Q=0} = \left(  \overline{\chi}^{++} \chi^{++},	\ \overline{\chi}^+_2 \chi^+_2,		\ \overline{\chi}^+_1 \chi^+_1,
			\ \frac{1}{\sqrt{2}} \, \widetilde{\chi}^0 \widetilde{\chi}^0,		\ \frac{1}{\sqrt{2}} \, \chi^0 \chi^0 	 		\right)^T, 	\\
	& \Psi^{S=1}_{Q=0} = \left(  \overline{\chi}^{++} \chi^{++},	\ \overline{\chi}^+_2 \chi^+_2,		\ \overline{\chi}^+_1 \chi^+_1	\right)^T ,	\\
	& \Psi_{Q=1} = \left(	\overline{\chi}^+_2 \chi^{++},		\ \overline{\chi}^+_1 \chi^{++},		\ \widetilde{\chi}^0 \chi^+_2,  
			\ \chi^0 \chi^+_2,			\ \widetilde{\chi}^0 \chi^+_1,			\ \chi^0 \chi^+_1		  \right)^T, 	\\
	& \Psi^{S=0}_{Q=2} =  \left(	\widetilde{\chi}^0 \chi^{++},	 	\ \chi^0 \chi^{++}		 \right)^T.
\end{align}
\end{subequations}

The explicit form of potential matrices for the neutral spinless pairs writes:
\begin{equation}
	\label{eq:C4:V_S0_Q0}
	V^{S=0}_{Q=0} = 	
	\begin{pmatrix}
	\Delta_{2,2}   -4\mathcal{A}   -a^Z_{++} \mathcal{Z}		& -\frac{3}{2} c_+^2	\mathcal{W}		& -\frac{3}{2} s_+^2	\mathcal{W}		& 0		&0		\\
	-\frac{3}{2} c_+^2 \mathcal{W}			& \Delta_{+_2,+_2}   -\mathcal{A}   -a^Z_{+_2} \mathcal{Z}		& -4 s_+^2 c_+^2 \mathcal{Z}
					& \frac{-1}{\sqrt{2}}  a_+^w (c_+) \mathcal{W}				& \frac{-1}{\sqrt{2}}  a_+^w (-c_+) \mathcal{W}		\\
	-\frac{3}{2} s_+^2 \mathcal{W}			& -4 s_+^2 c_+^2 \mathcal{Z}			& \Delta_{+_1,+_1}   -\mathcal{A}   -a^Z_{+_1} \mathcal{Z}
					& \frac{-1}{\sqrt{2}}  a_+^w (-s_+) \mathcal{W}			& \frac{-1}{\sqrt{2}}  a_+^w (s_+) \mathcal{W}	\\
	0				& \frac{-1}{\sqrt{2}}  a_+^w (c_+) \mathcal{W}				& \frac{-1}{\sqrt{2}}  a_+^w (-s_+) \mathcal{W}			
					& 2\Delta_{\widetilde{0},\widetilde{0}}^{\mathbb{C}4}		& -\mathcal{Z}			\\
	0				& \frac{-1}{\sqrt{2}}  a_+^w (-c_+) \mathcal{W}			& \frac{-1}{\sqrt{2}}  a_+^w (s_+) \mathcal{W}		& -\mathcal{Z}		& 0
	\end{pmatrix}
\end{equation}

where the mass difference is given by 
$\Delta_{2,2} = 2 (\Delta_{++} + 2 \Re \delta_0)$, 
$\Delta_{+_2,+_2} = 2 (\Delta_+ + 2 \Re \delta_0)$, 
$\Delta_{+_1,+_1} = 2 (\Delta_- + 2 \Re \delta_0)$, and
$\Delta_{\widetilde{0},\widetilde{0}}^{\mathbb{C}4} = 2 \Delta_{\widetilde{0}}^{\mathbb{C}4} = 8 \Re \delta_0$.

The neutral weak coupling reads 
$a^Z_{++} = \left( 4c_w^2 - 1 \right)^2$, 
$a^Z_{+_2} = \left( 2c_w^2 - 2 c_+^2 + 1 \right)^2$, and 
$a^Z_{+_1} = \left( 2c_w^2 + 2 c_+^2 - 1 \right)^2$; 
and charged coupling 
$a^w_+ (\mathfrak{c}) = \mathfrak{c}^2  -4\sqrt{3} \, \mathfrak{c} \, \sqrt{1-\mathfrak{c}^2} \cos \widehat{\lambda}_0 +3$ 
with trigonometric function $\mathfrak{c}$ being either cos or sin.

Amplitudes for annihilation of DM pairs follow from the matrix below:
\begin{equation}
	\Gamma^{S=0}_{Q=0} =	\frac{\pi\alpha_w^2}{32 m_0^2}	
	\begin{pmatrix}		 
	\Gamma_{22}		& \Gamma_{22,++} (c_+)		& \Gamma_{22,++} (s_+)		& \frac{1}{\sqrt{2}} \Gamma_{22,00}		& \frac{1}{\sqrt{2}} \Gamma_{22,00}		\\
	\Gamma_{22,++} (c_+)		& \Gamma_{++} (c_+, c_+)		& \Gamma_{++} (c_+, s_+)		& \frac{1}{\sqrt{2}} \Gamma_{++,00} (c_+)
					& \frac{1}{\sqrt{2}} \Gamma_{++,00} (c_+)		\\
	\Gamma_{22,++} (s_+)		& \Gamma_{++} (c_+, s_+)		& \Gamma_{++} (s_+, s_+)		& \frac{1}{\sqrt{2}} \Gamma_{++,00} (s_+)
					& \frac{1}{\sqrt{2}} \Gamma_{++,00} (s_+)		\\	
	\frac{1}{\sqrt{2}} \Gamma_{22,00}			& \frac{1}{\sqrt{2}} \Gamma_{++,00} (c_+)			& \frac{1}{\sqrt{2}} \Gamma_{++,00} (s_+)
					& \frac{1}{2} \Gamma_{00}				& \frac{1}{2} \Gamma_{00}		\\	
	\frac{1}{\sqrt{2}} \Gamma_{22,00}			& \frac{1}{\sqrt{2}} \Gamma_{++,00} (c_+)			& \frac{1}{\sqrt{2}} \Gamma_{++,00} (s_+)
					& \frac{1}{2} \Gamma_{00}				& \frac{1}{2} \Gamma_{00}	
	\end{pmatrix}	\,,
\end{equation}

Explicitly for annihilation into different di-boson channels, we get:
\begin{equation}
	\begin{array}{l}
	{G_{\gamma\gamma}}^{S=0}_{Q=0} =	\sqrt{2}  \, \alpha \left(  4, \ 1, \ 1, \ 0, \ 0 \right)^T	,	\\		
	{G_{ZZ}}^{S=0}_{Q=0} =( \alpha_w / 2\sqrt{2} c_w^2 )	
		\left(    ( 4c_w^2 -1)^2, \ 		( 2c_w^2 +1)^2	 -8c_w^2 c_+^2, \ 		( 2c_w^2 -1)^2  +8c_w^2 c_+^2, \ 	\frac{1}{\sqrt{2}}, \ 	\frac{1}{\sqrt{2}} \ 	\right)^T,	\\	
		{G_{WW}}^{S=0}_{Q=0}	= ( \alpha_w / 2)		\left(  3, \ 		3 +4c_+^2	, \ 	7 -4c_+^2, \ 	\frac{7}{\sqrt{2}}, \ 		\frac{7}{\sqrt{2}} \ 	\right)^T.		
	\end{array}
\end{equation}

where $\Gamma_{22}  \equiv	t_w^4	+18 t_w^2		+99$; 
$\Gamma_{++} (\mathfrak{c}, \mathfrak{s})  \equiv	t_w^4	+2 (4 \mathfrak{c}^2 -3) (4 \mathfrak{s}^2 -3) t_w^2		+ (8 \mathfrak{c}^2 -9) (8 \mathfrak{s}^2 -9)
										+ 2 (4 \mathfrak{c}^2 +3) (4 \mathfrak{s}^2 +3)$; 
$\Gamma_{00}  \equiv	t_w^4	+2 t_w^2		+99$; 
$\Gamma_{22,++} (\mathfrak{c})  \equiv	t_w^4	+6 (4 \mathfrak{c}^2 -3)  t_w^2		- (48 \mathfrak{c}^2 -99)$; 
$\Gamma_{22,00}  \equiv	t_w^4	-6 t_w^2		+51$; and 
$\Gamma_{++,00} (\mathfrak{c})  \equiv	t_w^4	-2 (4 \mathfrak{c}^2 -3)  t_w^2		+ (48 \mathfrak{c}^2 +51)$. 
with $\mathfrak{c}$ and $\mathfrak{s}$ denoting trigonometric functions.

The potential among the charged states can be written as:
\begin{align}
	\label{eq:C4:V_Q=1}
	& V_{Q=1} = 	\\
	& \begin{pmatrix}
	\underset{2,+_2 \ }{\Delta -2\mathcal{A}}  -a^Z_{2+_2} \mathcal{Z}		& - \underset{2+_2, 2+_1}{a^Z \qquad \mathcal{Z}} 
			& - \underset{2+,+0 \quad }{a^w (c_+, c_+)} \mathcal{W}		& -i \underset{2+,+0 \qquad }{ a^w (c_+, -c_+)} \mathcal{W}
			& \underset{2+,+0 \quad }{ a^w (s_+, c_+)} \mathcal{W}		& i \underset{2+,+0 \qquad }{ a^w (s_+, -c_+)} \mathcal{W}		\\
	- \underset{2+_2, 2+_1}{a^Z \qquad \mathcal{Z}}				& \underset{2,+_1 \ }{\Delta -2\mathcal{A}}  -a^Z_{2+_1} \mathcal{Z}
			& - \underset{2+,+0 \qquad }{ a^w (c_+, -s_+)} \mathcal{W}	& -i \underset{2+,+0 \quad }{ a^w (c_+, s_+)} \mathcal{W}
			& \underset{2+,+0 \qquad }{ a^w (s_+, -s_+)} \mathcal{W}		& i \underset{2+,+0 \quad }{ a^w (s_+, s_+)} \mathcal{W}			\\
	- \underset{2+,+0 \qquad }{ a^{w*} (c_+, c_+)} \mathcal{W}			& - \underset{2+,+0 \qquad }{ a^{w*} (c_+, -s_+)} \mathcal{W}
			& \Delta_{+_2, \widetilde{0}}			& -i a^Z_{+_2 0} \ \mathcal{Z}			& 0			& -2i s_+ c_+ \mathcal{Z}		\\
	i \underset{2+,+0 \ \qquad }{ a^{w*} (c_+, -c_+)} \mathcal{W}			& i \underset{2+,+0 \qquad }{ a^{w*} (c_+, s_+)} \mathcal{W}
			& i a^Z_{+_2 0} \ \mathcal{Z}			& \Delta_{+_2, 0}	& 2i s_+ c_+ \ \mathcal{Z}			& 0				\\
	\underset{2+,+0 \qquad }{ a^{w*} (s_+, c_+)} \mathcal{W}			& \underset{2+,+0 \qquad }{ a^{w*} (s_+, -s_+)} \mathcal{W}
			& 0		& -2i s_+ c_+ \mathcal{Z}		& \Delta_{+_1, \widetilde{0}}			& -i a^Z_{+_1 0} \ \mathcal{Z}		\\
	-i \underset{2+,+0 \ \qquad }{ a^{w*} (s_+, -c_+)} \mathcal{W}			& -i \underset{2+,+0 \qquad }{ a^{w*} (s_+, s_+)} \mathcal{W}
			& 2i s_+ c_+ \mathcal{Z}				& 0				& i a^Z_{+_1 0} \ \mathcal{Z}		& \Delta_{+_1, 0}
	\end{pmatrix}			\nonumber
\end{align}

The mass difference terms read 
$\Delta_{2,+_2} = \Delta_{++} +\Delta_+   + 4 \Re \,\delta_0$, 
$\Delta_{2,+_1} = \Delta_{++} +\Delta_-   + 4 \Re \,\delta_0$, 
$\Delta_{+_2, \widetilde{0}} =\Delta_+   + 6 \Re \,\delta_0$, 
$\Delta_{+_2, 0} =\Delta_+   + 2 \Re \,\delta_0$, 
$\Delta_{+_1, \widetilde{0}} =\Delta_-   + 6 \Re \,\delta_0$, 
$\Delta_{+_1, 0} =\Delta_-   + 2 \Re \,\delta_0$.

The neutral weak couplings are given by 
$a^Z_{2+_2} =	\left( 4c_w^2 -1 \right)	\left( 2c_w^2 -2c_+^2 +1 \right)$, 
$a^Z_{2+_1} =	\left( 4c_w^2 -1 \right)	\left( 2c_w^2 +2c_+^2 -1 \right)$, 
$a^Z_{2+_2, 2+_1} =	2 s_+ c_+ \left( 4c_w^2 -1 \right)$, 
$a^Z_{+_2 0} =		2c_w^2 -2c_+^2 +1$, and
$a^Z_{+_1 0} =		2c_w^2 +2c_+^2 -1$; and charged coupling is 
$ a^w_{2+, +0} (\mathfrak{c}, \mathfrak{s})	=	(\sqrt{6} /4 ) \ \mathfrak{c}	\left( 2 \mathfrak{s} e^{-i \widehat{\lambda}_0}		-\sqrt{ 3 (1- \mathfrak{s}^2 } \right)$.

Annihilation matrix takes the form:
\begin{align}
	& \Gamma^{S=0}_{Q=1} =	\frac{\pi\alpha_w^2}{16 m_0^2}		\\	
	& \begin{pmatrix}		 
	\Gamma_{2+} (c_+, c_+)		& - \Gamma_{2+} (c_+, s_+)	& \underset{2+, 0+ \quad}{\Gamma \ (c_+, c_+)}		& i \underset{2+, 0+ \qquad}{\Gamma \ (c_+, -c_+)}									& \underset{2+, 0+ \qquad}{\Gamma \ (c_+, -s_+)}		& i \underset{2+, 0+ \quad}{\Gamma \ (c_+, s_+)}		\\
	- \Gamma_{2+} (c_+, s_+)		& \Gamma_{2+} (s_+, s_+)	& - \underset{2+, 0+ \quad}{\Gamma \ (s_+, c_+)}		& -i \underset{2+, 0+ \qquad}{\Gamma \ (s_+, -c_+)}
							& - \underset{2+, 0+ \qquad}{\Gamma \ (s_+, -s_+)}		& -i \underset{2+, 0+ \quad}{\Gamma \ (s_+, s_+)}		\\
	\underset{2+, 0+ \qquad}{\Gamma^* \ (c_+, c_+)}			& - \underset{2+, 0+ \qquad}{\Gamma^* \ (s_+, c_+)}	& \Gamma_{+0} (c_+, c_+)		
							& i \Gamma_{+0} (c_+, -c_+)				& \Gamma_{+0} (c_+, -s_+)			& i \Gamma_{+0} (c_+, s_+)	\\
	-i \underset{2+, 0+ \qquad}{\Gamma^* \ (c_+, -c_+)}			& i \underset{2+, 0+ \qquad}{\Gamma^* \ (s_+, -c_+)}		& -i \Gamma^*_{+0} (c_+, -c_+)											& \Gamma_{+0} (-c_+, -c_+)				& -i \Gamma_{+0} (-c_+, -s_+)			& \Gamma_{+0} (-c_+, s_+)	\\
	\underset{2+, 0+ \qquad}{\Gamma^* \ (c_+, -s_+)}			& - \underset{2+, 0+ \qquad}{\Gamma^* \ (s_+, -s_+)}		& \Gamma^*_{+0} (c_+, -s_+)
							& i \Gamma^*_{+0} (-c_+, -s_+)			& \Gamma_{+0} (-s_+, -s_+)			& i \Gamma_{+0} (-s_+, s_+)	\\
	-i \underset{2+, 0+ \qquad}{\Gamma^* \ (c_+, s_+)}			& i \underset{2+, 0+ \qquad}{\Gamma^* \ (s_+, s_+)}		& -i \Gamma^*_{+0} (c_+, s_+)												& \Gamma^*_{+0} (-c_+, s_+)				& -i \Gamma^*_{+0} (-s_+, s_+)			& \Gamma_{+0} (s_+, s_+)	
	\end{pmatrix}	\,,	\nonumber		
\end{align}

and annihilation into each channel is described by:
\begin{equation}
	\begin{array}{ll}
	{G_{\gamma W^-}}^{S=0}_{Q=1} =( s_w \alpha_w / 2 )	
		& \left(    3\sqrt{6} c_+ e^{\frac{i}{2} \widehat{\lambda}_0}, \ 	-3 \sqrt{6} s_+ e^{\frac{i}{2} \widehat{\lambda}_0}, \
		2 c_+ e^{-\frac{i}{2} \widehat{\lambda}_0}	- \sqrt{3} s_+ e^{\frac{i}{2} \widehat{\lambda}_0}, 		\right.	\\
		& \left.		-i (	 2 c_+ e^{-\frac{i}{2} \widehat{\lambda}_0}	+ \sqrt{3} s_+ e^{\frac{i}{2} \widehat{\lambda}_0})	, \ 
		-2 s_+ e^{-\frac{i}{2} \widehat{\lambda}_0}	- \sqrt{3} c_+ e^{\frac{i}{2} \widehat{\lambda}_0}, \ 		
		i (	 2 s_+ e^{-\frac{i}{2} \widehat{\lambda}_0}	- \sqrt{3} c_+ e^{\frac{i}{2} \widehat{\lambda}_0} )		\right)^T,	\\	
	{G_{ZW^-}}^{S=0}_{Q=1}	= ( \alpha_w / 2 c_w)		
		&\left(	\sqrt{6} (3c_+^2 -1) c_+ e^{\frac{i}{2} \widehat{\lambda}_0} , \ 	- \sqrt{6} (3c_+^2 -1) s_+ e^{\frac{i}{2} \widehat{\lambda}_0} , 	\right.	 \\
		& 2 (c_+^2 -1) c_+ e^{-\frac{i}{2} \widehat{\lambda}_0}		- \sqrt{3} (c_+^2 +1) s_+ e^{\frac{i}{2} \widehat{\lambda}_0} , \ 
		-i [2 (c_+^2 -1) c_+ e^{-\frac{i}{2} \widehat{\lambda}_0} 		+ \sqrt{3} (c_+^2 +1) s_+ e^{\frac{i}{2} \widehat{\lambda}_0} ] ,	\\
		&  \left.	-2 (c_+^2 -1) s_+ e^{-\frac{i}{2} \widehat{\lambda}_0}		- \sqrt{3} (c_+^2 +1) c_+ e^{\frac{i}{2} \widehat{\lambda}_0} , \ 
		i [2 (c_+^2 -1) s_+ e^{-\frac{i}{2} \widehat{\lambda}_0} 		- \sqrt{3} (c_+^2 +1) c_+ e^{\frac{i}{2} \widehat{\lambda}_0} ] 		\right)^T.		
	\end{array}
\end{equation}

where $\Gamma_{2+} (\mathfrak{c}, \mathfrak{s})	\equiv	6 \, a_{3,3}$;
$\Gamma_{2+, 0+}  (\mathfrak{c}, \mathfrak{s})		\equiv	\sqrt{6}  \, \mathfrak{c}	
	\left(		2\, a_{3,+} \mathfrak{s} \, e^{-i \widehat{\lambda}_0}		+\sqrt{3} \, a_{3,-} \sqrt{1-\mathfrak{s}^2}	\right)$; and 
$ \Gamma_{+0} (\mathfrak{c}, \mathfrak{s})	\equiv	4 \, a_{+,+} \mathfrak{c} \mathfrak{s}	+3\, a_{-,-} \sqrt{1-\mathfrak{c}^2} \sqrt{1-\mathfrak{s}^2}
	+ 2 \sqrt{3} a_{+,-}		\left(	\mathfrak{c} \sqrt{1-\mathfrak{s}^2} e^{i \widehat{\lambda}_0}		+ \mathfrak{s} \sqrt{1-\mathfrak{c}^2} e^{-i \widehat{\lambda}_0}	\right)$; 
with $ a_{\mathfrak{p}, \mathfrak{q}} =	\mathfrak{p} \mathfrak{q} s_w^2	+(\mathfrak{p} c_w^2 -1 ) (\mathfrak{q} c_w^2 -1 ) / c_w^2$.

Potential and annihilation matrices that describe the doubly-charged states are:
\begin{equation}
	V^{S=0}_{Q=2} =	
	\begin{pmatrix}
		\Delta_{++} +6 \Re \delta_0					& \frac{i}{2} \left( 4c_w^2 -1 \right) \mathcal{Z}		\\
		- \frac{i}{2} \left( 4c_w^2 -1 \right) \mathcal{Z}			& \Delta_{++} +2 \Re \delta_0
	\end{pmatrix},			\qquad
	\Gamma^{S=0}_{Q=2} =		\frac{3 \pi\alpha_w^2}{4 m_0^2}
	\begin{pmatrix}
		1	& -i 	\\		i 	& 1
	\end{pmatrix}	
\end{equation}

In case of spin-one pairs, the potential between neutral states can be written as:
\begin{equation}
	V^{S=1}_{Q=0} =	
	\begin{pmatrix}
		\Delta_{2,2}   -4\mathcal{A}   -a^Z_{++} \mathcal{Z}		& -\frac{3}{2} c_+^2	\mathcal{W}		& -\frac{3}{2} s_+^2	\mathcal{W}		\\
		-\frac{3}{2} c_+^2 \mathcal{W}			& \Delta_{+_2,+_2}   -\mathcal{A}   -a^Z_{+_2} \mathcal{Z}		& -4 s_+^2 c_+^2 \mathcal{Z}		\\
		-\frac{3}{2} s_+^2 \mathcal{W}			& -4 s_+^2 c_+^2 \mathcal{Z}			& \Delta_{+_1,+_1}   -\mathcal{A}   -a^Z_{+_1} \mathcal{Z}
	\end{pmatrix}
\end{equation}
		
Note that this is the same as spinless $V^{S=0}_{Q=0}$, \ref{eq:C4:V_S0_Q0} except for the identical pairs $\chi^0 \chi^0$ and $\widetilde{\chi}^0 \widetilde{\chi}^0$ rows and columns that are removed due to Pauli exclusion.

Pair annihilation into the SM fermions and scalars in this sector can be formulated as:
\begin{equation}
	\Gamma^{S=1}_{Q=0} =		\frac{\pi\alpha_w^2}{32 m_0^2}			
	\begin{pmatrix}		 
		41 t_w^4  +225							& 41 t_w^4  +75 \left( 4c_+^2 -3 \right)		& 41 t_w^4  +75 \left( 4s_+^2 -3 \right)		\\
		41 t_w^4  +75 \left( 4c_+^2 -3 \right)			& 41 t_w^4  +25 \left( 4c_+^2 -3 \right)^2		& 41 t_w^4  +25 \left( 4c_+^2 -3 \right)  \left( 4s_+^2 -3 \right)		\\
		41 t_w^4  +75 \left( 4s_+^2 -3 \right)			& 41 t_w^4  +25 \left( 4c_+^2 -3 \right)  \left( 4s_+^2 -3 \right)			& 41 t_w^4  +25 \left( 4s_+^2 -3 \right)^2
	\end{pmatrix}		 
\end{equation}

The same potential \ref{eq:C4:V_Q=1} forms spin-one charged pairs of dark matter, but the annihilation matrix in this case is given by:
\begin{align}
	& \Gamma^{S=1}_{Q=1} =	\frac{25 \pi\alpha_w^2}{32 m_0^2}		\\	
	& \begin{pmatrix}
	6 \, c_+^2			& -6 \, c_+ s_+			& \underset{2+, 0+ \quad}{\Gamma' \, (c_+, c_+)}	 	& i \underset{2+, 0+ \qquad}{\Gamma' \, (c_+, -c_+)}
					& \underset{2+, 0+ \qquad}{\Gamma' \, (c_+, -s_+)}				& i \underset{2+, 0+ \quad}{\Gamma' \, (c_+, s_+)}		\\
	-6 \, c_+ s_+		& 6 \, s_+^2			& - \underset{2+, 0+ \quad}{\Gamma' \, (s_+, c_+)}		& -i \underset{2+, 0+ \qquad}{\Gamma' \, (s_+, -c_+)}
					& - \underset{2+, 0+ \qquad}{\Gamma' \, (s_+, -s_+)}		& -i \underset{2+, 0+ \quad}{\Gamma' \, (s_+, s_+)}		\\
	\underset{2+, 0+ \qquad}{\Gamma'^* \, (c_+, c_+)}			& - \underset{2+, 0+ \qquad}{\Gamma'^* \, (s_+, c_+)}		& \Gamma'_{+0} (c_+, c_+)	
					& i \Gamma'_{+0} (c_+, -c_+)			& \Gamma'_{+0} (c_+, -s_+)			& i \Gamma'_{+0} (c_+, s_+)	\\				
	-i \underset{2+, 0+ \qquad}{\Gamma'^* \, (c_+, -c_+)}		& i \underset{2+, 0+ \qquad}{\Gamma'^* \, (s_+, -c_+)}		& -i \Gamma'^*_{+0} (c_+, -c_+)									& \Gamma'_{+0} (-c_+, -c_+)			& -i \Gamma'_{+0} (-c_+, -s_+)			& \Gamma'_{+0} (-c_+, s_+)	\\
	\underset{2+, 0+ \qquad}{\Gamma'^* \, (c_+, -s_+)}			& - \underset{2+, 0+ \qquad}{\Gamma'^* \, (s_+, -s_+)}		& \Gamma'^*_{+0} (c_+, -s_+)
					& i \Gamma'^*_{+0} (-c_+, -s_+)		& \Gamma'_{+0} (-s_+, -s_+)			& i \Gamma'_{+0} (-s_+, s_+)	\\
	-i \underset{2+, 0+ \qquad}{\Gamma'^* \, (c_+, s_+)}			& i \underset{2+, 0+ \qquad}{\Gamma'^* \, (s_+, s_+)}		& -i \Gamma'^*_{+0} (c_+, s_+)									& \Gamma'^*_{+0} (-c_+, s_+)			& -i \Gamma'^*_{+0} (-s_+, s_+)		& \Gamma'_{+0} (s_+, s_+)	
	\end{pmatrix}		\nonumber
\end{align}

where 
$ \Gamma'_{+0} (\mathfrak{c}, \mathfrak{s})	\equiv	4 \, \mathfrak{c} \mathfrak{s}	+3\, \sqrt{1-\mathfrak{c}^2} \sqrt{1-\mathfrak{s}^2}
	+ 2 \sqrt{3}	\left(	\mathfrak{c} \sqrt{1-\mathfrak{s}^2} e^{i \widehat{\lambda}_0}		+ \mathfrak{s} \sqrt{1-\mathfrak{c}^2} e^{-i \widehat{\lambda}_0}	\right)$;
and 	\\
$\Gamma'_{2+, 0+}  (\mathfrak{c}, \mathfrak{s})		\equiv	\sqrt{2}  \, \mathfrak{c}	
	\left(		2 \sqrt{3} \, \mathfrak{s} \, e^{-i \widehat{\lambda}_0}		+3 \, \sqrt{1-\mathfrak{s}^2}	\right)$. 
		
It should be reminded that Landau-Yang's theorem limits the annihilation final sates to the pairs of fermions and scalars, as a result, doubly-charged states are not allowed in spin-one sector.

In what follows we take the coupling $\lambda_0$ to be real.				
								
In figure \ref{fig:C2eft:Xnv_m}, we show the annihilation cross-section of $\chi^0\chi^0$ pair into $W^+W^-$ and $\gamma\gamma$ as a function of dark matter mass. In the sequence of resonances, the peeks occur due to addition of the enhancement factors corresponding to the ground state energy of DM bound states \cite{BS}. In addition, one can find dips in the spectrum which are due to \emph{Ramsauer-Townsend effect}. In the mass intervals with repulsive effective potential, the destructive interference of Sommerfeld factors causes a cancellation between the terms in \ref{eq:Xn_AB} resulting in the suppression of the cross-section \cite{Ramsauer_Townsend}.

Sommerfeld effect has enhanced the annihilation rate by orders of magnitude even for processes like $\chi^0 \chi^0 \to \gamma \gamma$ which are not allowed in perturbative framework at tree-level.

In general, the quadret cross-section is larger than that of the doublet model. That is owing to the fact that by increasing the dimension, number of terms summed up to compose annihilation amplitude \ref{eq:Xn_AB} grows. In addition, the non-perturbative coefficients and other quantum numbers forming these terms become greater. To exemplify this, consider the $\gamma\gamma$ channel 
$\sigma_{\gamma\gamma} = (\pi \alpha^2 / 2 m^2_0) \left| 4 d_{\overline{++}++}	 +d_{\bar{+}_1 +_1}	+d_{\bar{+}_2 +_2} \right|^2$ 
which has 8 additional terms compared to the doublet case. Not only these terms have larger Sommerfeld factors, but they can also be enhanced up to a factor of 16 due to the doubly charged state $\overline{\chi}^{++} \chi^{++}$.

This time, in figure \ref{fig:C2eft:Xnv_v} we illustrate the behaviour of annihilation cross section for the $\chi^0\chi^0$ state as a function of velocity. As explained in the previous subsection, the cross-section keeps growing until a threshold which depends on DM mass, and then gets to a constant value at small velocities. The fluctuations in the cross-section before plateau at non-relativistic regime is the result of quantum interference between five different Sommerfeld enhancement factors reinforcing or diminishing each other.


\section{Constraints}

Having discussed the theoretical aspects of electroweak dark matter hypothesis, we turn into detailed phenomenological analysis of the model in the coming sections. At first thermal relic abundance is precisely calculated to find the thermal value of DM mass. Then we examine the gamma ray constraints form near the MW black hole, the galactic centre, dwarf galaxies and finally monochromatic lines.


\subsection{Relic Density}

\begin{figure} [t] 				
	\begin{subfigure}{.5\linewidth}
		\centering
		\includegraphics[width=\linewidth]{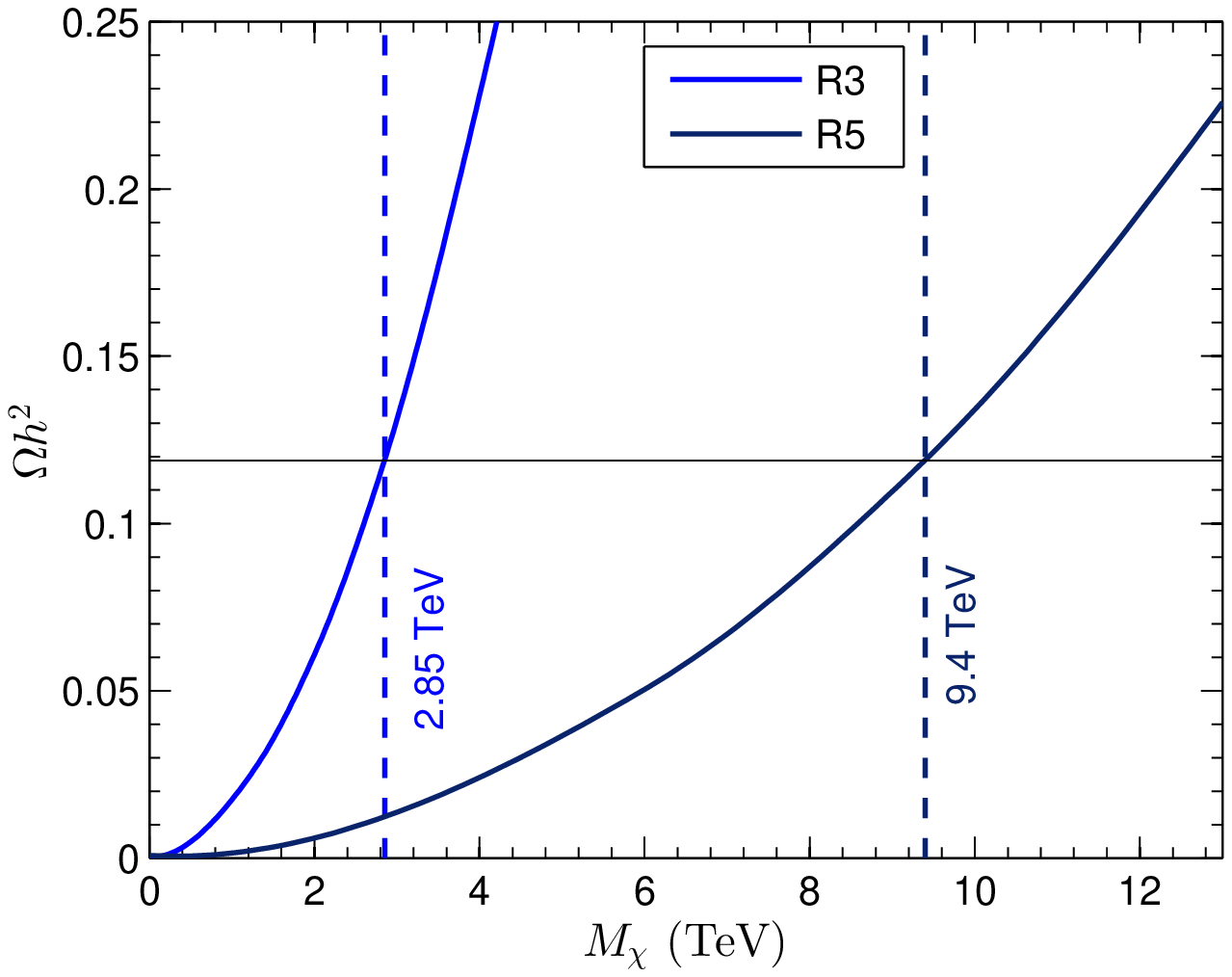}
		\subcaption{}
		\label{fig:Relic_R}
	\end{subfigure}%
	\begin{subfigure}{.5\linewidth}
		\centering
		\includegraphics[width=\linewidth]{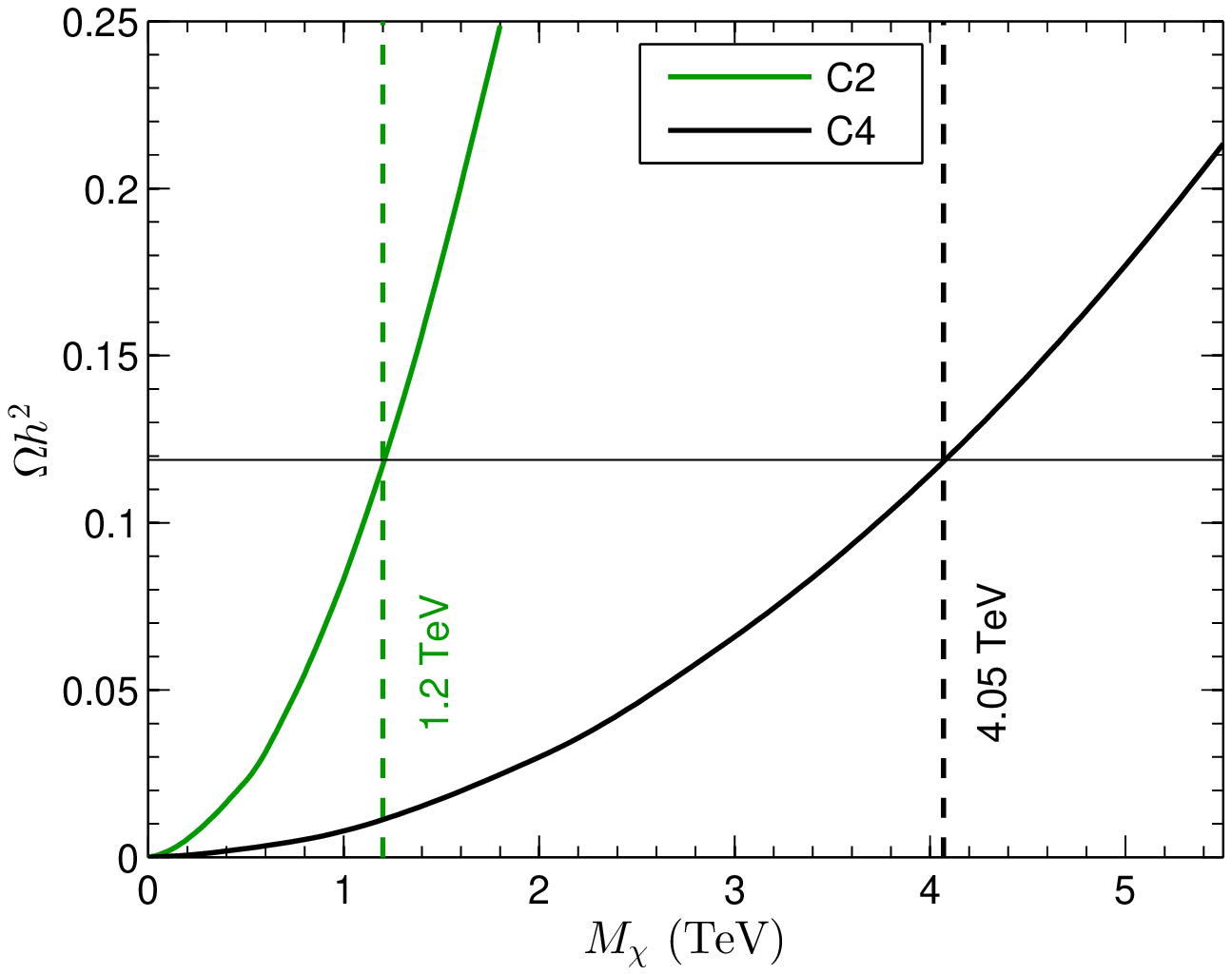}
		\subcaption{}
		\label{fig:Relic_C}
	\end{subfigure}
	\caption{Relic abundance of different EWDM modules which includes real triplet (blue) and quintuplet (navy) in the right panel \ref{fig:Relic_R} as well as pseudo-real doublet (green) and quadret (black) in the right panel \ref{fig:Relic_C} versus DM mass. The horizontal line refers to the observed value of dark matter abundance $\Omega h^2 = 0.1188\pm00010$ \cite{Planck}. EWDM freeze-out mass is the point where this horizontal line crosses the relic curve. This value is shown by a vertical dashed line for each representation.}
	\label{fig:Relic}
\end{figure}

Dark matter number density is affected by annihilation cross-section $\sigma_{ij}$ of different dark sector particles $\chi_i \chi_j$ into EW gauge boson pairs. In order to obtain today's abundance for an EWDM $n$-tuplet, we need to solve a set of $n$ coupled Boltzmann equations \cite{Boltzmann}. Since all other components will ultimately decay into DM candidate, it suffices to compute the total number density of odd particles $N =  \sum_{i=1}^{n} N_i$. The Boltzmann equation for $N$ is obtained by summing over all the coupled equations for individual dark particles. Decay and annihilation between different species cannot affect the total density as they are cancelled out in this sum.

Assuming that dark particles maintain their relative density long after thermal freeze-out $N_i/N \approx N^{eq}_i / N^{eq}$, it writes \cite{Coann}:
\begin{equation}
	\label{eq:BE}
	\frac{\mathrm{d}N}{\mathrm{d}t} +3HN =		\sum_{i,j=1}^n 		\langle \sigma_{ij} v\rangle		\left( N_i^{\mathrm{eq}} N_j^{\mathrm{eq}} - N_i N_j \right)
						\ \approx 		\langle \sigma_{\mathrm{eff}} v\rangle		\left( N_{\mathrm{eq}}^2 - N^2 \right)	\ .
\end{equation}	

where $H=\pi \sqrt{g_\mathrm{eff} /30} \, T^2/M_\mathrm{pl}$ is the Hubble constant, and equilibrium number density defined as $N_i^{eq} = (g_i /2\pi^2) m_i^2 T K_2(m_i /T) \approx g_i (m_i T / 2\pi )^\frac{3}{2} e^{-m_i/T}$ with $g_\mathrm{eff}$ being the effective degrees of freedom and $g_i$ number of internal degrees of freedom for the $i^\mathrm{th}$ particle. 
\footnote{In the non-relativistic limit  $x \gg 1$, one can use the asymptotic expansion of the modified Bessel function of the second kind $K_n(x) \approx e^{-x} \sqrt{\pi /2x}$ \cite{Bessel}.}
  The thermally averaged cross-section times velocity $\langle \sigma v\rangle$, in the degenerate mass case, is calculated through the integral \cite{Boltzmann, v_rel, *Xn_avg}:
\begin{equation}
	\langle \sigma_{ij} v_{ij} \rangle 	\equiv 		\int_0^1  \mathcal{P}(v_{ij}) \,\sigma_{ij} v_{ij} 	\mathrm{d} v_{ij}
				=	\frac{1}{8 m_0^4 T 	K_2^2 \left( \frac{m_0}{T} \right)}	
					\int_{4m_0^2}^\infty		\sigma_{ij} (\mathfrak{s}) \sqrt{\mathfrak{s}}	\left( \mathfrak{s} -4m_0^2 \right)	
							K_1 \left( \frac{\sqrt{\mathfrak{s}}}{T} \right)	\mathrm{d} \mathfrak{s}
\end{equation}

The quantity is integrated over the relativistic velocity distribution \\
$\mathcal{P}(v_{ij}) =	x \, \gamma_{ij}^3 ( \gamma_{ij} -1)		\sqrt{\gamma_{ij} +1}	\, K_1 \left( \sqrt{2} \, x  \sqrt{\gamma_{ij} +1} \right)	/\sqrt{2} K_2^2 (x)$
where $x = m_0 /T$ is inverse temperature, 
$K_\mathfrak{n} (x)$ the modified Bessel function of the second kind of order $\mathfrak{n}$, 
and relative Lorentz factor $\gamma_{ij} = 1/ \sqrt{ 1 -v_{ij}}$ 
is a function of the relative velocity $ v_{ij} =	\sqrt{ (\textbf{v}_i -\textbf{v}_j)^2  -(\textbf{v}_i \times \textbf{v}_j)^2 }	/ (1 -\textbf{v}_i . \textbf{v}_j)$.

Alternatively the integration can be performed over Mandelstam variable $\mathfrak{s} \equiv (p_i + p_j)^2$ 
by change of variable $ v_{ij} = \mathfrak{s} \sqrt{\mathfrak{s} -4m_0^2} / (\mathfrak{s} -2m_0^2)$.

Considering the ratio of number density of the $i^\mathrm{th}$ component to the total density at equilibrium $R_i \equiv \left(N_i/N\right)^{\mathrm{eq}}   = g_i B_i / \sum_j g_j B_j$, with relative Boltzmann factor being $B_i = (1+\mathcal{D}_i)^{3/2} \, e^{-x\mathcal{D}_i}$, and relative mass difference $\mathcal{D}_i = (m_i -m_0)/m_0$; then the effective cross-section can be defined as \cite{Coann, KK2}: 
\begin{equation}
	\label{eq:Xn_eff}
	\sigma_{\mathrm{eff}} \equiv 	\sum_{ij+\mathrm{cc}}  R_i R_j \sigma_{ij}		
							= \frac{1}{ \left( \sum_i g_i B_i \right)^2 }		\sum_{ij+\mathrm{cc}} g_i g_j \sigma_{ij} \,B_i B_j	\,,
\end{equation}

with degrees of freedom $g_i=2$ for a fermion, and summation is performed over all odd particles $\chi_i$ and anti-particles $\bar{\chi}_i$.

In the degenerate limit $\mathcal{D}_i \approx 0$, the relative Boltzmann factors approach the maximum value at unity $B_i \to 1$. So, coannihilation can increase the cross-section, and hence decrease the abundance by a large factor. In the regime of degeneracy, effective cross-section \ref{eq:Xn_eff} reduces to:
\begin{equation}
	\label{eq:Xn_eff_degen}
	\sigma_{\mathrm{eff}} = 	\frac{1}{n_\chi^2}		\sum_{i\ge j} \mathcal{C}_{i,j} \mathcal{C}_{i,\overline{j}} \,\sigma_{\{i,j\}+\mathrm{cc}} 	\,,
\end{equation}

here, $n_\chi$ is the total number of particles and antiparticles within the multiplet, and sum only runs over particles. $\sigma_{\{i,j\}+\mathrm{cc}}$ indicates the annihilation cross-section for any of the pairs $\chi_i \chi_j$, $\chi_j \chi_i$, $\bar{\chi}_i \bar{\chi}_j$ or $\bar{\chi}_j \bar{\chi}_i$. 

$ \mathcal{C}_{i,j} = 2/(1+ \delta_{i,j})$ is combinatoric factor: If $i \ne j$, then $\sigma_{\{i,j\}}$ appears twice in the sum \ref{eq:Xn_eff}. So $ \mathcal{C}_{i,j} =1$ if incoming particles are identical, and gets a factor of 2 otherwise. If the annihilating state is not self conjugated then $\sigma_{ij}+ \mathrm{cc}$ is presented two times in the summation. Therefore $\mathcal{C}_{i,\overline{j}} =1$ for a neutral state $Q=0$, and 2 for a charged pair.

Effective cross section for different representations of EWDM can be written as follows. For the real triplet:
\begin{equation}
	{\sigma_{\mathrm{eff}}}^{\mathbb{R}3} =	\frac{1}{9}	
		\left[ \sigma_{00}^{S=0}		+ 2 \left( \sigma_{\{\ -,+ \}}^{S=0}  + \sigma_{\{ -,+ \}}^{S=1} \right)
			+ 4 \left( \sigma_{\{ 0,+ \} + \mathrm{cc} }^{S=0} + \sigma_{\{ 0,+ \} + \mathrm{cc} }^{S=1} \right)
			+ 2 \, \sigma_{ ++\, + \mathrm{cc} }^{S=0} 	 \right]	\,,
\end{equation}

The real quintuplet:
\begin{align}
	{\sigma_{\mathrm{eff}}}^{\mathbb{R}5} &=	\frac{1}{25}	\left[ \sigma_{00}^{S=0}		
			+ 2 \left( \sigma_{\{\ -,+ \}}^{S=0}  + \sigma_{\{ -,+ \}}^{S=1} \right)	+ 2 \left( \sigma_{\{\ --,++ \}}^{S=0}  + \sigma_{\{ --,++ \}}^{S=1} \right)
			+ 4 \left( \sigma_{\{ 0,+ \} + \mathrm{cc} }^{S=0} + \sigma_{\{ 0,+ \} + \mathrm{cc} }^{S=1} \right)		\right.	\\	\nonumber
			&+ \left.	4 \left( \sigma_{\{ -,++ \} + \mathrm{cc} }^{S=0} + \sigma_{\{ -,++ \} + \mathrm{cc} }^{S=1} \right)
			+ 4 \, \sigma_{\{ 0,++ \} + \mathrm{cc} }^{S=0}	 	+ 2 \, \sigma_{ ++\, + \mathrm{cc} }^{S=0} 		 \right]	\,,
\end{align}

for the pseudo-real doublet:
\begin{equation}
	{\sigma_{\mathrm{eff}}}^{\mathbb{C}2} =	\frac{1}{16}	\left[ \sigma_{00}^{S=0} +  \sigma_{\widetilde{0}\widetilde{0}}^{S=0}		
			+ 2 \left( \sigma_{\{\bar{+},+\}}^{S=0}  + \sigma_{\{\bar{+},+\}}^{S=1} \right)
			+ 4 \left( \sigma_{\{0,+\} + \mathrm{cc}}^{S=0} + \sigma_{\{0,+\} + \mathrm{cc}}^{S=1} \right)
			+ 4 \left( \sigma_{\{\widetilde{0},+\} + \mathrm{cc}}^{S=0} + \sigma_{\{\widetilde{0},+\} + \mathrm{cc}}^{S=1} \right)	 \right]	\,,
\end{equation}

and finally the pseudo-real quadruplet:
\begin{align}
	{\sigma_{\mathrm{eff}}}^{\mathbb{C}4} &=	\frac{1}{64}	\left[ \sigma_{00}^{S=0} +  	\sigma_{\widetilde{0}\widetilde{0}}^{S=0}		
			+ 2 \left( \sigma_{\{\bar{+}_1,+_1\}}^{S=0}  	+ \sigma_{\{\bar{+}_1,+_1\}}^{S=1} \right)		
			+ 2 \left( \sigma_{\{\bar{+}_2,+_2\}}^{S=0}  	+ \sigma_{\{\bar{+}_2,+_2\}}^{S=1} \right)
			+ 2 \left( \sigma_{\{\bar{2},2\}}^{S=0}  		+ \sigma_{\{\bar{2},2\}}^{S=1} \right)	\right. 		\\	\nonumber
			&+ \left. 	4 \left( \sigma_{\{0,+_1\} + \mathrm{cc}}^{S=0}		 + \sigma_{\{0,+_1\} + \mathrm{cc}}^{S=1} \right)
			+  4 \left( \sigma_{\{\widetilde{0},+_1\} + \mathrm{cc}}^{S=0} 	+ \sigma_{\{\widetilde{0},+_1\} + \mathrm{cc}}^{S=1} \right)
			+ 4 \left( \sigma_{\{0,+_2\} + \mathrm{cc}}^{S=0}		 + \sigma_{\{0,+_2\} + \mathrm{cc}}^{S=1} \right)	\right.		\\	\nonumber
			&+  \left.	4 \left( \sigma_{\{\widetilde{0},+_2\} + \mathrm{cc}}^{S=0} 	+ \sigma_{\{\widetilde{0},+_2\} + \mathrm{cc}}^{S=1} \right)
			+ 4 \left( \sigma_{\{\bar{+}_1,2\} + \mathrm{cc}}^{S=0}		 + \sigma_{\{\bar{+}_1,2\} + \mathrm{cc}}^{S=1} \right)
			+ 4 \left( \sigma_{\{\bar{+}_2,2\} + \mathrm{cc}}^{S=0}		 + \sigma_{\{\bar{+}_2,2\} + \mathrm{cc}}^{S=1} \right)	\right.	\\	\nonumber
			&+ \left.	4 \, \sigma_{\{0,2\} + \mathrm{cc}}^{S=0} 	+ 4 \,\sigma_{\{\widetilde{0},2\} + \mathrm{cc}}^{S=0} 	 \right]	\,.
\end{align}

To factor out the expansion of the universe, in practice, the Boltzmann equation \ref{eq:BE} is solved for the comoving number density $Y \equiv N/s$ with respect to $x$ \cite{Boltzmann, Neutralino}. 
\begin{equation}
	\frac{\mathrm{d}Y}{\mathrm{d} x} =		\frac{s \langle \sigma_\mathrm{eff} v \rangle }{xH}	
		\left( 1 -\frac{x}{3 h_\mathrm{eff}} 	\frac{\mathrm{d} h_\mathrm{eff}}{\mathrm{d} x} \right)	\left( Y_{\mathrm{eq}}^2	-Y^2 \right)
\end{equation}

where entropy density reads 
$s = (2 \pi^2 /45) h_\mathrm{eff} T^3$ 
with $ h_\mathrm{eff}$ effective degrees of freedom in entropy, and equilibrium yield defined as 
$Y_\mathrm{eq} = ( 45 / 2 \pi^2 ) \ n_\chi x^2 K_2 (x) /  h_\mathrm{eff}$.

Figure \ref{fig:Relic} plots the relic density for different multiplets of electroweak dark matter as a function of mass. The last free parameter in the EWDM theory is mass. It can be fixed by matching the predicted relic abundance to today's measured value of $\Omega h^2 = 0.1188\pm00010$ \cite{Planck}. After doing so, the thermal value of dark matter mass is determined to be 1.20 TeV for pseudo-real doublet, 4.05 TeV for quadruplet, 2.85 TeV for real triplet, and 9.40 TeV for quintuplet.

As explained by WIMP miracle, EWDM which interacts at weak force strength can provide the observed abundance through freeze-out mechanism as long as it has a mass in TeV range.

Relic density of EWDM is however considerably lower than a usual WIMP particle. Because the effective cross-section which is essentially inversely proportional to the abundance has been largely enhanced by non-perturbative effects (\ref{eq:Xn_AB}) as well as contribution from coannihilation of heavier components (\ref{eq:Xn_eff_degen}).


\subsection{Indirect Detection}

\begin{figure} [t] 				
	\begin{subfigure}{.5\linewidth}
		\centering
		\includegraphics[width=\linewidth]{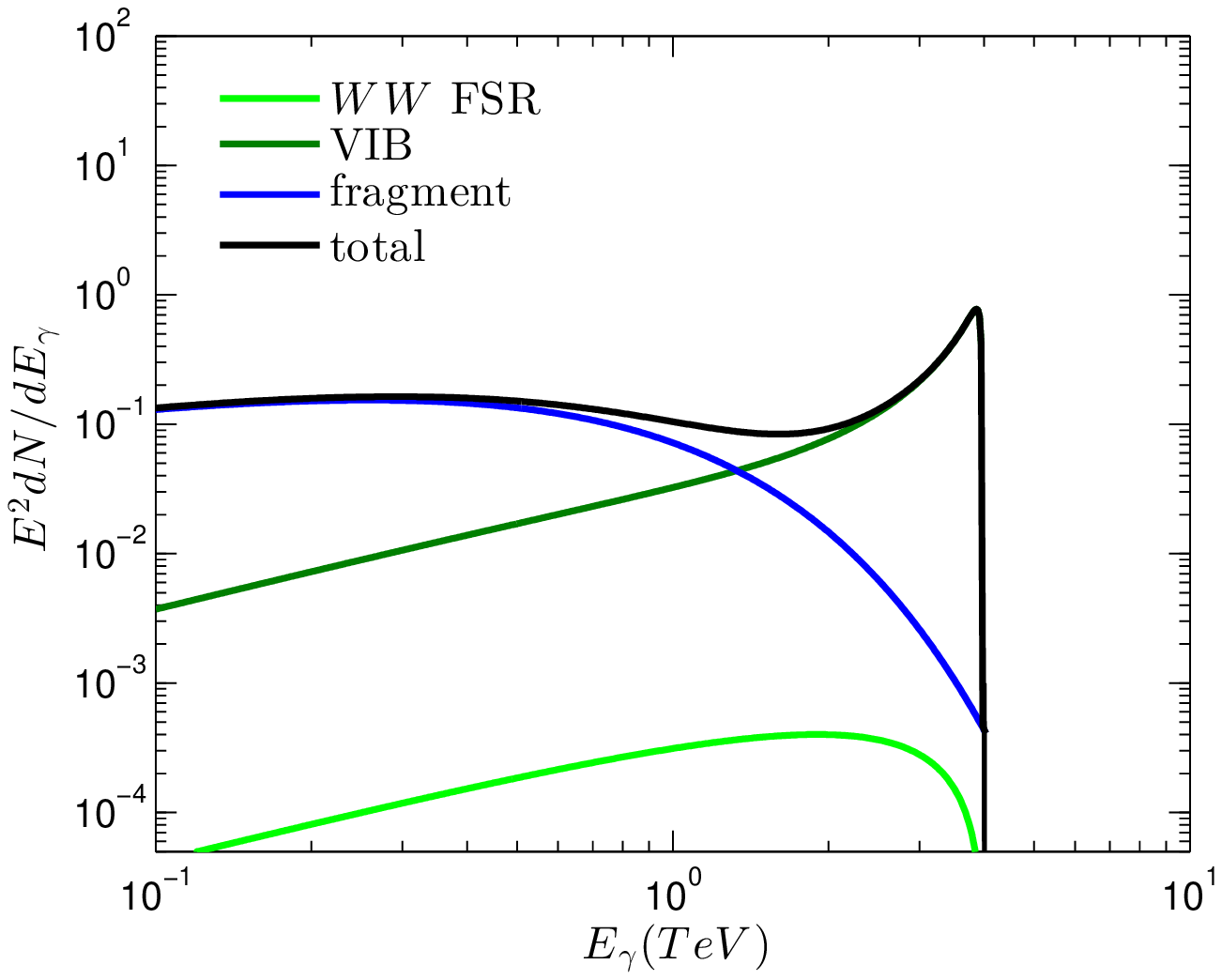}
		\subcaption{}
		\label{fig:E2dNdE_C4eft_Noline}
	\end{subfigure}%
	\begin{subfigure}{.5\linewidth}
		\centering
		\includegraphics[width=\linewidth]{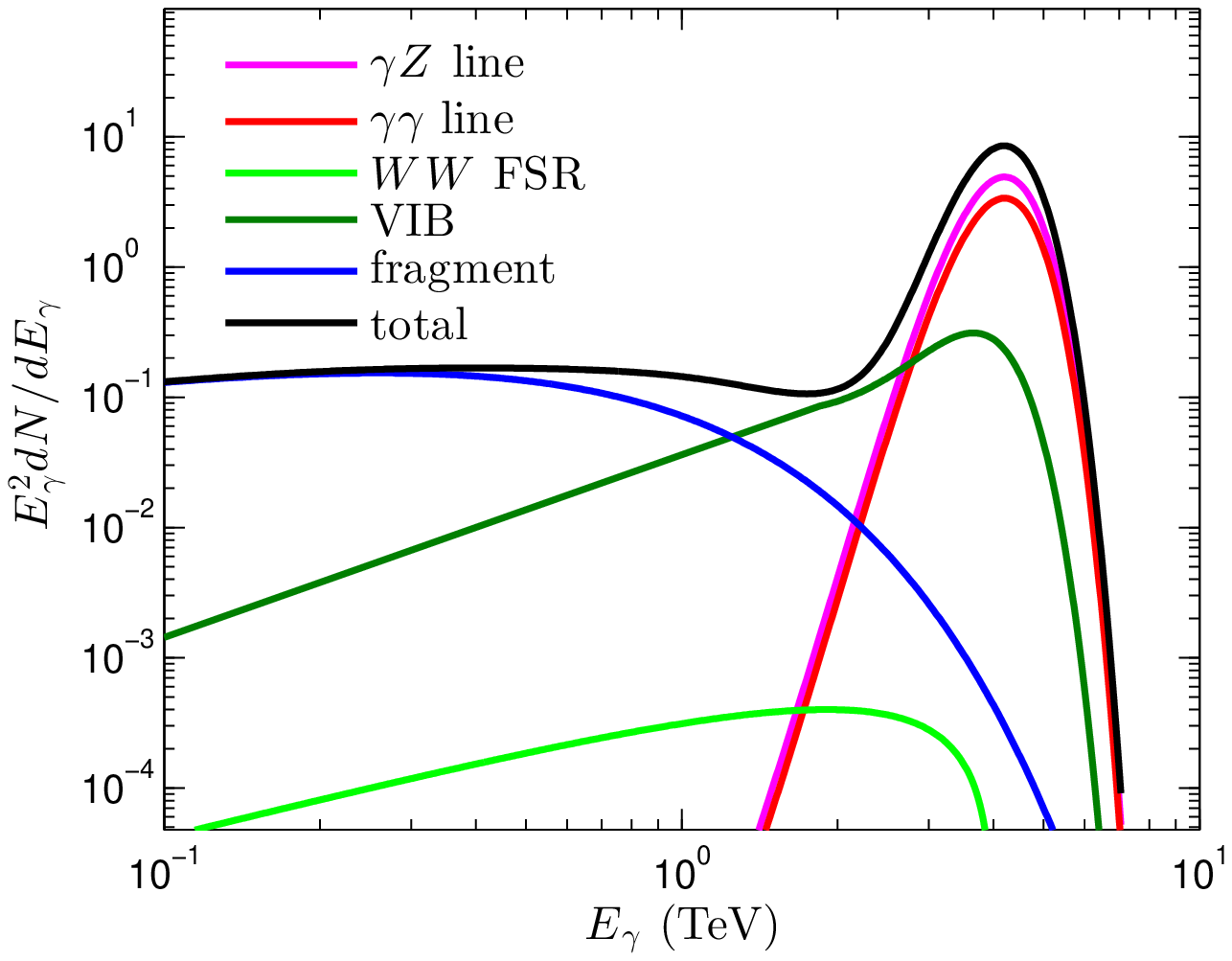}
		\subcaption{}
		\label{fig:E2dNdE_C4eft_Line}
	\end{subfigure}
	\caption{Weighted differential photon multiplicity for complex quadret with freeze-out mass of 4.05 TeV (black). Also shown, contribution from $W^+W^-$ and $ZZ$ fragmentation (blue), VIB (dark green), $W^+W^-$ FSR (light green), $\gamma\gamma$ line (red) and $\gamma Z$ line (magenta). The branching ratios for diboson channels read $\mathcal{B}_{W^+W^-}$ = 0.05, $\mathcal{B}_{ZZ}$  = 0.35,  $\mathcal{B}_{\gamma Z}$ = 0.45 and  $\mathcal{B}_{\gamma\gamma}$ = 0.15. Right panel (\ref{fig:E2dNdE_C4eft_Line}) depicts the same spectra, as seen by a detector with resolution of 15\%.}
	\label{fig:E2dNdE_C4eft}
\end{figure}

There are obvious advantages associated with use of gamma-rays in indirect searches for dark matter. Unlike charged particles that are easily affected by magnetic fields, photon radiation is not deflected, and thus can help locate the signal source. Apart from this, gamma-rays have negligible energy losses, and preserve spectral information \cite{Gamma_ray}.
 
Electroweak dark matter can annihilate, and produce photon rays which their flux spectrum reveals significant information about the nature of DM interactions.

In this work we use \emph{Navarro-Frenk-White (NFW)} as a common example of cuspy profiles which are mainly extracted from the structure formation N-body simulation results \cite{NFW1, NFW2}:
\begin{equation}
\label{eq:NFW}
	\rho(r)= 	\frac{\rho_s}{ (r/r_s)^\gamma	 \left[ 1+ (r/r_s)^\alpha \right] ^{(\beta-\gamma)/\alpha} }
\end{equation}

where $r$ is the distance from the point of observation to the Galactic Centre (GC), the scale density $\rho_s =$ 0.184 GeV/cm\textsuperscript{3}, and $r_s$ is the scale radius which is taken as  24.42 kpc, using the original NFW parameterisation $(\alpha, \beta, \gamma) = (1, 3, 1)$.
\footnote{ DM profiles are scaled so that first of all, the local DM density of  $\rho_\odot \sim 0.3 \ GeV/cm^3$ is reproduced at the location of the solar system \cite{DM_density}, and secondly, the total DM content within 60 kpc radius of the Milky Way (MW) satisfies $M_{60} = 4.7 \times 10^{11}\ M_\odot$ \cite{60kpc}.}

Prompt photon \emph{differential flux} due to dark matter annihilation with velocity averaged cross-section $\langle \sigma v \rangle$, in the direction along $\Delta\Omega$ solid angle, is given by   \cite{PPPC4DMID}:
\begin{equation}
	\frac{\mathrm{d}\Phi_\gamma}{\mathrm{d}E_\gamma }	=	\frac{1}{4\pi} \, \bar{J} \Delta\Omega \, \frac{\langle \sigma v \rangle}{m_0^2}
								 \sum_f  \mathcal{B}_f 	\left(   \frac{\mathrm{d}N_\gamma}{\mathrm{d}E_\gamma }   \right)_f
\end{equation}

where the sum runs over all possible annihilation channels $f$ with corresponding branching ratio $\mathcal{B}_f$ and photon multiplicity $(\mathrm{d}N_\gamma / \mathrm{d}E_\gamma )_f \equiv \left( \mathrm{d}\sigma_{f\gamma}/\mathrm{d}E_\gamma \right) / \sigma_f$.

The \emph{J-factor} $J (\Omega)$ depends on the DM spatial distribution, and defined as:
\begin{equation}
\label{eq:J}
 J(\Omega) =	\int_{los} \rho^2 (l) \ dl
 \end{equation}
 
 where integration is performed over the line of sight (los).
\footnote{In the galactic coordinate, the galactocentric distance $r$ is related to the radial distance to the point of observation $s$ through $r=\sqrt{s^2 +r_\odot^2 -2sr_\odot \cos b \cos \ell}$ where $b$ is the latitude, $\ell$ longitude, and the solar distance to GC is assumed to be $r_\odot \approx$ 8.5 kpc \cite{Galactic_coord}.}
In practice, we need the integrated flux over the detector angular acceptance, or observation window $\Delta\Omega$, and should use the averaged J-factor $\bar{J} = \int_{\Delta\Omega} J(\Omega') d\Omega' /\Delta\Omega$ \cite{Angular_Dsn}.

Spectral shape of gamma-ray in Electroweak model of dark matter mainly consists of two parts. First a mono-chromatic line at the end of spectrum $E_{\gamma}=m_0$, produced by DM annihilation into two photons $\chi\chi \to \gamma \gamma$; in addition to the narrow normalised Breit-Wigner line centred at $E_{Z\gamma} = m_0  ( 1 -m_Z^2 / 4m_0^2)$ \cite{Gravitino}, due to annihilation channel $\chi\chi \to Z \gamma$. These sharp spectral features provide powerful signatures to distinguish DM emission from the astronomical background.

The second contribution is a soft continuum of secondary photons from the fragmentation of  Electroweak gauge bosons pairs produced via $\chi\chi \to W^-W^+$ and $\chi\chi \to ZZ$. We use values tabulated in \cite{PPPC4DMID} where Monte Carlo simulations were performed using PYTHIA \cite{PYTHIA}. 

In addition to the mentioned two body annihilations, there is another source of gamma-ray emission from the three body processes \cite{3body} known as \emph{internal bremsstrahlung}. In FSR (Final State Radiation) process $\chi\chi \to W^+W^-\gamma$, the multiplicity of photon directly radiated from the final state $W^+W^-$ is given by \cite{FSR}:
\begin{equation}
\left(   \frac{dN_\gamma}{dE_\gamma }   \right)_{FSR} = 		\frac{ \alpha }{ \pi }	\frac{ 1-x }{x}	\ln \frac{ 4(1-x) }{\epsilon^2}
\end{equation}

where $x \equiv E_\gamma/m_0$, and $\epsilon = m_w/m_0$. It is also possible that photon is emitted from the virtual intermediate charged dark particle $\chi^q$, a process referred to as Virtual Internal Bremsstrahlung (VIB), the corresponding multiplicity, in the degenerate mass limit, reads \cite{VIB1, VIB2}:
\begin{multline}
\left(   \frac{dN_\gamma}{dE_\gamma }   \right)_{VIB} =		\frac{ 2\alpha }{ \pi x (1-x) }
	\left[		2 ( 1 -x +x^2 )^2  \ln \left( 2/\epsilon \right)				\right.	\\
	\left.		-\frac { 8 - 24x + 42x^2 - 37x^3 + 16x^4 - 3x^5 }{  (2-x )^3 }   \ln (1-x)	+\frac{ 4 - 12x  + 19x^2  - 22x^3  + 20x^4  - 10x^5 + 2x^6 }{ (2-x)^2 }		\right]
\end{multline}

It features a peak close to the end point of the spectrum as $x\to1$ that gets more pronounced if DM mass increases $\epsilon \to0$. In this limit the whole initial energy of $2m_0$ is absorbed by the radiated photon leaving very soft external particles in the final state which leads to \emph{Sudakov} logarithmic enhancement of the cross-section. In other words, for large DM masses, final state bosons start to behave like massless photons resulting in a mechanism similar to QED infrared divergence \cite{Sudakov_EW}.

The detector resolution function is best approximated by a Gaussian distribution with a width equal to the energy resolution of the instrument. The spectrum observed by such a detector is obtained through convolution of the original signal and the Gaussian resolution function. For example, the $\gamma\gamma$ and $\gamma Z$ spectral lines would be seen as a Gaussian distributions with a mean at line energies $E_{\gamma\gamma}$ and $E_{\gamma Z}$, normalised to 1 and 2 respectively.

Figure \ref{fig:E2dNdE_C4eft} illustrates the differential photon multiplicity $E_\gamma^2 \mathrm{d}N / \mathrm{d}E_\gamma$ for pseudo-real quadruplet with thermal mass of about 4.05 TeV. Different contributions i.e. fragmentation, FSR, VIB and line signals are appropriately weighted by branching ratios. In right panel \ref{fig:E2dNdE_C4eft_Line} we take into account the definite detector resolution of 15\%. It can be noticed that $\gamma\gamma$ and $\gamma Z$ monochromatic lines are observed as Gaussian distribution. After smearing out with resolution function, the line-like contribution from $W^+W^-$ radiation can also enhance the spectral peak.
 
 In the remaining sections of this paper, we analyse different parts of the predicted emission spectrum of EWDM against a range of observed data in order to set constraint on the annihilation rate. We employ measured photon flux near the MW black hole, diffuse emission in the inner galaxy, observations of dwarf satellites, and gamma-ray line searches, one after the other, to test our proposed dark matter model.
 
Our aim is not to artificially adjust the parameters to provide the best fit to the data, but rather a realistic comparison of the predictions of the EWDM theory with experimental results.


\subsubsection{Around Black Hole}

\begin{figure} [t] 				
	\begin{subfigure}{.5\linewidth}
		\centering
		\includegraphics[width=\linewidth]{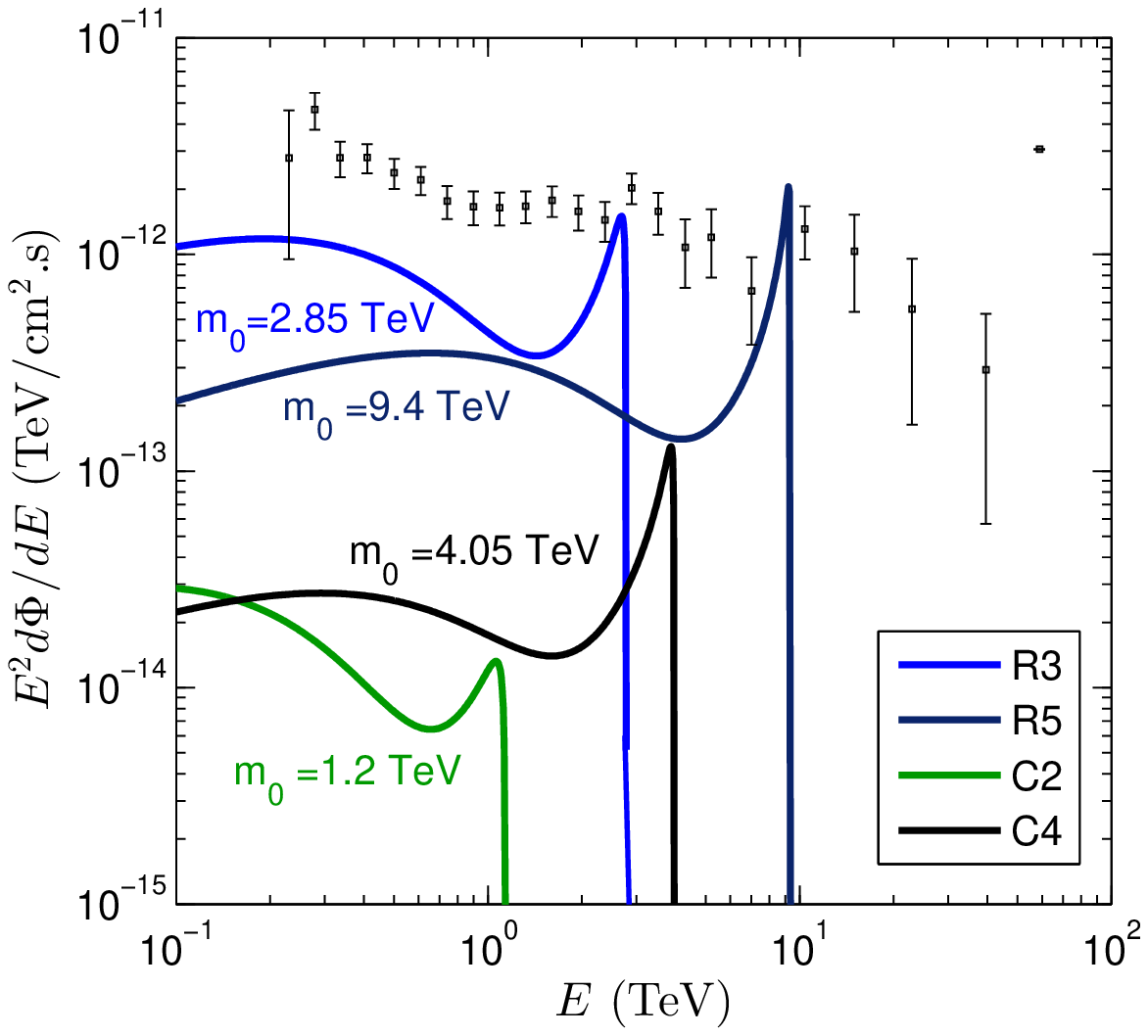}
		\subcaption{}
		\label{fig:BH_Noline}
	\end{subfigure}%
	\begin{subfigure}{.5\linewidth}
		\centering
		\includegraphics[width=\linewidth]{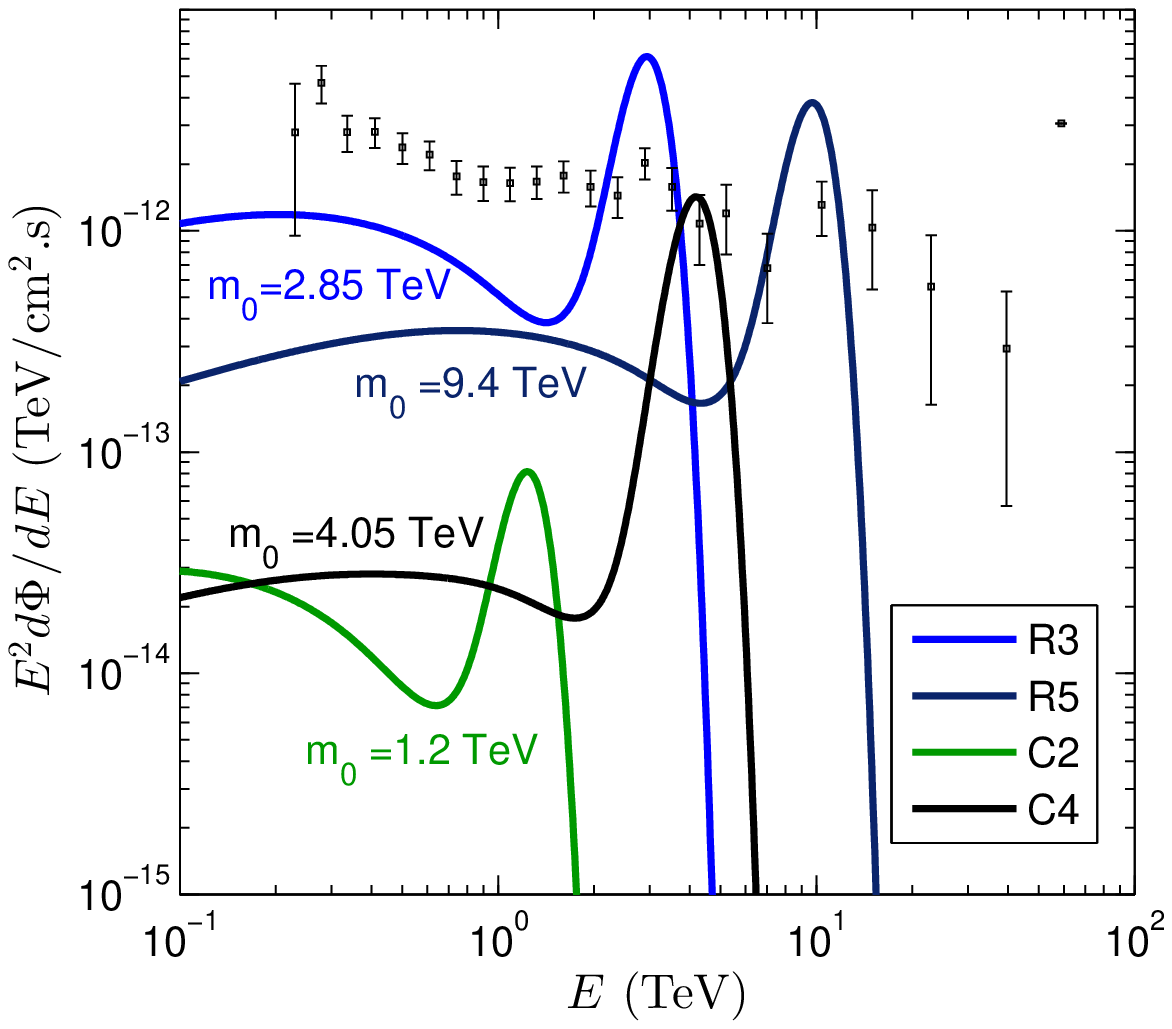}
		\subcaption{}
		\label{fig:BH_Line}
	\end{subfigure}
	\caption{Left \ref{fig:BH_Noline}: The differential gamma-ray flux for four fermionic multiplets of EWDM namely real triplet (R3) in blue, real quintet (R5) in navy, complex doublet (C2) in green, and complex quadret (C4) in black, ; based on their freeze-out masses respectively 2.85 TeV, 9.4 TeV, 1.2 TeV, and 4.05 TeV with corresponding annihilation cross-section of $1.3\times10^{-24}$ cm\textsuperscript{3}/s, $1.65\times10^{-24}$ cm\textsuperscript{3}/s, $1.25\times10^{-26}$ cm\textsuperscript{3}/s, and $2.2\times10^{-25}$ cm\textsuperscript{3}/s. The measurement data by HESS collaboration and relevant error-bars \cite{HESS_SgrA} are also plotted. Right \ref{fig:BH_Line}: The same spectra as in left panel, seen by a detector with resolution of \%15.}
	\label{fig:BH}
\end{figure}

Supermassive black holes (SMBHs) in the centre of galaxies can induce concentration of dark matter density. Since the flux \ref{eq:J} is proportional to the square of the density, the annihilating DM can be detected as a source of gamma-ray radiation in individual objects around the SMBH \cite{SMBH}.

HESS collaboration released detailed data about gamma-ray diffuse emission from a spherically symmetric region around SgrA*. The emission spectrum is observed in a ring centred at the black hole with inner radius of $0.15^\circ$ and outer radius $0.45^\circ$ corresponding to a field of view of $\Delta\Omega = 1.4\times10^{-4}$ sr. Based on the angular profile of the source, the J1745-290 can be considered as the origin of emission \cite{HESS_SgrA}.

The photon flux could possibly originate from the decay of cosmic rays protons accelerated by SgrA* PeVatron colliding with ambient gas \cite{HESS_SgrA}. Alternatively the detected spectrum might be interpreted as the result of absorption due to the presence of a strong infrared magnetic field close to the GC \cite{absorption}.

In this section, we compare the detected radiation by the $\gamma$-ray flux generated from annihilating dark matter with a steeped inner slope distribution towards the GC. In EWDM model a regular cuspy NFW profile \ref{eq:NFW} suffixes to explain the observed emission.

The predicted flux for all four available fermionic modules of EWDM is illustrated in figure \ref{fig:BH_Noline} against the data measured by HESS with corresponding error bars. 
The cross-sections are calculated at the freeze-out masses for each DM representation. NFW density distribution yields a normalised $\bar{J}$ value of 1887 averaged over the region of observation. Figure \ref{fig:BH_Line} shows the same spectrum as observed by a detector with resolution of \%15. They feature large peaks centred around the DM masses as a result of Gaussian smearing of $\gamma\gamma$, $\gamma Z$ and Bremsstrahlung spectral lines.

This profile brings the annihilation signal for the real EWDM to the level of sensitivity detected by the collaboration while still allowing room for secondary emissions like inverse Compton scattering \cite{BH_Celine} and other astronomical background mechanisms especially in the lower energy part of the spectrum.

As to the pseudo-real representations, clearly current detectors are not sensitive enough to observe the flux generated. Additional astronomical effects like a DM density spike accreted by super-massive black hole can boost the gamma-ray signal to the required level \cite{Spike}.


\subsubsection{Inner Galaxy}

\begin{figure} [t] 				
	\begin{subfigure}{.5\linewidth}
		\centering
		\includegraphics[width=\linewidth]{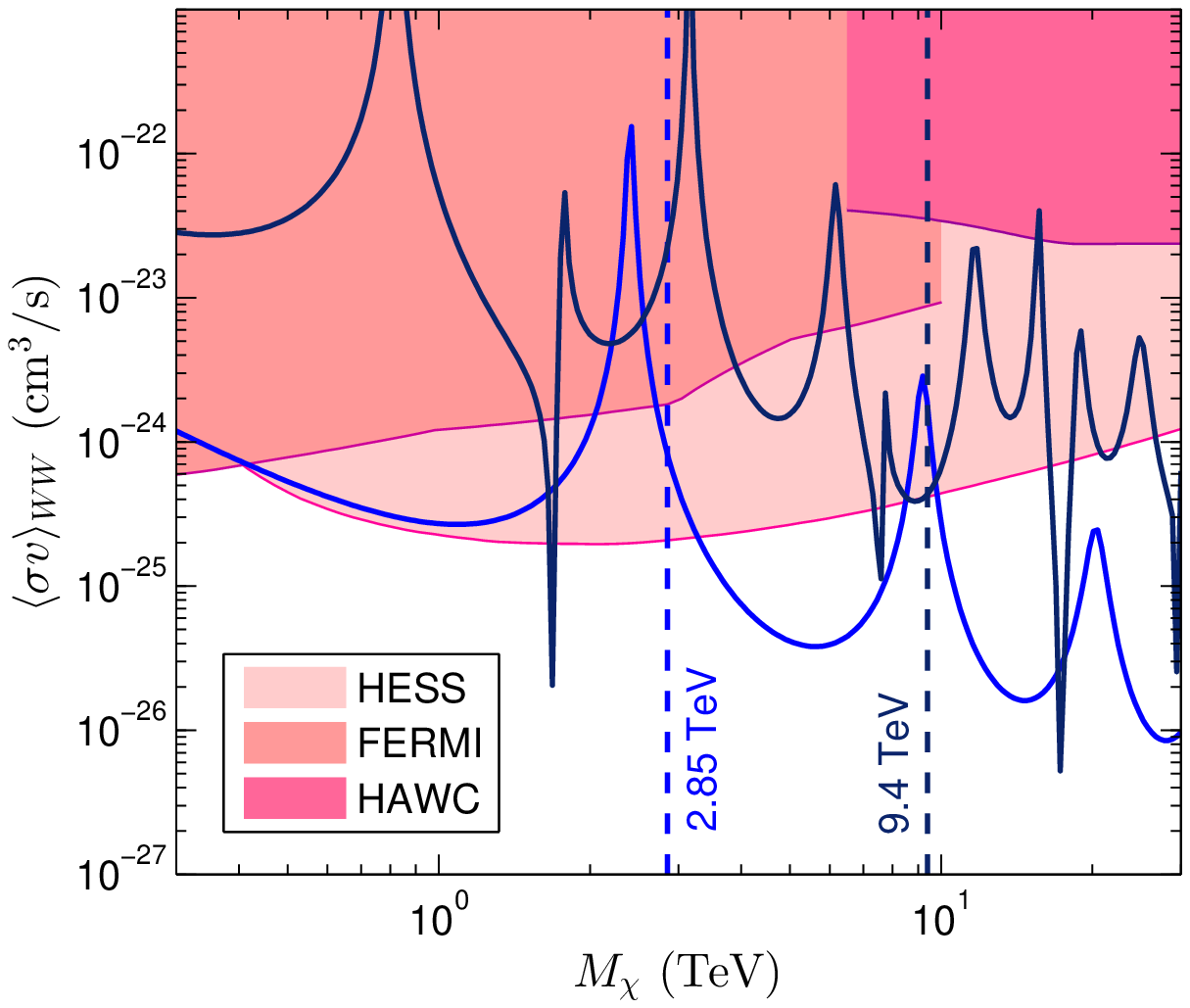}
		\subcaption{}
		\label{fig:Diffuse_R}
	\end{subfigure}%
	\begin{subfigure}{.5\linewidth}
		\centering
		\includegraphics[width=\linewidth]{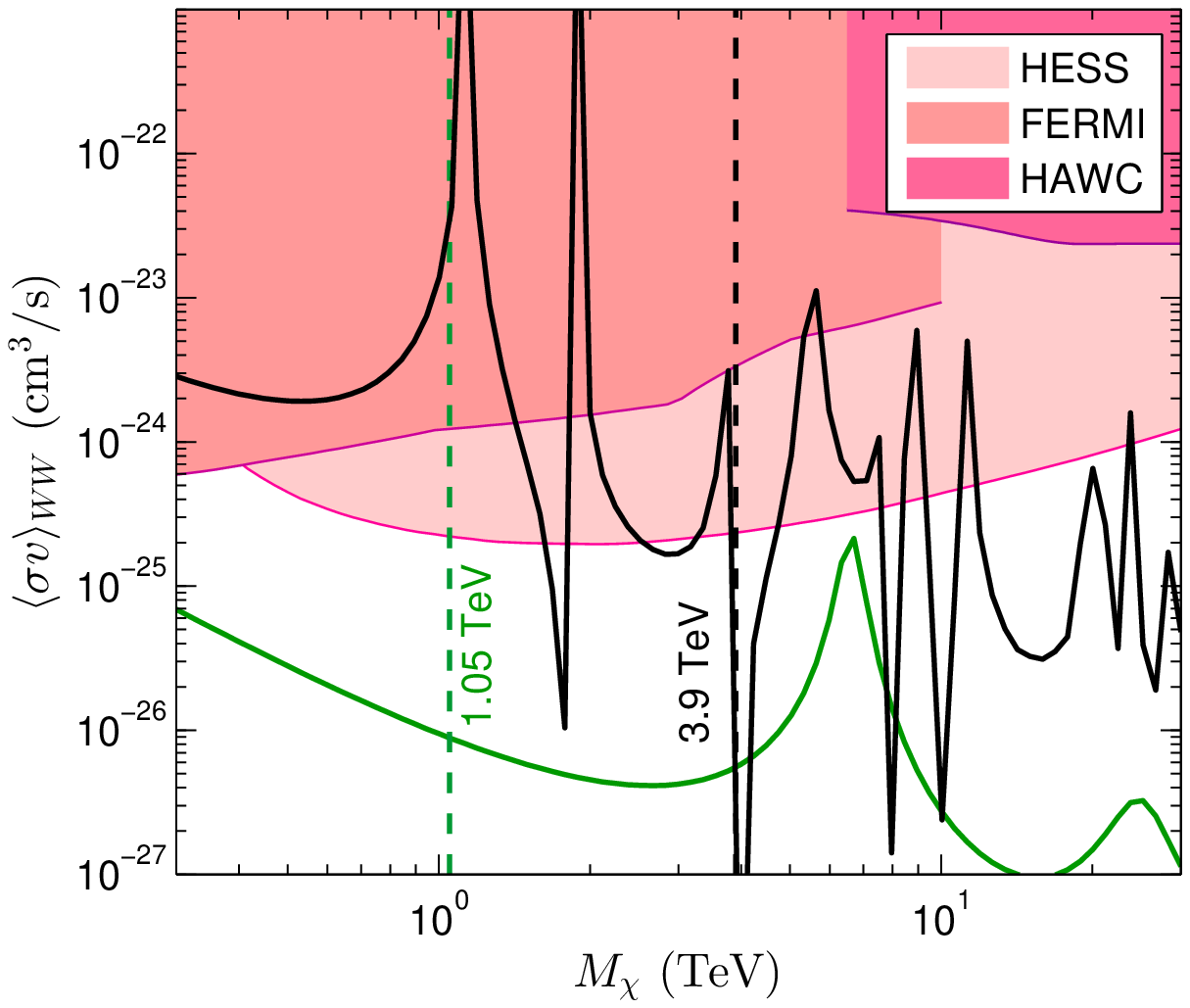}
		\subcaption{}
		\label{fig:Diffuse_C}
	\end{subfigure}
	\caption{Constraints on predictions of the real (\ref{fig:Diffuse_R}) and pseudo-real (\ref{fig:Diffuse_C}) models of EWDM from inner galaxy observations. The predicted annihilation cross-sections into $W^+W^-$ channel are plotted in solid line for real triplet (blue), real quintet (navy), complex doublet (green) and complex quadret (black). The overlaid lines from top to bottom indicate the observed limit for HAWC \cite{HAWC_bubble}, Fermi \cite{Fermi_halo}, and HESS \cite{HESS_GC} data using NFW profile. Where the cross-section curves enter the shaded areas, the corresponding mass values are excluded. Thermal DM masses are displayed by vertical dashed lines.}
	\label{fig:Diffuse}
\end{figure}

The Milky Way GC is the brightest source for photon-based probes of dark matter as it is the closest target and has a large dark matter density. In fact, dark matter searches in this region have placed strong bounds on DM annihilation cross section. Nevertheless, the challenge is presence of  large astrophysical backgrounds at a wide range of energies \cite{Jennifer}. The main background emissions in the Galactic Centre region are point sources, isotropic background, pion decay, inverse Compton scattering, Bremsstrahlung, and especially Synchrotron radiation \cite{20cm}.

In this work, conservative constraints are imposed on DM signal, and the astronomical background is not modelled. So, we only require that dark matter predicted emissions do not exceed the observed flux.

Dark matter annihilation signal have been placed by HESS from observation of the galactic centre accumulated over 10 years. The experiment searches the 300 parsec of the inner MW corresponding to a radius of $1^\circ$ around the dynamical centre excluding the plane within latitude $|b| < 0.3^\circ$ to avoid astronomical emissions \cite{HESS_GC}.

At distances away from the GC, in the MW halo, there is decline in the radiated photon density; however, this has the benefit of reduction in the background. In addition DM, spatial distribution is more clear in the halo \cite{Halo}.

Fermi LAT set constraints on dark matter annihilation in the diffuse spectrum from the Milky Way halo. They devised an optimised strategy to find the Region of Interest (RoI) in the inner galaxy where dark matter contribution to the observed flux is maximised in comparison with the astronomical foreground \cite{Fermi_halo}.

In addition to diffuse emission and isotropic backgrounds, there is another source of excess emission in the Inner Galaxy known as \emph{Fermi bubbles}. There are two bubbles above and below the Galactic Centre with longitudinal width of about 40 degrees. They have an almost uniform gamma-ray intensity with a sharp edge. Bubbles are thought to be the result of huge energy injection into the GC in the past like accretion onto the massive black hole or starburst \cite{Fermi_bubble}.

HAWC presented the limits on DM annihilation cross section in the Northern Fermi Bubble region from 15 month of observation. The data covers a wide energy-window from TeV to PeV scale dark matter \cite{HAWC_bubble}.

In figure \ref{fig:Diffuse}, we set constraining bounds on the EWDM annihilation cross-section in comparison with HESS, Fermi-LAT and HAWC gamma-ray data obtained from the observation of different areas in the inner galaxy.

In this analysis, we choose NFW profile to parametrise dark matter density distribution.

The general observation is that real EWDM hypothesis (\ref{fig:Diffuse_R}) is more constrained by experimental data compared to the complex scenario(\ref{fig:Diffuse_C}). It can be seen that essentially the whole spectrum of the real 5-plet with expectation of some islands is ruled out. For the real triplet, the mass interval below 3 TeV including the freeze-out mass of 2.85 TeV is excluded mainly by HESS measurements; however, for larger values of mass the model is still valid.

About the pseudo-real 4-plet, while observation data disfavours small mass values, the model is more successful in the large mass region of the spectrum. The thermal cross-section lies orders of magnitude below the imposed constraints, due to the deep trough caused by Ramsauer-Townsend effect. The pseudo-real doublet annihilation cross-section is clearly far below the current observational bounds. Future telescopes like CTA \cite{CTA} will get closer to the sensitivity required to detect signals from this complex model.


\subsubsection{dwarf Spheroidal satellite galaxies (dSph):}

\begin{figure} [t] 				
	\begin{subfigure}{.5\linewidth}
		\centering
		\includegraphics[width=\linewidth]{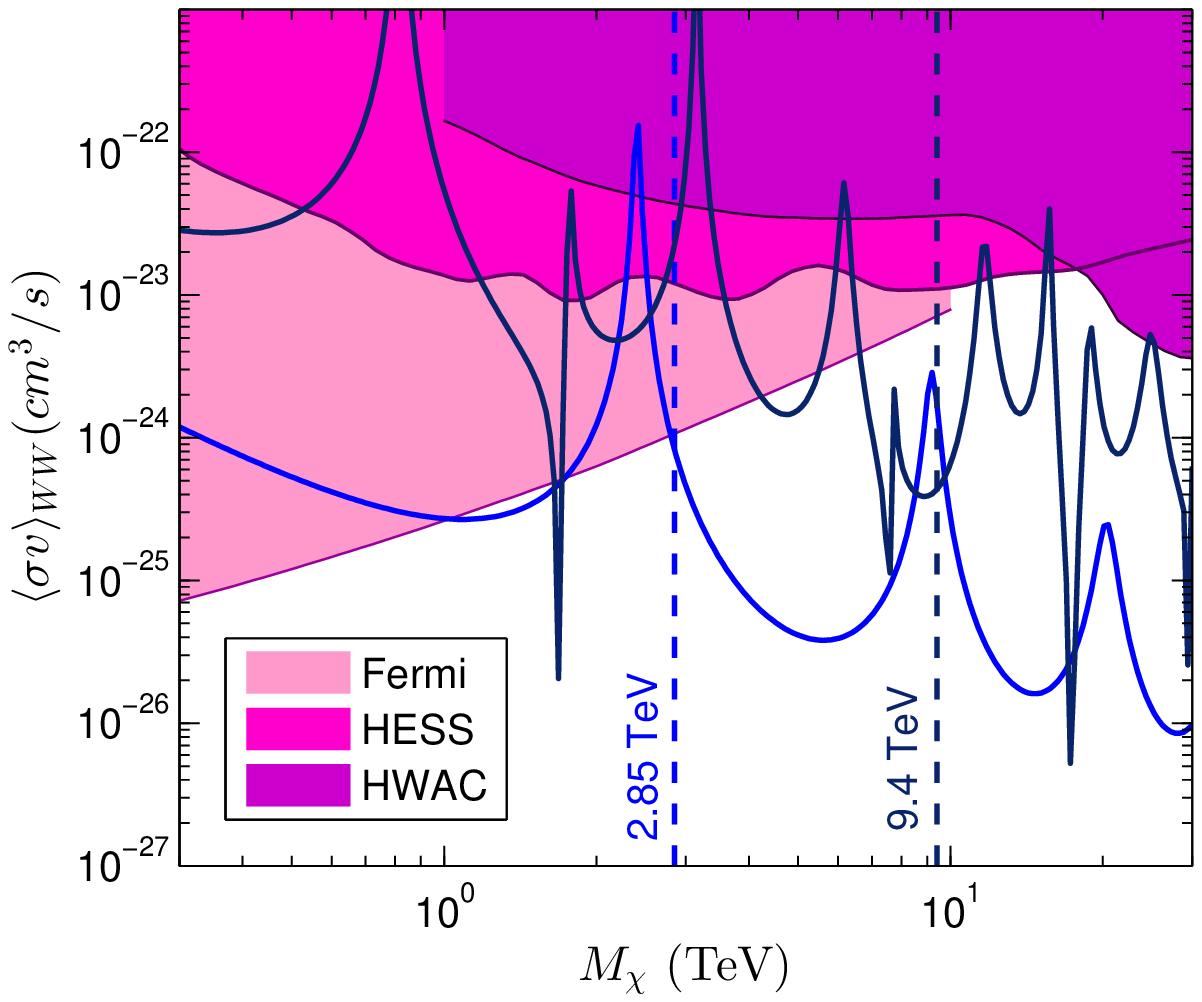}
		\subcaption{}
		\label{fig:dSph_R}
	\end{subfigure}%
	\begin{subfigure}{.5\linewidth}
		\centering
		\includegraphics[width=\linewidth]{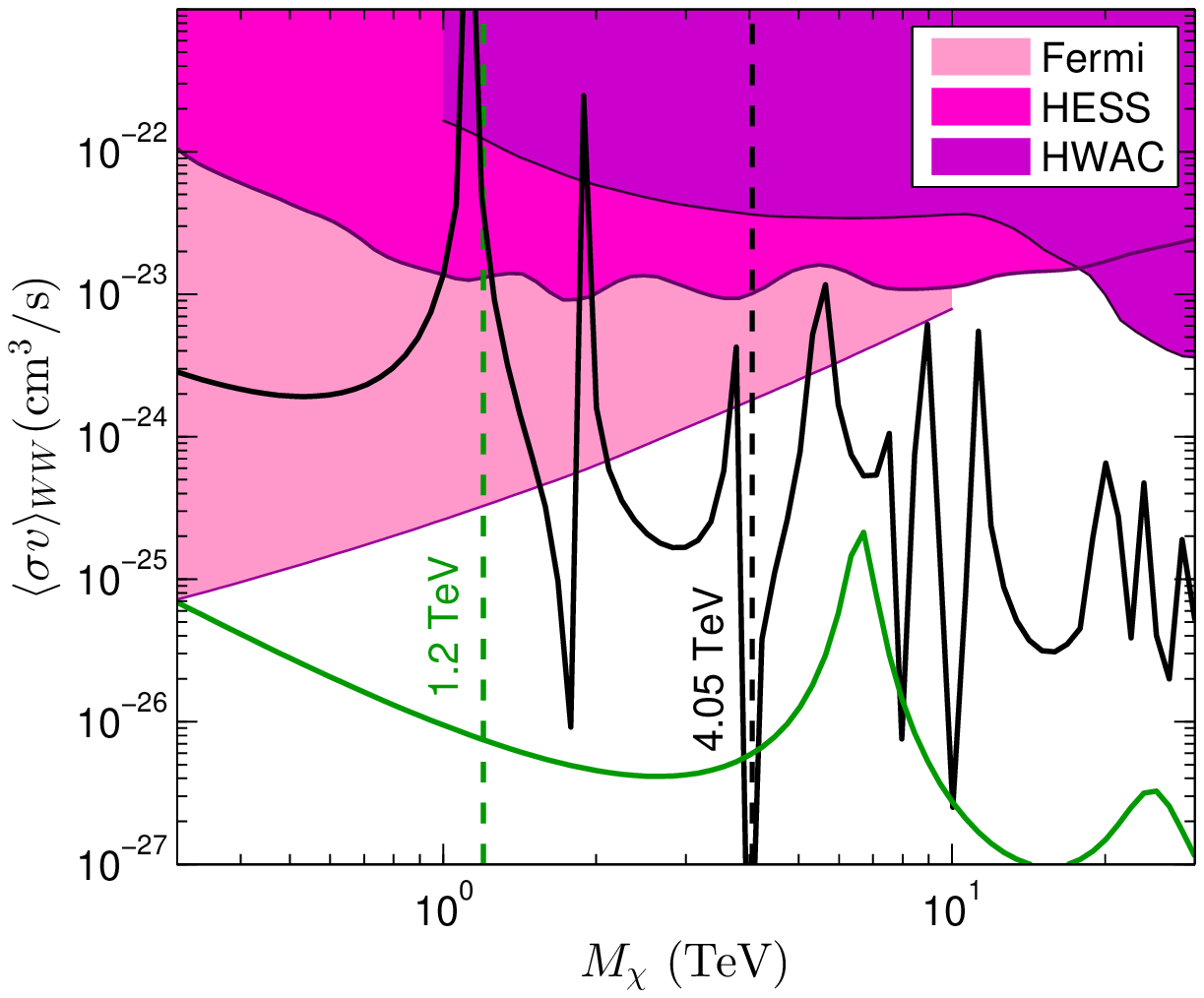}
		\subcaption{}
		\label{fig:dSph_C}
	\end{subfigure}
	\caption{Constraints on annihilation cross-section into $W^+W^-$ channel for the real representations of EWDM namely triplet (blue) and quintet (navy) on the left (\ref{fig:dSph_R}); as well as the pseudo-real models doublet (green) and quadret (black) on the right (\ref{fig:dSph_R}) from dwarf satellite galaxies obtained by Fermi \cite{Fermi_dSph}, HESS \cite{HESS_dSph} and HAWC \cite{HAWC_dSph} collaborations. The shaded area is excluded by the respective experiments. Dashed vertical lines correspond to the thermal mass of each model.}
	\label{fig:dSph}
\end{figure}

Satellite galaxies bound to the Milky Way are believed to be promising candidates for indirect detection of dark matter. Located around 100 kpc from the earth, they can provide significant gamma-ray flux for experiments. Due to lack of non-thermal astrophysical processes, intense star formation, gas or dust, dSph's are almost free of any astronomical background emissions. Satellites are known to be dark matter dominated with high mass-to-luminosity ratios \cite{dSph}.

In this section we compare our predicted continuum of gamma ray with experimental data from three observatories: Fermi-LAT which provided the strongest constraints on low mass DM up to 10 TeV through a combined analysis of 15 satellite galaxies \cite{Fermi_dSph}. HESS observed five ultra-faint dSph's for very high energy emissions up to 60 TeV not accessible to Fermi-LAT \cite{HESS_dSph}. HAWC complemented their work by searching for the continuum emission from 11 satellites in the energy window from 1 to 100 TeV \cite{HAWC_dSph}.

Figure \ref{fig:dSph} compares the limits obtained by the three experimental collaborations against EWDM cross-section continuous spectrum for $W^+W^-$ channel, omitting the contribution from the spectral lines.

Although the low mass intervals are excluded especially by FERMI data, the high energy region including the thermal masses are below the current constraints. The pseudo-real models are less restricted and almost all the TeV range of the quadruplet and the full spectrum of doublet are allowed by observational constraints.

Estimation of dark matter distribution in dwarf galaxies suffer from uncertainties which impact on the value determined for J-factor. 
In classical dSph's with large stellar tracers of the gravitational potential, systematic uncertainties are dominant which are strongly linked to the choice of kinematic analyses as well as relevant astrophysical assumptions \cite{classical_dSph}. For instance Jeans framework assumes spherical symmetry, dynamical equilibrium, and constant velocity anisotropy which might not be true in the physical system \cite{Nonspherical_dSph}.
However, in ultra-faint satellites, due to sparsity of stellar kinematic sample data sets, statistical uncertainties tend to dominate the errors \cite{Jeans_dSph}. To accommodate this problem, one can relax the limits in figure \ref{fig:dSph} by an order of magnitude, as an average effect of uncertainties \cite{Contamination_dSph, MDM_2015}.


\subsubsection{Gamma-ray Line}

\begin{figure} [t] 				
	\begin{subfigure}{.5\linewidth}
		\centering
		\includegraphics[width=\linewidth]{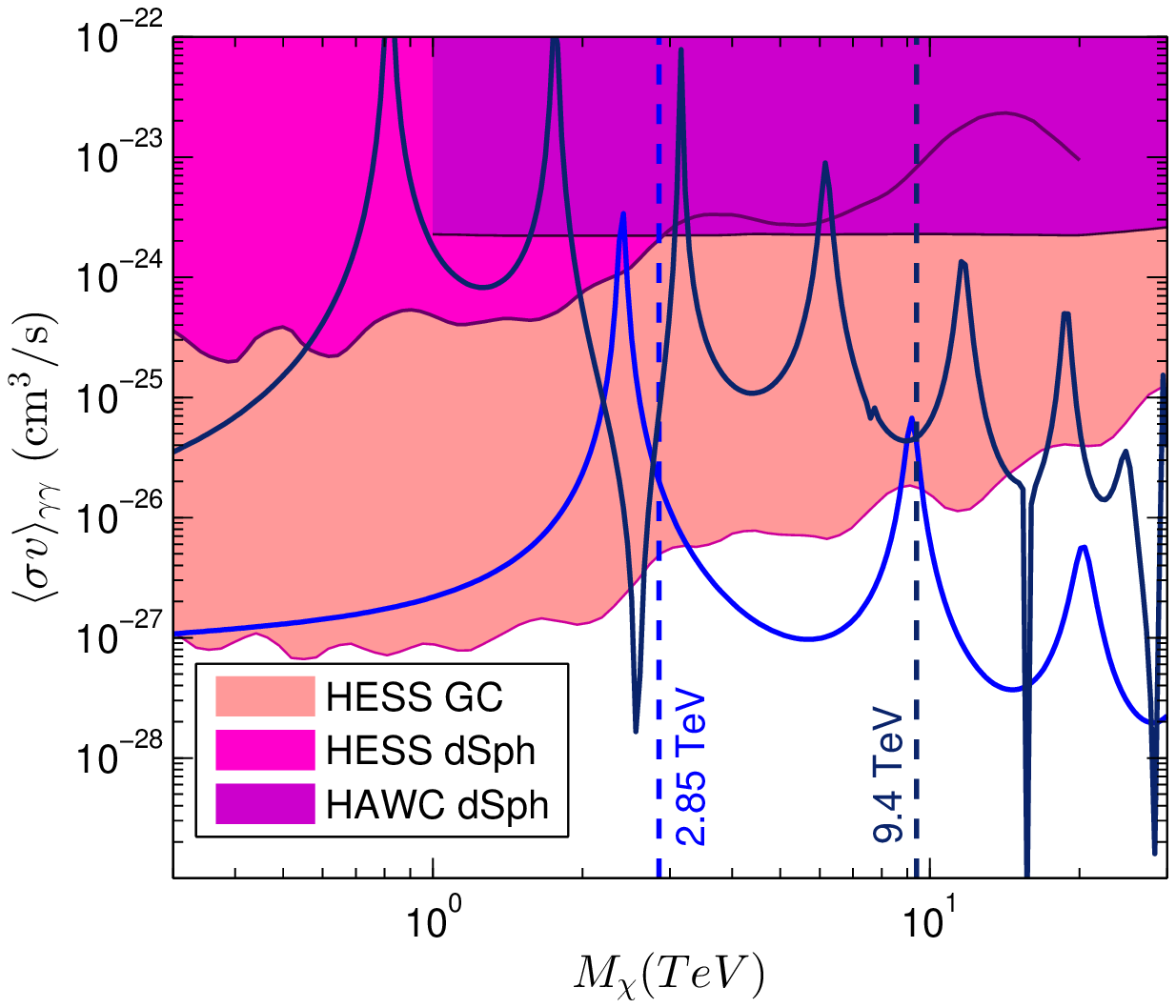}
		\subcaption{}
		\label{fig:Line_R}
	\end{subfigure}%
	\begin{subfigure}{.5\linewidth}
		\centering
		\includegraphics[width=\linewidth]{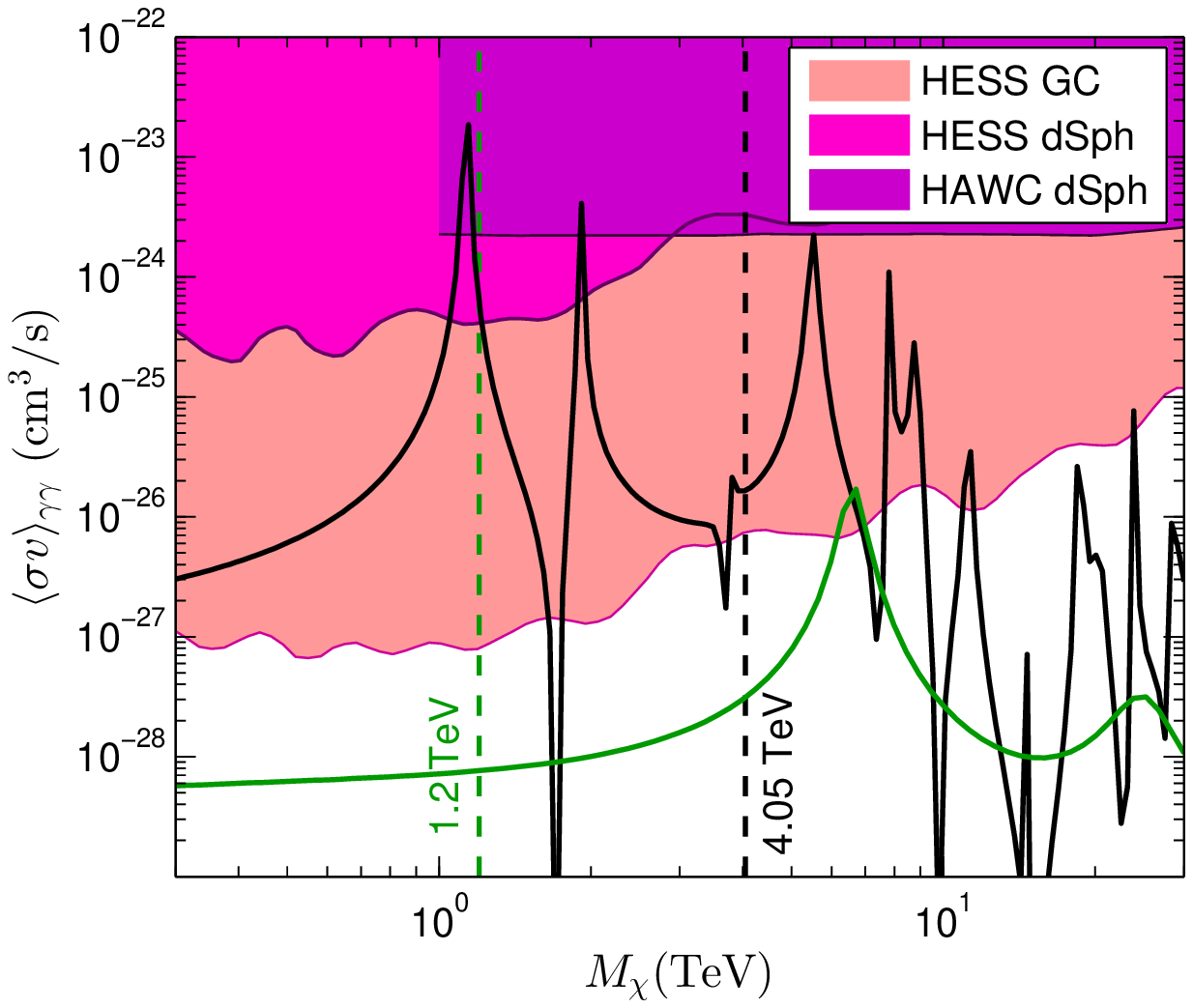}
		\subcaption{}
		\label{fig:Line_C}
	\end{subfigure}
	\caption{Spectral line bounds from dwarf satellites observed by HAWC \cite{HAWC_line_dSph} and HESS \cite{HESS_line_dSph}, together with constraints from GC detected by HESS \cite{HESS_GC_Line} on computed cross-section for different modules of EWDM including the real triplet (blue) and quintuplet (navy) (\ref{fig:Line_R}) in addition to the pseudo-real doublet (green) and quadruplet (black) (\ref{fig:Line_C}). The shaded areas are excluded by the relevant experiments. The vertical dashed lines show the freeze-out masses for corresponding model.}
	\label{fig:Line}
\end{figure}

At gamma-ray energies no other known astronomical phenomenon can generate monochromatic spectrum. So detecting a spectral line would be a smoking gun evidence for existence of a DM particle with a mass centred at the energy of monochromatic radiation \cite{Line_Photino, Line}. As explained, in TeV-scale dark matter theories like EWDM, Sommerfeld effect greatly enhances the flux produced from annihilation into $\gamma\gamma$ and $\gamma Z$ lines at tree-level, compared to the soft continuum of other final state products. This makes probing this spectral feature with outstanding energy resolution a favourable strategy in heavy dark matter models. 	

The galactic centre is a promising target for dark matter searches due to high dark matter content and relative proximity to the earth. The flux limits obtained from this region, imposes strongest constraints on DM velocity averaged cross-section into monochromatic emission.

The ground-based HESS telescope presented a search for mono-energetic photons within the central galaxy from ten years of observation for the DM mass spanning from 300 GeV to 70 TeV. The gamma-ray events are selected from an annuli of $0.1^\circ$ widths centred at the GC, while the galactic plane with latitudes from $-0.3^\circ$ to $0.3^\circ$ is excluded to avoid emission from other astrophysical sources \cite{HESS_GC_Line}.

In addition, due to high mass-to-light ratio and thus large dark matter concentration, proximity compared to other cosmic sources, and absence of emissions from non-thermal processes; dwarf satellite galaxies are considered to be good targets for searching DM gamma-ray signals \cite{dSph2}.

HESS collaboration has observed five dSph systems namely Fornax, Coma Berenices, Sculptor, Carina and Sagittarius for around 130 hours in total. The energy resolution of the instrument in TeV energy range is crucial to detect the sharp signals in  annihilation channel. The J-factor is computed using dynamical mass models fit to the kinematic data of suitable stellar tracers for each dwarf satellite \cite{HESS_line_dSph}.

HAWC observatory extended the energy range up to 100 TeV, providing the strongest bounds on spectral lines from satellites in the high mass domain. They published the search results for 10 dSph's that are Bootes I, Canes Venatici I, Canes Venatici II, Coma Berenices, Hercules, Leo I, Leo II, Leo IV, Segue I and Sextans using 1038 days of data. To determine J-factor, NFW profile is used with different astronomical parameters for each dSph \cite{HAWC_line_dSph}.

Figure \ref{fig:Line} compares the annihilation cross-section proposed by different representations of EWDM with gamma ray line constraints in dwarf galaxies searched by HESS and HAWC, in addition to the Milky Way centre measured again by HESS using NFW parametrisation of the DM density profile.

In general, the pseudo-real models have lower spectral line cross-section than the real cases, thus affected less by the observational bounds.

While, with exception of some small islands, electroweak DM is not limited by dwarf satellite experiments, it can be seen that the stringent bounds are imposed from the inner galaxy. The low mass parts of the spectrum including thermal masses are essentially ruled out by GC constraints. However, the high energy portions larger than 3 TeV, 15 TeV and 7 TeV are spared for the real triplet, quintet and the complex quartet respectively.

Calculated cross-section for the complex doublet is around two orders of magnitude lower than HESS limits, keeping this model safe from current ID results.

Exclusion limits on averaged cross-section in the MW central region strongly depends on the spatial distribution of dark matter density. If instead of a peaked profile like NFW, a cored DM profile such as the Burkert is employed, then the constraints on cross-section will be weakened up to three orders of magnitude \cite{Cuspy_Cored} which will reopen the small energy intervals. Indeed existence of a DM core in the GC is not only contradicting the N-body simulation results but in many cases thought to be closer to the observations \cite{Cored_1, Cored_2}.

Mass of electroweak dark matter is not determined only through freeze-out phenomenon. There is a variety of production mechanisms like moduli field decay \cite{Moduli_Decay}, B-ball decay \cite{BBB} and reheating thermal histories \cite{Reheating} which give rise to a range of DM masses while still constituting the current abundance. So predictions of the EWDM model is not limited to the freeze-out mass and the entire spectral range needs to be collocated against experimental constraints.


\section{Conclusion}

The Standard Model can be extended by adding a fermion or scalar multiplet charged under the $SU(2)_L \times U(1)_Y$ gauge group, so that successful features of the SM electroweak sector are minimally impacted. Real representations of EWDM are allowed dark matter candidates and were studied in the literature before; However, the pseudo-real representations are excluded by direct detection results due to the tree-level coupling to nuclei through the exchange of Z boson.

In this report, we generalised the pseudo-real EWDM framework by introducing a dimension five operator which couples the multiplet to Higgs boson, with a special focus on the fermionic quadruplet case. This mechanism revitalised the pseudo-real theory as the mentioned effective term splits the pseudo-Dirac dark matter into two Majorana states, therefore eliminating the tree-level Z-mediated interaction with nuclei.

It was discussed that in TeV mass scale and low velocities, long-range forces mediated by EW gauge vectors cause non-perturbative effects. We provided the complete formulae for the potential and annihilation amplitudes of all valid fermionic modules, and accurately computed the predicted cross-section in this framework. Consequently annihilation amplitudes has to be multiplied by Sommerfeld enhancement factors which significantly increase the cross-section both in the current universe and during the freeze-out, raising the value assigned to DM mass based on today's abundance.

Assuming EWDM to be a thermal relic, we computed the dark matter mass for all available multiplets after carefully solving the Boltzmann equation. However, DM could also be produced by other non-thermal scenarios, so one needs to take the full mass spectrum into account for phenomenological predictions.

We derived constraints on different representations of electroweak dark matter through a variety of gamma ray probes. Continuum of photon emission provides promising indirect detection prospects particularly for the pseudo-real models of EWDM. Almost the whole multi-TeV mass range including the thermal mass of the quadret, and essentially the entire spectrum of the doublet remain safe from the astrophysical constraints. However, in comparison, the real modules are more restricted by these observational data.

Mono-energetic line bounds from the galactic centre prove to be the stringiest test of EWDM theory. If dark matter yield is produced through the freeze-out mechanism then the cuspy density profiles like NFW or Einasto are disfavoured, and one needs to consider a DM core in the inner MW similar to Burkert or Isothermal profiles. This restriction is nevertheless relaxed if we assume relic abundance is generated via other production phenomena.

In conclusion, the electroweak model of dark matter is still allowed and even further motivated according to the most recent gamma-ray probes. Remarkable progress can be made by future astrophysical observations.


\section{acknowledgement}

I would like to acknowledge Celine Boehm, Filippo Sala, Alexander Pukhov, and Thomas Hahn for their discussions and contributions to this project.

We used micrOMEGAs 5.0.4 \cite{micrOMEGAs} to compute the relic abundances, and FeynArts 3.10 \cite{FeynArts} for generating Feynman diagrams.


\appendix
\section{Lagrangian and Feynman rules}

\label{app:Feynman}

The complete interaction Lagrangian describing the couplings of the pseudo-real multiplet with the SM particles can be decomposed as:
\begin{equation}
	\mathcal{L}_\mathrm{int}	=	\mathcal{L}_A 	+\mathcal{L}_Z 	+\mathcal{L}_W 	+\mathcal{L}_H
\end{equation}

Interactions of dark sector particles and gauge fields are derived from expansion of the kinetic term in the general Lagrangian \ref{eq:L_C}.
Electromagnetic interactions are given by the Lagrangian:
\begin{equation}
	\mathcal{L}_A	 =	-e \left[		\frac{n}{2}	\, \overline{\chi}^\frac{n}{2} \gamma^\mu \chi^\frac{n}{2} 
					+	\sum_{q = -n/2 +1} ^{n/2-1}	q \left( 	\overline{\chi}^q_1 \gamma^\mu \chi^q_1	+\overline{\chi}^q_2 \gamma^\mu \chi^q_2	\right) 	\right]	A_\mu 	\,,
\end{equation}

Interactions mediated by the neutral weak boson can be written as:
\begin{align}
	\mathcal{L}_Z &=		- \frac{g_w}{2c_w} 	\left\{ 	\vphantom{\sum_{q = -n/2 +1} ^{n/2-1}}
					\left( nc_w^2 -1 \right) \overline{\chi}^\frac{n}{2} \gamma^\mu \chi^\frac{n}{2}	+i \overline{\widetilde{\chi}^0} \gamma^\mu \chi^0		\right.	\\	\nonumber
			& \left.	+ \sum_{q = -n/2 +1} ^{n/2-1}	 \left[ 	\left( 2c_w^2 q + \cos \phi_q \right)	\overline{\chi}^q_1 \gamma^\mu \chi^q_1	
					+	\left( 2c_w^2 q - \cos \phi_q \right)	 	\overline{\chi}^q_2 \gamma^\mu \chi^q_2 
					+	\sin \phi_q	\left(	\overline{\chi}^q_1 \gamma^\mu \chi^q_2	+\overline{\chi}^q_2 \gamma^\mu \chi^q_1	\right)	\right]	\right\} Z_\mu \,,
\end{align}

The odd particles couple to the charged weak gauges through:
\begin{align}
	\mathcal{L}_W &=		- \frac{g_w}{2\sqrt{2}}	\left\{	
			2 \sqrt{n-1} \, e^{\frac{i}{2} \widehat{\lambda}_0}	\left[ c_{\frac{n}{2} -1}	 \,	\overline{\chi}^{\frac{n}{2} -1}_2 \gamma^\mu \chi^\frac{n}{2}	
									- s_{\frac{n}{2} -1} \,	\overline{\chi}^{\frac{n}{2} -1}_1 \gamma^\mu \chi^\frac{n}{2}		\right]	\right.	\\	\nonumber
		&+	\frac{1}{\sqrt{2}}		\left(   n \ e^{-\frac{i}{2} \widehat{\lambda}_0} c_+   
									- \sqrt{ n^2 -4} \ e^{\frac{i}{2} \widehat{\lambda}_0} s_+ \right)	\overline{\widetilde{\chi}^0} \gamma^\mu \chi_2^+
			- \frac{i}{\sqrt{2}} 	\left(   n \ e^{-\frac{i}{2} \widehat{\lambda}_0} c_+   
									+ \sqrt{ n^2 -4} \ e^{\frac{i}{2} \widehat{\lambda}_0} s_+ \right)	\overline{\chi^0} \gamma^\mu \chi_2^+  		\\	\nonumber
		&  - \frac{1}{\sqrt{2}}	\left(   n \ e^{-\frac{i}{2} \widehat{\lambda}_0} s_+   
									+ \sqrt{ n^2 -4} \ e^{\frac{i}{2} \widehat{\lambda}_0} c_+ \right)		\overline{\widetilde{\chi}^0} \gamma^\mu \chi_1^+
			+ \frac{i}{\sqrt{2}}	\left(   n \ e^{-\frac{i}{2} \widehat{\lambda}_0} s_+   
									- \sqrt{ n^2 -4} \ e^{\frac{i}{2} \widehat{\lambda}_0} c_+ \right)	\overline{\chi^0} \gamma^\mu \chi_1^+ 		\\	\nonumber
		& +	\sum_{q=2}^{\frac{n}{2} -1}	\left[		\left(	\sqrt{ n^2 -4(q-1)^2}	\, s_{q-1} s_q
									- \sqrt{ n^2 -4q^2} \, c_{q-1} c_q	\right)	\overline{\chi}^{q-1}_1 \gamma^\mu \chi^q_1	\right.	\\	\nonumber
		& \qquad	+ 	\left(	\sqrt{ n^2 -4(q-1)^2} \,		 c_{q-1} c_q	
									- \sqrt{ n^2 -4q^2} \,	 s_{q-1} s_q	\right)	\overline{\chi}^{q-1}_2 \gamma^\mu \chi^q_2		\\	\nonumber
		& \qquad	-		\left(	\sqrt{ n^2 -4(q-1)^2}	\, s_{q-1} c_q
									+ \sqrt{ n^2 -4q^2} \,	 c_{q-1} s_q	\right)	\overline{\chi}^{q-1}_1 \gamma^\mu \chi^q_2		\\	\nonumber
		& \qquad	-		\left(	\sqrt{ n^2 -4(q-1)^2}	\, c_{q-1} s_q
									+ \sqrt{ n^2 -4q^2} \,	 s_{q-1} c_q	\right)	\overline{\chi}^{q-1}_2 \gamma^\mu \chi^q_1		
										\left. \left.		\right]	\vphantom{\overline{\chi}^{\frac{n}{2} -1}_2}	\right\}		W^-_\mu 	+ \mathrm{cc} \,,
\end{align}

In special case of the complex doublet, the charged weak boson interactions read:
\begin{equation}
	\mathcal{L}_W^{\mathbb{C}2} =	-\frac{g_w}{2}	\left(	\overline{\widetilde{\chi}^0} \gamma^\mu \chi^+		
						-i \,	\overline{\chi^0} \gamma^\mu \chi^+	\right)	W^-_\mu 		+ \mathrm{cc}\,,
\end{equation}

Non-renormalisable interactions with Higgs boson can be cast as:
\begin{align}
	\mathcal{L}_H &=	-\Delta^{(\mathrm{t})}_{\frac{1}{2} (n-1)}	\left(	\frac{2}{\nu} h 	+ \frac{1}{\nu^2} h^2	\right)	\overline{\chi}^\frac{n}{2}	\chi^\frac{n}{2}		\\
		& - \left(	\frac{2}{\nu} h 	+ \frac{1}{\nu^2} h^2	\right)		\left[	\left(	\frac{1}{2} \,\Delta^{(\mathrm{t})}_{-\frac{1}{2}} 	-\Re \delta_0	\right)	\overline{\chi^0} \chi^0	
			+ \left(	\frac{1}{2} \,\Delta^{(\mathrm{t})}_{-\frac{1}{2}} 	+\Re \delta_0	\right)	\overline{\widetilde{\chi}^0}		\widetilde{\chi}^0		
			+ 	2i \, \Im \delta_0		\overline{\chi^0}	\widetilde{\chi}^0	\right]							\\	\nonumber
		& - \sum_{q=1}^{\frac{n}{2} -1}		\left(	\frac{2}{\nu} h 	+ \frac{1}{\nu^2} h^2	\right)		\left[	
			\left(	\Delta^{(\mathrm{t})}_{q-\frac{1}{2}} 	-2 |\delta_q| \sin \phi_q	\right)		\overline{\chi}^q_1  \chi^q_1	
			+ \left(	\Delta^{(\mathrm{t})}_{q-\frac{1}{2}} 	+2 |\delta_q| \sin \phi_q	\right)		\overline{\chi}^q_2  \chi^q_2
			+ 2 	|\delta_q| \cos \phi_q	\left(	\overline{\chi}^q_1  \chi^q_2	-	\overline{\chi}^q_2  \chi^q_1	\right)	\right]
\end{align}

Using the interaction Lagrangian above, one can readily obtain the Feynman rules for the coupling of the complex EWDM with gauge fields and scalars of the SM.

Feynman rule for the coupling of the charged particles to photon can be written as:
\begin{equation}
	C^\mu ( \chi^q_i, \chi^q_j, A )	=	-i e q \,\gamma^\mu \delta_{ij}
\end{equation}

For vertices involving neutral electroweak $Z$ boson, one has:
\begin{align}
	C^\mu ( \chi^\frac{n}{2}, \chi^\frac{n}{2}, Z )	&=	- \frac{i }{2} \frac{g_w}{c_w}	\left(n c_w^2 -1 \right) \gamma^\mu	\\		\nonumber
	C^\mu ( \chi^q_1, \chi^q_1, Z )	&=				- \frac{i }{2} \frac{g_w}{c_w}	\left( 2 c_w^2 q +\cos \phi_q \right) \gamma^\mu	\\		\nonumber
	C^\mu ( \chi^q_2, \chi^q_2, Z )	&=				- \frac{i }{2} \frac{g_w}{c_w}	\left( 2 c_w^2 q -\cos \phi_q \right) \gamma^\mu	\\		\nonumber
	C^\mu ( \chi^q_1, \chi^q_2, Z )	&=				- \frac{i }{2} \frac{g_w}{c_w}	\sin \phi_q \gamma^\mu	\\							\nonumber
	C^\mu (\, \overline{\chi}^0, \widetilde{\chi}^0, Z )	&=		- C^\mu (\, \overline{\widetilde{\chi}^0}, \chi^0, Z ) =			- \frac{g_w}{2c_w} \gamma^\mu					
\end{align}

Note that $Z$ boson can change the flavour of the odd particles with the same electric charge.

Interactions of the fields with different charges are mediated by $W^-$ gauge boson through:
\begin{align}
	C^\mu ( \overline{\chi}^{n/2-1}_2, \chi^\frac{n}{2}, W^- )	&=		- \frac{i }{\sqrt{2}}	\sqrt{n-1}	\,e^{\frac{i}{2} \widehat{\lambda}_0}	c_{\frac{n}{2} -1}	\,g_w \gamma^\mu 	\\	\nonumber
	C^\mu ( \overline{\chi}^{n/2-1}_1, \chi^\frac{n}{2}, W^- )	&=		\frac{i }{\sqrt{2}}	\sqrt{n-1}	\,e^{\frac{i}{2} \widehat{\lambda}_0}	s_{\frac{n}{2} -1}	\,g_w \gamma^\mu 	\\	\nonumber
	C^\mu ( \overline{\chi}^{q-1}_1, \chi^q_1, W^- )	&=		\frac{i }{2\sqrt{2}}	\left(	\sqrt{ n^2 -4q^2} \, c_{q-1} c_q		-\sqrt{ n^2 -4(q-1)^2}	\, s_{q-1} s_q	\right)	\,g_w \gamma^\mu 	\\	\nonumber
	C^\mu ( \overline{\chi}^{q-1}_2, \chi^q_2, W^- )	&=		\frac{i }{2\sqrt{2}}	\left(	\sqrt{ n^2 -4q^2} \, s_{q-1} s_q		-\sqrt{ n^2 -4(q-1)^2}	\, c_{q-1} c_q	\right)	\,g_w \gamma^\mu 	\\	\nonumber
	C^\mu ( \overline{\chi}^{q-1}_1, \chi^q_2, W^- )	&=		\frac{i }{2\sqrt{2}}	\left(	\sqrt{ n^2 -4q^2} \, c_{q-1} s_q		+\sqrt{ n^2 -4(q-1)^2}	\, s_{q-1} c_q	\right)	\,g_w \gamma^\mu 	\\	\nonumber
	C^\mu ( \overline{\chi}^{q-1}_2, \chi^q_1, W^- )	&=		\frac{i }{2\sqrt{2}}	\left(	\sqrt{ n^2 -4q^2} \, s_{q-1} c_q		+\sqrt{ n^2 -4(q-1)^2}	\, c_{q-1} s_q	\right)	\,g_w \gamma^\mu 	\\	\nonumber
	C^\mu ( \overline{\chi^0}, \chi^+_1, W^- )			&=		\frac{1}{4}		\left(   n \ e^{-\frac{i}{2} \widehat{\lambda}_0} s_+   	
						- \sqrt{ n^2 -4} \ e^{\frac{i}{2} \widehat{\lambda}_0} c_+ \right)		\,g_w \gamma^\mu 	\\	\nonumber
	C^\mu ( \overline{\widetilde{\chi}^0}, \chi^+_2, W^- )		&=		- \frac{i}{4}				\left(   n \ e^{-\frac{i}{2} \widehat{\lambda}_0} c_+   
						- \sqrt{ n^2 -4} \ e^{\frac{i}{2} \widehat{\lambda}_0} s_+ \right)		\,g_w \gamma^\mu 	\\	\nonumber
	C^\mu ( \overline{\chi^0}, \chi^+_2, W^- )			&=		- \frac{1}{4}		\left(   n \ e^{-\frac{i}{2} \widehat{\lambda}_0} c_+   	
						+ \sqrt{ n^2 -4} \ e^{\frac{i}{2} \widehat{\lambda}_0} s_+ \right)		\,g_w \gamma^\mu 	\\	\nonumber
	C^\mu( \overline{\widetilde{\chi}^0}, \chi^+_1, W^- )		&=		\frac{i}{4}				\left(   n \ e^{-\frac{i}{2} \widehat{\lambda}_0} s_+   
						+ \sqrt{ n^2 -4} \ e^{\frac{i}{2} \widehat{\lambda}_0} c_+ \right)		\,g_w \gamma^\mu 
\end{align}

For the comple doublet we have:
\begin{align}
	C^\mu ( \overline{\chi^0}, \chi^+, W^- )		&=		- \frac{1}{2}	\,g_w \gamma^\mu 	\\	\nonumber
	C^\mu ( \overline{\widetilde{\chi}^0}, \chi^+, W^- )		&=		- \frac{i}{2}	\,g_w \gamma^\mu 
\end{align}

Factorising out the root of unity 
$C^\mu  \equiv -i C' \gamma^\mu$, 
Feynman rules for the conjugate charged field $W^+$ can be easily obtained from:
\begin{equation}
	C' (\, \overline{\chi}^q_i, \chi^{q-1}_j, W^+ )	= 	C'^* (\, \overline{\chi}^{q-1}_j, \chi^q_i, W^- )
\end{equation}

Feynman rules for non-renormalisable couplings to the Higgs boson has the form:
\begin{align}
	C ( \overline{\chi}^\frac{n}{2}, \chi^\frac{n}{2}, h )		&=		-2 \frac{i}{\nu}	\,\Delta^{(\mathrm{t})}_{\frac{1}{2} (n-1)}
			=		\frac{i}{4}	\left( n-1 \right)		\nu	\frac{\lambda_c}{\Lambda}			\\	\nonumber
	C ( \overline{\chi}^q_1, \chi^q_1, h )		&=		-2 \frac{i}{\nu}	\left(	\Delta^{(\mathrm{t})}_{q-\frac{1}{2}} 	-2 |\delta_q| \sin \phi_q	\right)
			=		\frac{i}{2}	\nu	\left[	\left( q -\frac{1}{2} \right)	\frac{\lambda_c}{\Lambda}		+ \sqrt{n^2 -4q^2}	\sin \phi_q	\left| \frac{\lambda_0}{\Lambda} \right|	\right]	\\	\nonumber
	C ( \overline{\chi}^q_2, \chi^q_2, h )		&=		-2 \frac{i}{\nu}	\left(	\Delta^{(\mathrm{t})}_{q-\frac{1}{2}} 	+2 |\delta_q| \sin \phi_q	\right)
			=		\frac{i}{2}	\nu	\left[	\left( q -\frac{1}{2} \right)	\frac{\lambda_c}{\Lambda}		- \sqrt{n^2 -4q^2}		\sin \phi_q	\left| \frac{\lambda_0}{\Lambda} \right|	\right]	\\	\nonumber
	C ( \overline{\chi}^q_2, \chi^q_1, h )		&=			- C ( \overline{\chi}^q_1, \chi^q_2, h )	=	4 \frac{i}{\nu}	|\delta_q|	\cos \phi_q
			=	\frac{i}{2}	\nu		\sqrt{n^2 -4q^2}	\cos \phi_q	\left| \frac{\lambda_0}{\Lambda} \right|		\\	\nonumber
	C ( \overline{\chi^0}, \chi^0, h )		&=		- \frac{i}{\nu}		\left(	\Delta^{(\mathrm{t})}_{-\frac{1}{2}}		- 2 \Re \,\delta_0		\right)
			=	- \frac{i}{4} \nu	\left(	\frac{\lambda_c}{2 \Lambda}	- n \Re \frac{\lambda_0}{\Lambda}	\right)		\\	\nonumber
	C ( \overline{\widetilde{\chi}^0}, \widetilde{\chi}^0, h )		&=		- \frac{i}{\nu}		\left(	\Delta^{(\mathrm{t})}_{-\frac{1}{2}}		+ 2 \Re \,\delta_0		\right)
			=	- \frac{i}{4} \nu	\left(	\frac{\lambda_c}{2 \Lambda}	+ n \Re \frac{\lambda_0}{\Lambda}	\right)				\\	\nonumber
	C ( \overline{\chi^0}, \widetilde{\chi}^0, h )		&=		- C ( \overline{\widetilde{\chi}^0}, \chi^0, h )		=		\frac{4}{\nu}	\Im \,\delta_0
			=	\frac{n}{2}	\,\nu		\,\Im \frac{\lambda_0}{\Lambda}
\end{align}

Quartic non-renormalisable interactions of two Higgs bosons and two DM particles can be cast as:
\begin{align}
	C ( \overline{\chi}^\frac{n}{2}, \chi^\frac{n}{2}, h, h )		&=		- \frac{i}{\nu^2}	\,\Delta^{(\mathrm{t})}_{\frac{1}{2} (n-1)}
			=		\frac{i}{8}	\left( n-1 \right)		\frac{\lambda_c}{\Lambda}			\\	\nonumber
	C ( \overline{\chi}^q_1, \chi^q_1, h, h )		&=		- \frac{i}{\nu^2}	\left(	\Delta^{(\mathrm{t})}_{q-\frac{1}{2}} 	-2 |\delta_q| \sin \phi_q	\right)
			=		\frac{i}{4}	\left[	\left( q -\frac{1}{2} \right)	\frac{\lambda_c}{\Lambda}		+ \sqrt{n^2 -4q^2}	\sin \phi_q	\left| \frac{\lambda_0}{\Lambda} \right|	\right]	\\	\nonumber
	C ( \overline{\chi}^q_2, \chi^q_2, h, h )		&=		- \frac{i}{\nu^2}	\left(	\Delta^{(\mathrm{t})}_{q-\frac{1}{2}} 	+2 |\delta_q| \sin \phi_q	\right)
			=		\frac{i}{4}	\left[	\left( q -\frac{1}{2} \right)	\frac{\lambda_c}{\Lambda}		- \sqrt{n^2 -4q^2}		\sin \phi_q	\left| \frac{\lambda_0}{\Lambda} \right|	\right]	\\	\nonumber
	C ( \overline{\chi}^q_2, \chi^q_1, h, h )		&=			- C ( \overline{\chi}^q_1, \chi^q_2, h, h )	=	2 \frac{i}{\nu^2}	|\delta_q|	\cos \phi_q
			=	\frac{i}{4}	\sqrt{n^2 -4q^2}	\cos \phi_q	\left| \frac{\lambda_0}{\Lambda} \right|		\\	\nonumber
	C ( \overline{\chi^0}, \chi^0, h, h )		&=		- \frac{i}{\nu^2}		\left(	\frac{1}{2}	\Delta^{(\mathrm{t})}_{-\frac{1}{2}}		- \Re \,\delta_0		\right)
			=	- \frac{i}{8} 		\left(	\frac{\lambda_c}{2 \Lambda}	- n \Re \frac{\lambda_0}{\Lambda}	\right)		\\	\nonumber
	C ( \overline{\widetilde{\chi}^0}, \widetilde{\chi}^0, h, h )		&=		- \frac{i}{\nu^2}		\left(	\frac{1}{2}	\Delta^{(\mathrm{t})}_{-\frac{1}{2}}		+ \Re \,\delta_0		\right)
			=	- \frac{i}{8} 		\left(	\frac{\lambda_c}{2 \Lambda}	+ n \Re \frac{\lambda_0}{\Lambda}	\right)				\\	\nonumber
	C ( \overline{\chi^0}, \widetilde{\chi}^0, h, h )		&=		- C ( \overline{\widetilde{\chi}^0}, \chi^0, h, h )		=		\frac{2}{\nu}	\Im \,\delta_0
			=	\frac{n}{4}		\,\Im \frac{\lambda_0}{\Lambda}
\end{align}


\section{Real Representation}
\label{app:R}

By definition a \emph{real} representation equals its complex conjugate representation, so dark matter in a real module of electroweak $SU(2)\times U(1)_y$ sector, should have zero hypercharge $y=0$. Note that since $t^{(3)}=q-y$, for the neutral component (the actual DM) weak isospin vanishes $t^{(3)}_0=0$. Therefore only odd dimensional multiplets of real dark matter exist \cite{EWDM_Essig}. A generic real dark multiplet has the form:
\begin{equation}
	X=\left( \chi^t, \ldots \chi^q,\dots \chi^0, \ldots \chi^{-t} \right)^T
\end{equation}

The $\frac{1}{2}(n+1)$ neutral component $\chi^0$ which is the stable DM particle is a Majorana fermion, while the other charged components $\chi^q$ are Dirac. Interaction of dark sector with SM is described by the Lagrangian:
\begin{equation}
\label{eq:L_R}
	\mathcal{L}_D= \frac{1}{2} \bar{X} \left( i \slashed{D} - m \right) X
\end{equation}

where covariant derivative reads 
$D_\mu = \partial_\mu 	+i e\,QA_\mu 	+i g_w c_w Q Z_\mu 	+  \frac{i}{\sqrt{2}}  g_w	\left( W^-_\mu T^- +W^+_\mu T^+ \right)$.

The mass splitting between the degenerate neutral and charged components is caused by the dark sector self energies via coupling to Z and W gauge bosons at one loop order. This radiative mass splitting takes the form~\cite{MDM}:
\begin{equation}
	\label{eq:Dm_R}
	\Delta_q^\mathbb{R}	\equiv	m_q -m_0		=	q^2 \Delta
\end{equation}

where $m_q$ is mass of the charged field $\chi^q$ and $m_0=m$ is dark matter $\chi^0$ mass. The splitting between the neutral and $\chi^\pm$ components is given by $\Delta = \alpha_w m_w \sin^2 (\theta_w/2) \approx$ 166 MeV \cite{dm_R}.

Interaction Lagrangian of DM particle $\chi^0$ in the real case is then given by:
\begin{subequations}
\begin{align}
	\mathcal{L}_D \supset 	&\ \frac{g_w}{4\sqrt{2}}	\sum_{q=-t+1}^t 	\sqrt{n^2 -(2q-1)^2}	\ \overline{\chi^{q-1}} \gamma^\mu \chi^q \ W^-_\mu	+ \ cc
	\label{eq:DM0_int}\\
						&+	\frac{1}{2}   g_w c_w q 	 \sum_{q=-t}^t 		 \overline{\chi^q} \gamma^\mu \chi^q \ Z	_\mu 	+ \ 
							\frac{1}{2}   eq	\sum_{q=-t}^t 		\overline{\chi^q} \gamma^\mu \chi^q \ A	_\mu													
						\label{eq:DMc_int}
\end{align}
\end{subequations}

where the first line \ref{eq:DM0_int} includes the dark matter-standard model couplings of the form \\
$g_w \sqrt{n^2 -1} / 4 \sqrt{2}		\left(		\overline{\chi^-} \gamma^\mu \chi^0 	+ \overline{\chi^0} \gamma^\mu \chi^+	\right) \ W^-_\mu +$ cc; 
and the second line \ref{eq:DMc_int} only contains the charged dark fields and SM interactions.

The main observation is the lack of any tree-level coupling of DM to Z-boson
\footnote{This is an expected result as real DM is a Majorana fermion. Recall that any vector bilinear of the form $\bar{\chi}^M \gamma^\mu \chi^M$ vanishes \cite{Zportal}.}
 It means that the constraints form DD experiments are relaxed in the case of real non-chiral DM. There is nevertheless a coupling to neutral weak gauge induced by one-loop corrections.

In the following sections, we provide the results for two fermionic real models of electroweak DM:


\subsection{Real Triplet}
\label{sec:R3}

\begin{figure} [t] 				
	\begin{subfigure}{.5\linewidth}
		\centering
		\includegraphics[width=\linewidth]{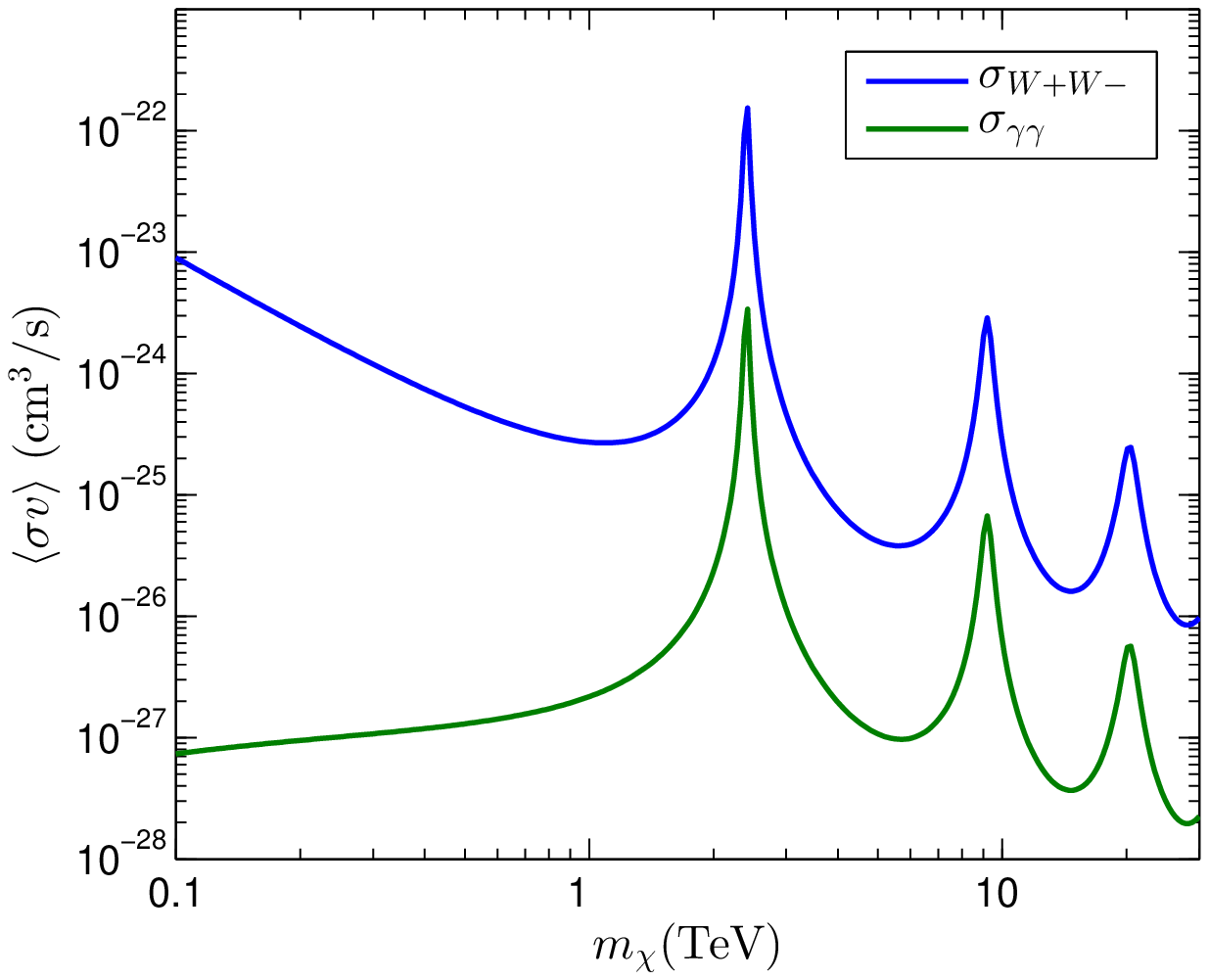}
		\subcaption{}
		\label{fig:R3:Xnv_m}
	\end{subfigure}%
	\begin{subfigure}{.5\linewidth}
		\centering
		\includegraphics[width=\linewidth]{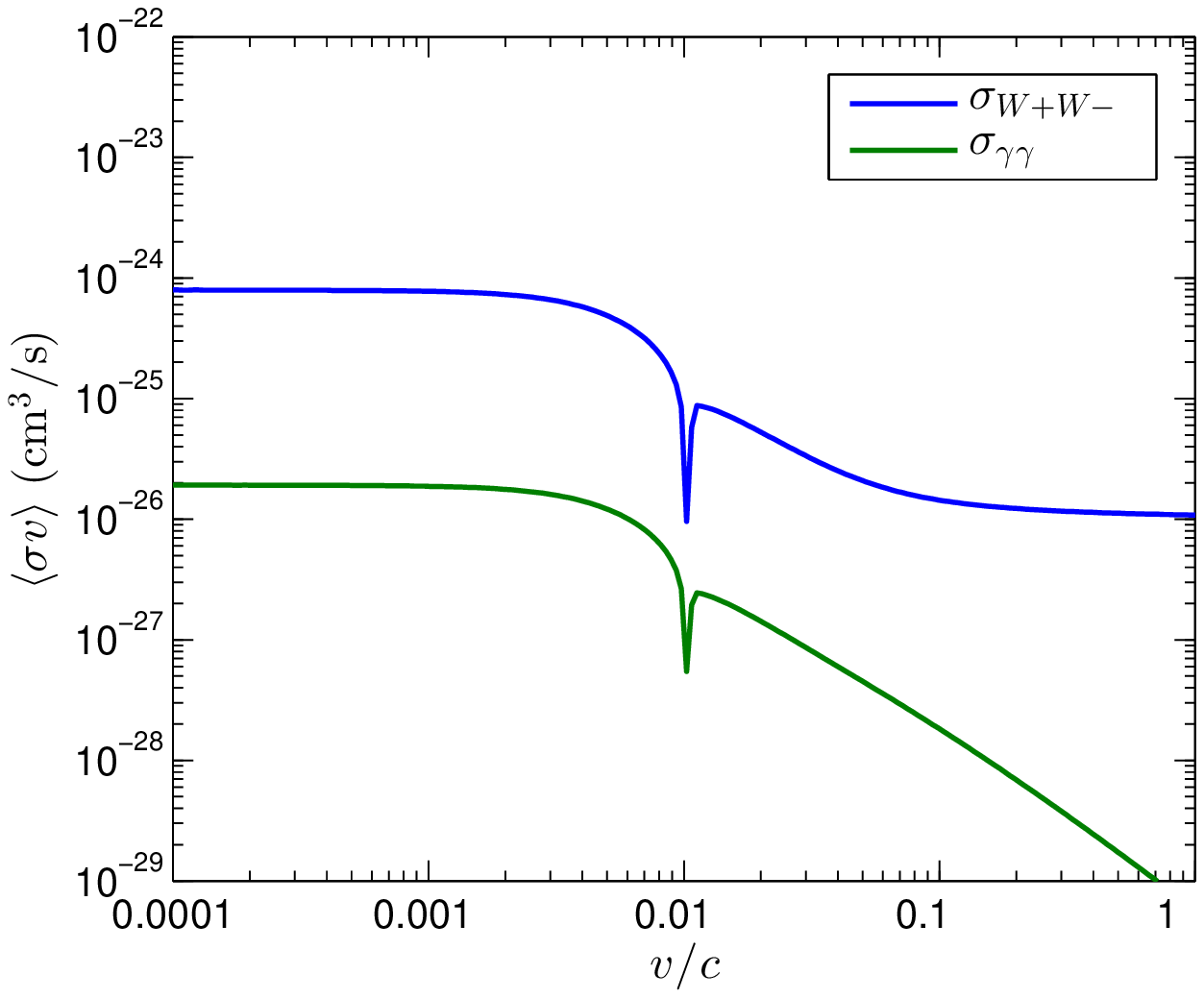}
		\subcaption{}
		\label{fig:R3:Xnv_v}
	\end{subfigure}
	\caption{Annihilation cross-section of $\chi^0 \chi^0$ state into independent modes $W^+W^-$ (blue) and $\gamma\gamma$ (red) in real triplet model, as function of mass at $v=10^{-3}c$  (\ref{fig:R3:Xnv_m}), and of DM velocity at $m_0 = 2.85$ TeV (\ref{fig:R3:Xnv_v}).}
	\label{fig:R3:Xnv}
\end{figure}

The EWDM triplet takes the form:
\begin{equation}
	X = \begin{pmatrix}		\chi^+	\\ 	\chi^0	\\	\chi^- 	\end{pmatrix}
\end{equation}

As an example, in the MSSM, in the limit where the lightest neutralino is a pure \emph{wino}, a similar triplet can be composed from wino-like Majorana sparticle and other charginos \cite{MSSM_phny}. For instance, this happens in scenarios of anomaly mediated supersymmetry breaking \cite{MSSM_anomaly}, models with heavy scalars \cite{MSSM_scalar} or in Split Supersymmetry \cite{MSSM_Split}.

The real triplet is well studied in the context of supersymmetric dark matter, and we present our results to calibrate them with other published works.

The following pair basis expands the five available sectors:
\begin{subequations}
\begin{align}
	\label{eq:R3_Q0}
	& \Psi^{S=0}_{Q=0} = 	\left(  \chi^- \chi^+, \ 	\frac{1}{\sqrt{2}} \, \chi^0 \chi^0  \right)^T 	\,,	\\
	& \Psi^{S=1}_{Q=0} = 	\chi^- \chi^+ 	\,,	\\
	& \Psi_{Q=1} =   		\chi^0 \chi^+ 	\,,	\\
	& \Psi^{S=0}_{Q=2} =	\frac{1}{\sqrt{2}} \,	\chi^+ \chi^+ 	\,.	
\end{align}
\end{subequations}

The neutral state in \ref{eq:R3_Q0} is formed from identical $\chi^0$ particles, and gets an extra normalisation factor.

Gauge interactions and mass differences are included in the following potentials:
\begin{equation}
	V^{S=0}_{Q=0} = 	
	\begin{pmatrix}		
		2 \Delta -\mathcal{V} 		& -\sqrt{2} \,\mathcal{W}		\\
		-\sqrt{2} \,\mathcal{W}		&0
	\end{pmatrix}	\,,
	\qquad
	V^{S=0}_{Q=1} = 	\Delta +\mathcal{W} \,,
	\qquad
	V^{S=0}_{Q=2} = 	2 \Delta +\mathcal{V}  \,,
\end{equation}

\begin{equation}
	V^{S=1}_{Q=0} = 	2 \Delta -\mathcal{V} \,,
	\qquad
	V^{S=1}_{Q=1} = 	\Delta -\mathcal{W} \,.
\end{equation}

where $\mathcal{V} \equiv 	\mathcal{A} +4 c_w^4 \mathcal{Z} =		\alpha/r	+\alpha_w c_w^2 \,e^{-M_z r}/ \,r$.

Tree-level annihilation cross-sections are obtained from:
\begin{equation}
	\Gamma^{S=0}_{Q=0} =	\frac{\pi\alpha_w^2}{2 m_0^2}	
		\begin{pmatrix}		 
			3			& \sqrt{2}		\\
			\sqrt{2}		& 2
		\end{pmatrix}	\,,
	\qquad
	\begin{array}{l}
		{G_{\gamma\gamma}}^{S=0}_{Q=0} =	2 \, \alpha \left( 1, \ 0 \right)^T	,	\\		
		{G_{WW}}^{S=0}_{Q=0}	= 		\alpha_w	\left( \sqrt{2},	\ 2  \right)^T,		
	\end{array}
\end{equation}

\begin{equation}
	\Gamma^{S=0}_{Q=1} =		\Gamma^{S=0}_{Q=2} =	\frac{\pi\alpha_w^2}{2 m_0^2} \,,
	\qquad
	\Gamma^{S=1}_{Q=0} =		\Gamma^{S=1}_{Q=1} =	\frac{25 \pi\alpha_w^2}{4 m_0^2} \,.
\end{equation}	
	
Figure \ref{fig:R3:Xnv_m} shows the annihilation cross-section of $\chi^0 \chi^0$ pair into two independent channels $W^+W^-$ and $\gamma\gamma$ as a function of DM mass. It is clear that Sommerfeld phenomenon has enhanced the value in the high mass part by a few orders of magnitude. In addition presence of the peaks in the spectrum is related to the binding energies required to form bound sates of EWDM.

The changes in annihilation cross-section with respect to DM velocity is illustrated in figure \ref{fig:R3:Xnv_v}. As expected, the annihilation rate increases as the velocity decreases in the relativistic region until it reaches to a steady value at $v \approx m_w/m_0$.


\subsection{Real Quintuplet}
\label{sec:R5}

\begin{figure} [t] 				
	\begin{subfigure}{.5\linewidth}
		\centering
		\includegraphics[width=\linewidth]{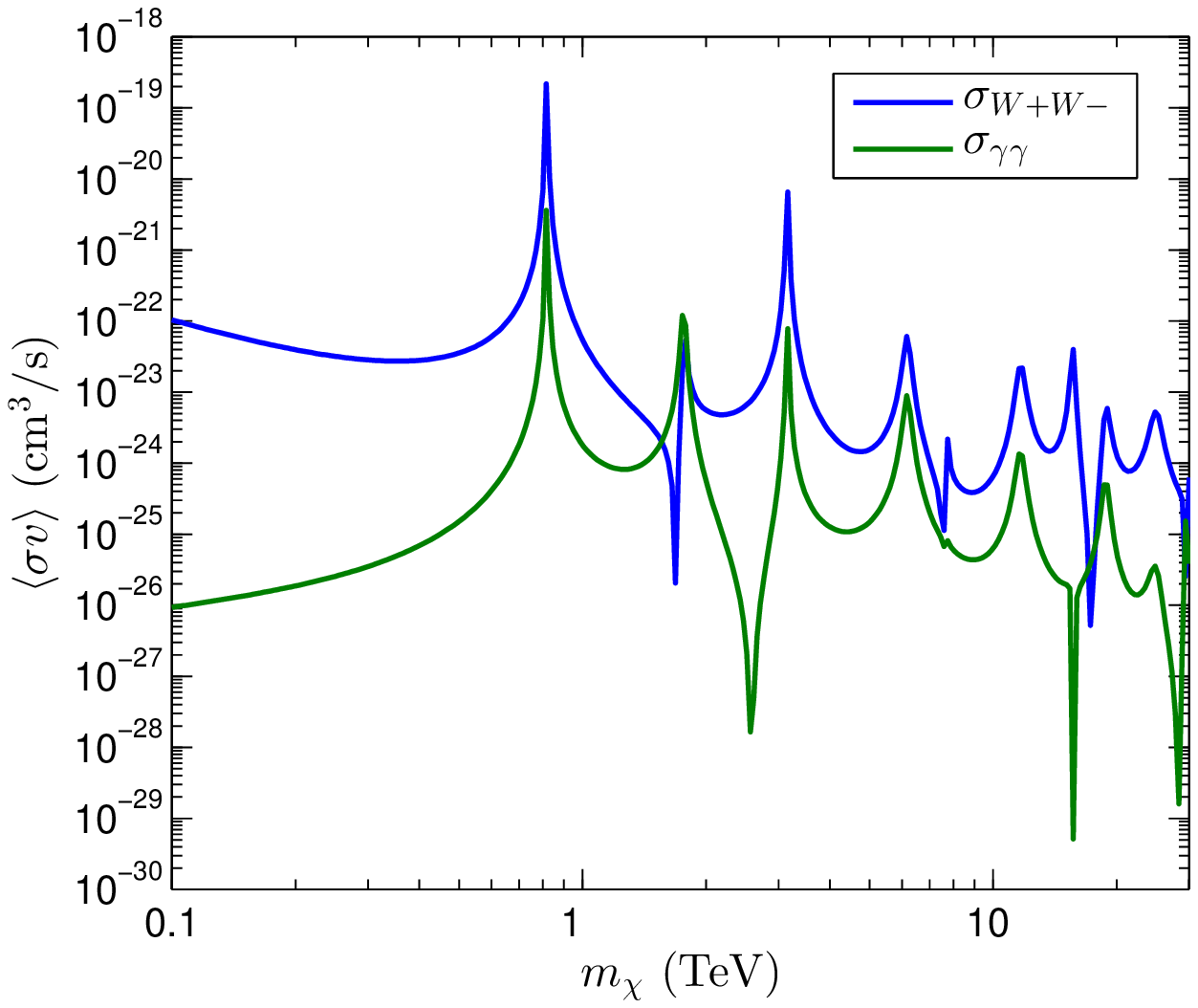}
		\subcaption{}
		\label{fig:R5:Xnv_m}
	\end{subfigure}%
	\begin{subfigure}{.5\linewidth}
		\centering
		\includegraphics[width=\linewidth]{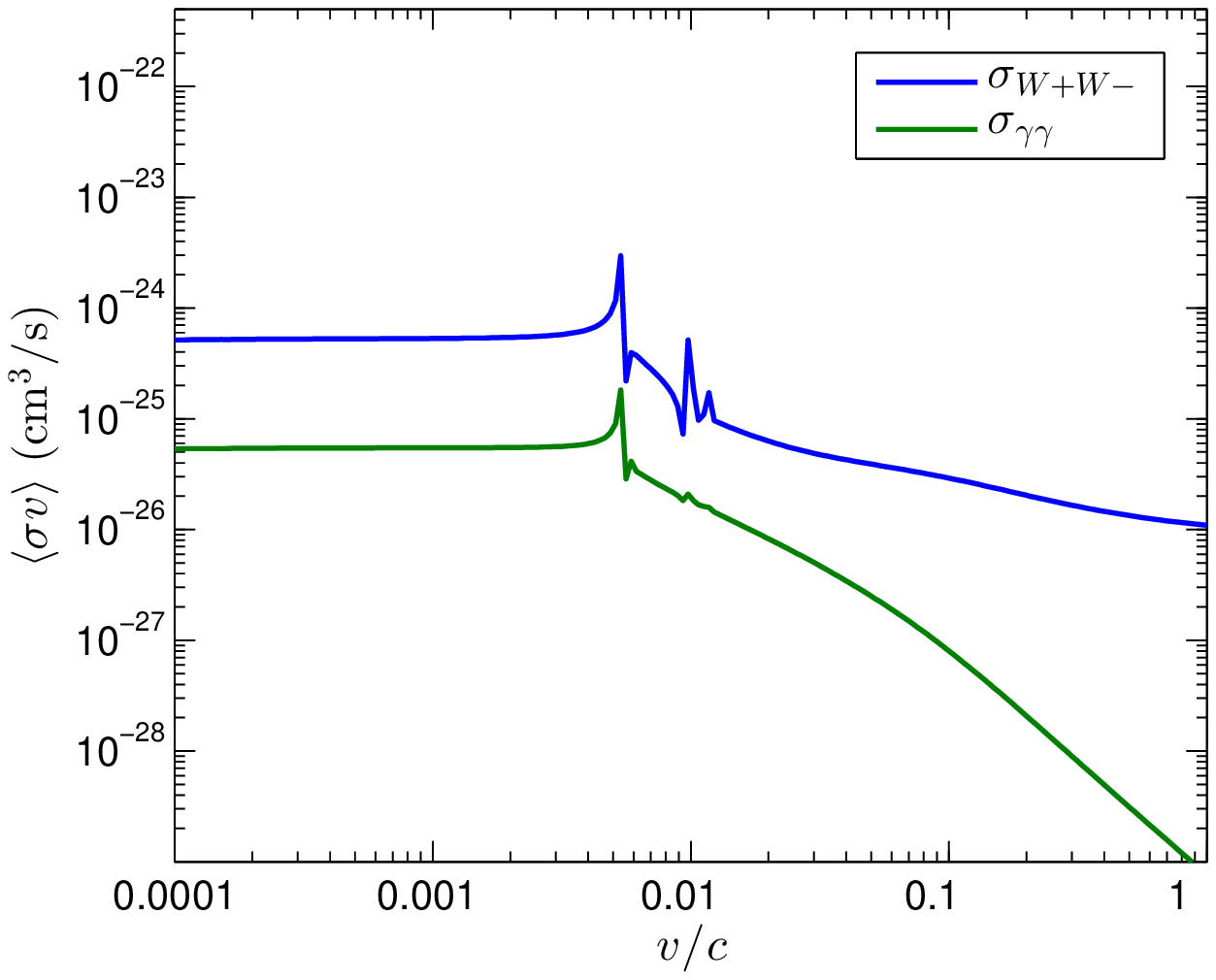}
		\subcaption{}
		\label{fig:R5:Xnv_v}
	\end{subfigure}
	\caption{Annihilation spectrum of the real quintuplet dark matter pair $\chi^0 \chi^0$ into $W^+W^-$ (blue) and $\gamma\gamma$ (red) vectors, with respect to DM mass at $v=10^{-3}c$  (\ref{fig:R5:Xnv_m}), and to the DM velocity at freeze-out mass of $m_0 = 9.4$ TeV (\ref{fig:R5:Xnv_v}).}
	\label{fig:R5:Xnv}
\end{figure}

The real quintet in terms of components can be written as:
\begin{equation}
	\begin{pmatrix}
	\chi^{++}\\	\chi^+\\	\chi^0\\	\chi^-\\	\chi^{--}	
	\end{pmatrix}
\end{equation}

This multiplet is the only fermionic DM candidate in \emph{Minimal Dark Matter} (MDM) framework. Due to gauge invariance and matter content of the SM, there is no Yukawa term to couple dark matter to Higgs and other SM particles. Even considering the effect of non-renormalisable operators, as a result of this accidental symmetry, MDM candidate has no open decay channel, and thus automatically remains stable without a discrete $Z2$ symmetry \cite{MDM, MDM_2009}.

In what follows, we review the theoretical results and phenomenological predictions for MDM 5let. 

The following basis is defined for different sectors in this representation:
\begin{subequations}
\begin{align}
	\label{eq:R5_Q0}
	& \Psi^{S=0}_{Q=0} =	\left(   \chi^{--} \chi^{++}, \ 	\chi^- \chi^+, \ 		\frac{1}{\sqrt{2}} \, 	\chi^0 \chi^0  \right)^T ,	\\
	& \Psi^{S=1}_{Q=0} =	\left(   \chi^{--} \chi^{++}, \ 	\chi^- \chi^+		\right)^T	,		\\
	& \Psi_{Q=1} =   		\left(	  \chi^- \chi^{++}, \ 		\chi^0 \chi^+ 	\right)^T	,		\\
	& \Psi^{S=0}_{Q=2} =   \left(	  \chi^0 \chi^{++}, \ 		\frac{1}{\sqrt{2}} \,	\chi^+ \chi^+ 	\right)^T 	.	
\end{align}
\end{subequations}

The potentials between different states read:
\begin{equation}
	V^{S=0}_{Q=0} = 	
	\begin{pmatrix}		
		8 \Delta -4 \mathcal{V} 		& -2\mathcal{W}				&0	\\
			-2\mathcal{W}			& 2\Delta -\mathcal{V}		& -3\sqrt{2} \,\mathcal{W}		\\
				0				& -3\sqrt{2} \,\mathcal{W}			&0
	\end{pmatrix}	\,,
\end{equation}

\begin{equation}
	V^{S=0}_{Q=1} = 	
	\begin{pmatrix}		
		5 \Delta -2 \mathcal{V}			& -\sqrt{6} \mathcal{W}			\\
		-\sqrt{6} \mathcal{W}			& \Delta +3 \mathcal{W}
	\end{pmatrix}	\,,
	\qquad
	V^{S=0}_{Q=2} = 	
	\begin{pmatrix}		
			4 \Delta  					& -2\sqrt{3} \mathcal{W}			\\
		-2\sqrt{3}  \mathcal{W}			& 2 \Delta +\mathcal{V} 
	\end{pmatrix}	\,,
\end{equation}

\begin{equation}
	V^{S=1}_{Q=0} = 	
	\begin{pmatrix}		
		8 \Delta -4 \mathcal{V} 		& -2\mathcal{W}			\\
			-2\mathcal{W}			& 2\Delta -\mathcal{V}
	\end{pmatrix}	\,,
	\qquad
	V^{S=1}_{Q=1} = 	
	\begin{pmatrix}		
		5 \Delta -2 \mathcal{V}			& -\sqrt{6} \mathcal{W}			\\
		-\sqrt{6} \mathcal{W}			& \Delta -3 \mathcal{W}
	\end{pmatrix}	\,.
\end{equation}

The s-wave tree-level annihilation cross section can be written as:		
\begin{equation}
	\Gamma^{S=0}_{Q=0} =	\frac{3\pi\alpha_w^2}{2 m_0^2}	
		\begin{pmatrix}		 
				12			& 6		& 2 \sqrt{2}		\\
				6			& 9		& 5 \sqrt{2}		\\
			 2 \sqrt{2}		& 5 \sqrt{2}		& 6
		\end{pmatrix}	\,,
	\qquad
	\begin{array}{l}
		{G_{\gamma\gamma}}^{S=0}_{Q=0} =	2 \, \alpha \left( 4,	\ 1,	 \ 0 \right)^T	,	\\		
		{G_{WW}}^{S=0}_{Q=0}	= 		\sqrt{2}	\alpha_w	\left( 2,	\ 5,	3\sqrt{2},	\right)^T,		
	\end{array}
\end{equation}	

\begin{equation}
	\Gamma^{S=0}_{Q=1} =	\frac{3\pi\alpha_w^2}{2 m_0^2}	
		\begin{pmatrix}		 
				6		& \sqrt{6}		\\
			\sqrt{6}			& 1		
		\end{pmatrix}	\,,
	\qquad
	\Gamma^{S=0}_{Q=2} =	\frac{3\pi\alpha_w^2}{2 m_0^2}	
		\begin{pmatrix}		 
				4		& 2\sqrt{3}		\\
			2\sqrt{3}			& 3		
		\end{pmatrix}	\,,	
\end{equation}	
			
\begin{equation}
	\Gamma^{S=0}_{Q=1} =	\frac{25 \pi\alpha_w^2}{4 m_0^2}	
		\begin{pmatrix}		 
			4		& 2		\\
			2		& 1		
		\end{pmatrix}	\,,
	\qquad
	\Gamma^{S=0}_{Q=2} =	\frac{25 \pi\alpha_w^2}{4 m_0^2}	
		\begin{pmatrix}		 
				2		& \sqrt{6}		\\
			\sqrt{6}			& 3		
		\end{pmatrix}	\,.	
\end{equation}	

Figure \ref{fig:R5:Xnv_m} shows the predicted annihilation cross-section of $\chi^0\chi^0$ into $W^-W^+$ and $\gamma\gamma$ final states as a function of DM mass. As usual, Sommerfeld phenomenon causes peaks and dips in the spectrum due to dark matter bound state formation and repulsive potentials in Ramsauer-Townsend effect, respectively. 

The right panel \ref{fig:R5:Xnv_v} depicts  $\chi^0\chi^0$ annihilation with respect to the DM velocity at thermal mass. As expected, non-perturbative effects are dominant in the low velocity regime.


\nocite{apsrev42Control}
\bibliography{EWDM_ref, EWDMnotes}		

\end{document}